\documentclass[10pt,aps,prc,floatfix,twocolumn,nofootinbib]{revtex4-1}

\usepackage[pdfpagelabels]{hyperref}
\usepackage{xcolor}

\definecolor{linkcolor}{rgb}{0,0,0.40} 
\hypersetup{%
    pdfsubject=Paper,
    pdfkeywords={nuclear physics} {Bayesian} {chiral EFT} {Compton} {experimental design},
    unicode = true, 
    breaklinks = true,
    colorlinks = true,
    linkcolor = linkcolor,
    citecolor = linkcolor,
    menucolor = linkcolor,
    urlcolor = linkcolor
}

\usepackage[labelfont={small},subrefformat=parens,caption=false]{subfig}
\usepackage{graphicx,amsmath,amssymb,bm}
\usepackage{amsfonts}
\usepackage[utf8]{inputenc}
\usepackage{verbatim}
\usepackage{float}

\usepackage{cancel}
\usepackage{multirow}

\usepackage{array}
\usepackage{xparse}

\usepackage{physics}
\usepackage{dsfont}
\usepackage{xspace}

\usepackage{cellspace}
\usepackage{dcolumn}
\newcolumntype{d}[1]{D{.}{.}{#1}}

\graphicspath{{./figures-article-compressed/}}

\makeatletter
\newcommand\newsubcommand[3]{\newcommand#1{#2\sc@sub{#3}}}
\def\sc@sub#1{\def\sc@thesub{#1}\@ifnextchar_{\sc@mergesubs}{_{\sc@thesub}}}
\def\sc@mergesubs_#1{_{\sc@thesub#1}}

\newcommand\newsupcommand[3]{\newcommand#1{#2\sc@sup{#3}}}
\def\sc@sup#1{\def\sc@thesup{#1}\@ifnextchar^{\sc@mergesups}{^{\sc@thesup}}}
\def\sc@mergesups^#1{^{\sc@thesup#1}}
\makeatother

\DeclareMathAlphabet{\mathbcal}{OMS}{cmsy}{b}{n}

\newcommand{\calO}{\mathcal{O}}

\DeclareMathOperator{\E}{\mathbb{E}}

\newcommand{\kernel}{\kappa}

\newcommand{\cbar}{\bar c}
\newcommand{\sdth}{\cbar}

\newcommand{\iid}{\text{i.i.d.}}

\newcommand{\discrcorr}[1]{R_{\delta #1}}

\newcommand{\MeV}{\, \text{MeV}}

\newcommand{\NkLO}[1]{\ensuremath{\mathrm{N}^{#1}\mathrm{LO}}\xspace}

\newcommand{\param}{\boldsymbol{\theta}}

\DeclareMathOperator{\GP}{\mathcal{GP}}

\newcommand{\ordervec}{\vec}
\newcommand{\inputdimvec}{}
\newcommand{\inputvec}{\mathbf}

\newsubcommand{\ckvec}{\ordervec{c}}{k}

\newsubcommand{\bkvec}{\ordervec{b}}{k}
\newcommand{\kinparvec}{\inputdimvec{x}}
\newcommand{\kinparvecset}{\inputdimvec{\inputvec{x}}}

\newsubcommand{\ckvecset}{\ordervec{\inputvec{c}}}{k}

\newsubcommand{\ckvecapprox}{\mathbf{c}'}{k}
\newsubcommand{\ckvecapproxset}{\mathbf{C}'}{k}

\newsubcommand{\bkvecapprox}{\mathbf{b}'}{k}
\newsubcommand{\bkvecset}{\mathbf{B}}{k}
\newsubcommand{\bkvecapproxset}{\mathbf{B}'}{k}

\newcommand{\genobs}{y}

\newsubcommand{\genobsvec}{\ordervec{\genobs}}{k}
\newsubcommand{\genobsvecset}{\ordervec{\inputvec{\genobs}}}{k}
\newcommand{\genobsset}{\inputvec{\genobs}}

\newcommand{\genobsexp}{\genobs_{\textup{exp}}}        

\newcommand{\genobsth}{\genobs_{\textup{th}}}          

\newcommand{\lecs}{\vec{a}}

\newsubcommand{\akvec}{\mathbf{a}}{k}

\newsubcommand{\akvecapprox}{\mathbf{a}'}{k}
\newsubcommand{\akvecset}{\mathbf{A}}{k}
\newsubcommand{\akvecapproxset}{\mathbf{A}'}{k}

{}  

\DeclareMathOperator{\pr}{pr} 
\newcommand{\given}{\,|\,}  

\newcommand{\normal}{\mathcal{N}}

\newcommand{\trans}{\intercal}

\newcommand{\chiEFT}{$\chi$EFT}

\newcommand{\genobsref}{\ensuremath{y_{\mathrm{ref}}}}

\def\diffd{\mathrm{d}}  

\DeclareDocumentCommand\differential{ o g d() }{ 
    \IfNoValueTF{#2}{
        \IfNoValueTF{#3}
            {\diffd\IfNoValueTF{#1}{}{^{#1}}}
            {\mathinner{\diffd\IfNoValueTF{#1}{}{^{#1}}\argopen(#3\argclose)}}
        }
        {\mathinner{\diffd\IfNoValueTF{#1}{}{^{#1}}#2} \IfNoValueTF{#3}{}{(#3)}}
    }
\DeclareDocumentCommand\dd{}{\differential} 

\newcommand{\pathd}{\mathcal{D}}  

\DeclareDocumentCommand\pathdifferential{ o g d() }{ 
    \IfNoValueTF{#2}{
        \IfNoValueTF{#3}
            {\pathd\IfNoValueTF{#1}{}{^{#1}}}
            {\mathinner{\pathd\IfNoValueTF{#1}{}{^{#1}}\argopen(#3\argclose)}}
        }
        {\mathinner{\pathd\IfNoValueTF{#1}{}{^{#1}}#2} \IfNoValueTF{#3}{}{(#3)}}
    }

\newcommand{\design}{\mathbf{d}}
\DeclareMathOperator*{\argmax}{arg\,max}
\newcommand{\shrinkage}{\mathcal{S}}
\newcommand{\omegalab}{\omega_{\text{lab}}}
\newcommand{\thetalab}{\theta_{\text{lab}}}

\newcommand{\dsigma}{\dd{\sigma}}
\newcommand{\diffcs}{\dsigma}  

\newcommand{\eg}{\textit{e.g.}\xspace}
\newcommand{\ie}{\textit{i.e.}\xspace}
 \newcommand{\verifyvalue}[1]{#1}

\newcommand{\mpi}{\ensuremath{m_\pi}}

\newcommand{\HIGS}{HI$\gamma$S}

\def\AnswerYes{y}
\def\ShowLabelsVersion{n}         
\ifx\ShowLabelsVersion\AnswerYes
   \usepackage[color]{showkeys} 
   \definecolor{refkey}{gray}{.5}   
   \definecolor{labelkey}{gray}{.5} 
   
\fi

\begin{document}

\title{
Designing Optimal Experiments: 
\texorpdfstring{\\}{} An Application to Proton Compton Scattering
}

\author{J.~A.~Melendez}
\email{melendez.27@osu.edu}
\affiliation{Department of Physics, The Ohio State University, Columbus, OH 43210, USA}

\author{R.~J.~Furnstahl}
\email{furnstahl.1@osu.edu}
\affiliation{Department of Physics, The Ohio State University, Columbus, OH 43210, USA}

\author{H.~W.~Grie\3hammer}
\email{hgrie@gwu.edu}
\affiliation{Institute for Nuclear Studies, Department of Physics, The George Washington University, Washington, DC 20052, USA}
\altaffiliation[Permanent address]{}
\affiliation{Department of Physics,  Duke University, Box 90305, Durham NC 27708, USA}
\affiliation{High Intensity Gamma-Ray Source, Triangle Universities
    Nuclear Laboratories, Box 90308, Durham NC 27708, USA}

\author{J.~A.~McGovern}
\email{judith.mcgovern@manchester.ac.uk}
\affiliation{School of Physics and Astronomy, The University of Manchester, Manchester M13 9PL, UK}

\author{D.~R.~Phillips}
\email{phillid1@ohio.edu}
\affiliation{Department of Physics and Astronomy and Institute of Nuclear and Particle Physics, Ohio University, Athens, OH 45701, USA}

\author{M.~T.~Pratola}
\email{mpratola@stat.osu.edu}
\affiliation{Department of Statistics, The Ohio State University, Columbus, OH 43210, USA}
\date{\today}

\begin{abstract}
  Interpreting measurements requires a physical theory, but the theory's
  accuracy may vary across the experimental domain.  To optimize experimental
  design, and so to ensure that the substantial resources necessary for modern
  experiments are focused on acquiring the most valuable data, both the theory
  uncertainty and the expected pattern of experimental errors must be
  considered.  We develop a Bayesian approach to this problem, and apply it to
  the example of proton Compton scattering.  Chiral Effective Field Theory
  ($\chi$EFT) predicts the functional form of the scattering amplitude for
  this reaction, so that the electromagnetic polarizabilities of the nucleon
  can be inferred from data.  With increasing photon energy, both experimental
  rates and sensitivities to polarizabilities increase, but the accuracy of
  $\chi$EFT decreases.  Our physics-based model of $\chi$EFT truncation errors
  is combined with present knowledge of the polarizabilities and reasonable
  assumptions about experimental capabilities at HI$\gamma$S and MAMI to
  assess the information gain from measuring specific observables at specific
  kinematics, \ie~to determine the relative amount by which new data
  are apt to shrink uncertainties.  The strongest gains would likely come from
  new data on the spin observables $\Sigma_{2x}$ and $\Sigma_{2x^\prime}$ at
  $\omega\simeq140$ to $200$\,MeV and $40^\circ$ to $120^\circ$. These would
  tightly constrain $\gamma_{E1E1}-\gamma_{E1M2}$. New data on the
  differential cross section between $100$ and $200$\,MeV and over a wide
  angle range will substantially improve constraints on
  $\alpha_{E1}-\beta_{M1}$, $\gamma_\pi$ and $\gamma_{M1M1}-\gamma_{M1E2}$.
  Good signals also exist around $160$\,MeV for $\Sigma_3$ and
  $\Sigma_{2z^\prime}$. Such data will be pivotal in the continuing quest to
  pin down the scalar polarizabilities and refine understanding of the spin
  polarizabilities.
\end{abstract}

\maketitle

\section{Introduction}

Six low-energy parameters known as polarizabilities characterize the response
of the nucleon to low-frequency light: the electric and magnetic dipole
polarizabilities $\alpha_{E1}$ and $\beta_{M1}$, and four spin
polarizabilities $\gamma_i$; see,
\eg,~refs.~\cite{Griesshammer:2012we,Holstein:2013kia, Hagelstein:2015egb} for
recent reviews.  Despite their fundamental importance to understanding the
proton and neutron, the value of only one combination is known with better
than 2\% accuracy, with current uncertainties for the rest varying from 10\%
to more than $100$\%.  Most recent values and uncertainties are collected in
ref.~\cite{Griesshammer:2015ahu} and references therein, and summarized in
table~\ref{tab:polarizability_info} below.  Recent advances in Chiral
Effective Field Theory (\chiEFT) have enabled precise quantitative predictions
of Compton scattering that take the polarizabilities as
inputs~\cite{Griesshammer:2012we,McGovern:2012ew,Griesshammer:2017txw,
  Margaryan:2018opu, Griesshammer:2013vga}.%
\footnote{In the \chiEFT\ we are using, the lowest-lying nucleonic resonance,
  the $\Delta(1232)$, is retained as an explicit degree of freedom.}  This new
ability to precisely trace the impact of these fundamental nucleon-structure
constants on experimental observables is opportune.  It comes at a time when
photon facilities of unprecedented luminosity and sensitivity are now
available.  These concurrent developments have inspired new experimental
campaigns to refine knowledge of nucleon polarizabilities; see, for example,
recent overviews in refs.~\cite{Martel:2019tgp, Martel:2017pln, Huber:2015uza,
  Ahmed:2020hux, Weller:2009zza}.

But not all measurements are created equal, and beam time is not cheap.  In
Nuclear Physics, as in many other advanced disciplines, the costs of running
an experiment include not only the workforce, time and money invested, but
also the opportunity cost of measurements that could have been carried out
with those same resources but were not.  Thus, when planning an experiment, it
is important to consider which data are most likely to provide the largest
information gain.

How, then, does one assess the impact of measurements that have yet to be
made?  One possibility is to simulate various possible experimental scenarios
and compute the extent to which each improves constraints on theory
parameters.  For example, ref.~\cite{Catacora-Rios:2019goa} recently
investigated the ability of data on proton- and neutron elastic scattering
from nuclei to constrain optical-model parameters.  In the context of Compton
scattering from the proton and neutron, ref.~\cite{Griesshammer:2017txw}
assessed the sensitivity of many different observables to the
polarizabilities, but provided only an initial theory perspective on those
sensitivities.  Neither of these papers used a single quantitative measure to
choose between competing experimental designs. Such a measure should assess
competing experimental designs in light of existing data and the feasibility
of new data, and include a rigorous assessment of theory uncertainties.

Here, we argue for the framework of Bayesian statistics---Bayesian
experimental design, in particular---as an enlightened way forward.
Experimental design, in the context of statistics, provides insight into the
allocation of scarce resources that have alternative uses.  This is a highly
practical field of study, with applications including engineering, biology,
environmental processes, computer experiments, and
psychology~\cite{chaloner1995bayesian,Liepe2013a,Ryan:2015aaa,Myung2009BayesianAO,vernon2010galaxy,Berliner2008,Cumming2009,farrow2006trade,santner2003design,currin1988bayesian,jones2016bayes}.
One begins by encoding as a utility function the goals of the experiment and
the constraints inherent in carrying it out.  Then, one considers the range of
possible future experimental measurements, and computes the expected utility
for each. The optimal design is then the one that maximizes the expected
utility function.  In this context, a \emph{design} refers to the choice of
observable, the kinematic points at which to measure it, and the relative
allocation of beam time.  With Bayesian experimental design, we incorporate
\emph{a priori} knowledge of the parameters we wish to constrain---in this
case the nucleon polarizabilities---via Bayesian priors.  The utility function
must also account for the accuracy future experiments will reasonably achieve,
and factor in the kinematic regimes where collecting data will likely be
excessively difficult.

As we emphasize in this work, accounting for theory uncertainties in such an
analysis is essential for a proper assessment of the optimal design. All
models are wrong, but some are useful~\cite{Box}.  Still fewer are wrong in a
way that is useful. Effective Field Theory calculations are carried out up to
a particular order in a systematically improvable expansion.  This predictable
character of an EFT's uncertainty permits quantitative answers regarding this
trade-off between increased experimental sensitivity and decreased theoretical
accuracy.  We address this trade-off using the physics-based Bayesian machine
learning model proposed in ref.~\cite{Melendez:2019izc}.  The known
order-by-order EFT predictions allow the algorithm to learn the convergence
pattern and how it is correlated in kinematic space, and then to use that
pattern to formulate a statistical model of the truncation error, thus
ensuring that high-energy data are not over-weighted.

\chiEFT\ incorporates all the physics of Compton scattering at photon energies
between $0$ and about $300\MeV$.  According to ref.~\cite{Pascalutsa:2002pi},
the \chiEFT\ Compton amplitude at a given order in the EFT expansion has a
theory truncation error approximately proportional to 
$(m_\pi/\Lambda_b)^{\nu/2}$. Here, $\Lambda_b$ is the breakdown scale of the
theory, and $\nu$ depends on, but is not necessarily equal to, the order of
the calculation (cf.\ sec.~\ref{sec:bayesian_methodology_compton} below).  A
na\"ive application of experimental design that ignored the theory uncertainty
might suggest running experiments at energies which are so high that \chiEFT\
is unreliable. That would lead to experiments with high precision, whose
information content is, however, very small, wasting scarce human and
financial resources.

Although we devote considerable space to the explanation of \chiEFT's
theoretical errors, it is of course mandatory to also account for experimental
realities. Our recommended experimental design should not involve kinematics
at which measurements are notoriously difficult. Difficulties can arise in two
different directions. The first is that beam time is limited. We account for
that experimental constraint by considering three levels of possible precision
of Compton data. These are intended to cover a range of plausible future
experiments. Since these results correspond to different numbers of photons on
target, they give us a sense of how knowledge of the different
polarizabilities will scale with beam intensity and experimental run time. The
second issue is that physical limitations can make it difficult to place
detectors at particular locations, or to run a certain machine at specific
energies. While these constraints are probably best assessed using a
facility-specific factor in the utility function, in this first study we
account for them crudely by precluding designs involving photon scattering
angles where physical limitations will make it difficult to place detectors.

One of the key aspects of our experimental-design approach---or indeed of any
Bayesian analysis---is that the underlying assumptions must be stated clearly
and quantitatively. In our case, all results are conditional upon the use of
$\chi$EFT to extract polarizabilities. We see this as a virtue of a Bayesian
approach: one is forced to be explicit about one's assumptions, and
acknowledge that any conclusions drawn are contingent upon them. In this
spirit we also point out that a full experimental-design analysis would begin
with an extraction of the polarizabilities from extant Compton data that uses
the Bayesian framework adopted here---including its model of theory
errors. This would establish the present knowledge of the polarizabilities,
\ie~what is known about them before any new experiments are performed. We do
not perform such an analysis in this work. Instead we use previous
determinations of polarizabilities in ChiEFT from
Refs.~\cite{McGovern:2012ew,Griesshammer:2015ahu} to form the prior that
encodes present knowledge of the polarizabilities. Our experimental-design
conclusions are thus indicative, not definitive; our main goal is to show the
power of Bayesian experimental design.

In the interest of accessibility to all readers, we try to ensure the sections
listed below are self-contained.  For example, an understanding of the EFT
truncation model should not be essential to understand the experimental
design, or the results of the analysis.
We begin in sec.~\ref{sec:basic-compton-facts} by recounting the relevant
facts of nucleon Compton scattering.  Next,
sec.~\ref{sec:bayesian_methodology_compton} describes the important results of
the Bayesian methodology employed in this work, including the model of EFT
truncation errors and experimental design, adapted to \chiEFT\ for Compton
scattering.  These methods are applied to Compton scattering on a proton for
various choices of experimental goals in
sec.~\ref{sec:results_experimental_design}.  Finally we conclude in
sec.~\ref{sec:summary_design}.  We provide details of our derivations in
app.~\ref{sec:experimental_design_details} and details of the truncation error
model in app.~\ref{sec:truncation_model_details_compton}.  Results for a
different prior on the polarizabilities, for different levels of experimental
precision and for the case of neutron Compton scattering are reserved for the
\verifyvalue{Supplemental Material}:
\verifyvalue{app.~\ref{sec:extra_compton_figures}.}  We provide all data and
codes needed to reproduce our results~\cite{BUQEYEgithub}.

\section{Basic Facts of Nucleon Compton Scattering}
\label{sec:basic-compton-facts}

We start with an enumeration of those aspects of Compton scattering on the
nucleon relevant for this presentation, to remind experts and introduce the
minimal necessary vocabulary for non-experts. Motivations, context and details
can be found in refs.~\cite{Griesshammer:2017txw, Griesshammer:2015ahu,
  McGovern:2012ew, Griesshammer:2012we} and elsewhere.

The primary physical degrees of freedom probed in Compton scattering up to
about $350\MeV$ are nucleons, pions and the $\Delta(1232)$ resonance.  The
relative importance of these changes with energy; at low energies, the process
is dominated by the Born terms: a point-nucleon with anomalous magnetic
moment, plus the $\pi^0$ $t$-channel coupling. As the energy increases,
effects from the pion cloud around the nucleon become apparent, and finally
the Delta resonance dominates over all other effects.  In the context of an
EFT, this means that the power counting changes with energy, something which
is explained in more detail in sec.~\ref{sec:compton_pc_rearrangement}.

At all these energies, the polarizability contributions are well-described by
six dipole polarizabilities which are labeled by the multipolarities of the
incoming and outgoing electromagnetic field.  In
ref.~\cite{Griesshammer:2017txw}, the following linear combinations were
identified as most convenient for exploring sensitivities while exploiting the
best available prior knowledge: the scalar (dipole) polarizabilities in the
combinations
\begin{align}
\alpha_{E1}\pm\beta_{M1} \, ,
\end{align}
and the mutually orthogonal spin-polarizability combinations
\begin{align}
    \gamma_{0,\pi} & \equiv -(\gamma_{E1E1} \pm \gamma_{M1M1} + \gamma_{E1M2} \pm \gamma_{M1E2}) \\
    \gamma_{E-} & \equiv \gamma_{E1E1} - \gamma_{E1M2} \\
    \gamma_{M-} & \equiv \gamma_{M1M1} - \gamma_{M1E2} \, .
\end{align}
These combinations map onto tight constraints on $\alpha_{E1}+\beta_{M1}$ and
$\gamma_0$ from sum rules. For the proton, these have error bars which are
better than those from direct Compton experiments.

Polarizabilities are fundamental hadron properties.  The scalar
polarizabilities are also important ingredients in, for example, the
proton-neutron mass
splitting~\cite{WalkerLoud:2012bg,Walker-Loud:2019qhh,Gasser:2015dwa,Gasser:2020mzy},
and the spin polarizabilities parametrize the response of the nucleon spin to
electromagnetic fields (such as the nucleonic Faraday effect).

Thirteen independent observables per nucleon parametrize the process when at
most two of the photon beam, nucleon target or recoil nucleon are
polarized. The (unpolarized) differential cross section $\diffcs$ (in
$\mathrm{nb}/\mathrm{sr}$) is larger than zero but otherwise unbounded. The
beam-target asymmetries $\Sigma_3$, $\Sigma_y$, $\Sigma_{1x}$, $\Sigma_{1z}$,
$\Sigma_{2x}$, $\Sigma_{2z}$, $\Sigma_{3y}$ and the polarization-transfer
observables from a polarized beam to the recoil nucleon $\Sigma_{1x^\prime}$,
$\Sigma_{1z^\prime}$, $\Sigma_{2x^\prime}$, $\Sigma_{2z^\prime}$,
$\Sigma_{3y^\prime}$ are ratios of differences over sums of rates and take
values between $-1$ and $1$.  Below the pion-production threshold
$\omega_\pi(\mathrm{lab})\approx150\MeV$, only six observables are non-zero:
$\diffcs$, $\Sigma_3$, $\Sigma_{2x}$, $\Sigma_{2z}$, $\Sigma_{2x^\prime}$,
$\Sigma_{2z^\prime}$.

The following data on these proton observables is available: about $420$
points of widely varying quality for the cross section (see extensive
discussions in refs.~\cite{Griesshammer:2012we, McGovern:2012ew}), about $120$
points for $\Sigma_3$~\cite{Blanpied:2001ae, Sokhoyan:2016yrc, Martel:2017pln,
  CollicottPhD}, $9$ for $\Sigma_{2x}$~\cite{Martel:2014pba, MartelPhD}, and
$10$ for $\Sigma_{2z}$~\cite{Martel:2017pln,Paudyal:2019mee}; no direct
neutron data exists.

\section{Bayesian Methodology in EFT}
\label{sec:bayesian_methodology_compton}

\subsection{Problems and Solutions of Design Strategy}
\label{sec:problems+solutions}

To estimate the best design strategy, our approach must incorporate two
distinct sources of uncertainty: (1) the EFT truncation error and (2) the
unknown measurements from future experiments, including their likely
measurement uncertainties.  Our Bayesian approach can handle both of these
problems in one coherent framework while being candid about our uncertainties.
The two problems and the solutions we propose are summarized as:
\begin{enumerate}
\item \emph{Problem:} \chiEFT\ must be truncated at a finite order, leading to
  a truncation error that is correlated in kinematic space. That is, we trust
  our theory more in some kinematic regimes than we do in others, and the
  discrepancy itself is a smooth function. This should be reflected when
  assessing how well the experimental data from these regimes constrain
  polarizabilities.
    
  \emph{Solution:} An estimate of the truncation error is found by summing
  over all plausible values for the higher-order terms in the EFT\@.  This
  results in a covariance matrix for the theory error that weights
  experimental data from trusted regimes more heavily than data from less
  trusted regimes~\cite{Wesolowski:2018lzj}.
    
\item \emph{Problem:} Given a choice of design, we still do not know the
  results of the yet-to-be-performed experiment. But such results are needed
  in order to estimate how well they would constrain the polarizabilities.
    
  \emph{Solution:} Bayesian experimental design considers all data that could
  plausibly be measured. For each of these we compute corresponding
  polarizability posteriors. The \emph{expected} utility, or worth, of such an
  experiment can then be judged by sampling a utility function over all the
  data possibilities that have been evaluated.
\end{enumerate}
Sampling is often computationally quite expensive; however, in our case, a
controlled approximation allows it to be done analytically, leading to a
simple and intuitive formula for the expected utility of an experiment.  In
the following subsections we describe in detail our approach to the problem of
truncation errors and experimental design.

\subsection{EFT Truncation Errors}
\label{sec:truncation_error_compton}

A Bayesian model of EFT truncation errors has been proposed and discussed
thoroughly in
refs.~\cite{Furnstahl:2015rha,Melendez:2017phj,Melendez:2019izc}.  Here we
recapitulate the main results of the convergence model that are relevant to
Compton scattering, and discuss how it must be modified to account for the
rearrangement of the power counting in the regime of the Delta resonance.  For
a much more thorough introduction to this model of EFT truncation errors, we
refer to ref.~\cite{Melendez:2019izc}.

Suppose we are interested in the prediction of an observable
$\genobs(\kinparvec)$ at some kinematic point $\kinparvec$.  Here,
$x \equiv \{\omega, \theta\}$ is the incident-photon energy $\omega$ and
scattering angle $\theta$ in the lab frame.  EFTs provide a hierarchy of
predictions $\{\genobs_n(\kinparvec;\lecs)\}$, with each order $n$ more
precise than the last.  These predictions depend on low-energy constants---the
polarizabilities%
\footnote{ In Compton scattering, these include the six nucleon
  polarizabilities in the combinations defined in
  sec.~\ref{sec:basic-compton-facts}: $\alpha_{E1}\pm\beta_{M1}$ and
  $\gamma_0,\,\gamma_\pi,\,\gamma_{E-},\,\gamma_{M-}$.  }---which we denote
collectively as a vector $\lecs$.  Let $k$ be the highest order at which the
complete EFT process has been calculated to date.  Then there is a theory
truncation error $\delta\genobs_k$ associated with all higher order terms left
out of the state-of-the-art EFT prediction.  Furthermore, if we are to compare
our predictions to experimental measurements $\genobsexp$, there is the
problem of experimental noise, $\delta\genobsexp$, to contend with.  In
ref.~\cite{Melendez:2019izc}, the authors assume the following relationship,
where theory and experimental uncertainties are independent:
\begin{align} \label{eq:errormodel_compton}
    \genobsexp(\kinparvec) = \genobs_k(\kinparvec;\lecs) + \delta\genobs_k(\kinparvec) + \delta\genobsexp(\kinparvec) \, .
\end{align}
Because $\delta\genobs_k(\kinparvec)$ and $\delta\genobsexp(\kinparvec)$ are
unknown, they are treated as random variables.  Given statistical models for
$\delta\genobs_k$ and $\delta\genobsexp$, we can use
eq.~\eqref{eq:errormodel_compton} to tell us the kinematics $x$ that will
result in the most stringent constraints on the polarizabilities $\lecs$.

We extend refs.~\cite{Furnstahl:2015rha,Melendez:2017phj,Melendez:2019izc} by
writing the observable expansion as
\begin{align} \label{eq:observable_expansion_compton}
    \genobs_k(\kinparvec;\lecs) = \genobsref(\kinparvec) \sum_{n=0}^{k} c_n(\kinparvec;\lecs) Q^{\nu_n(\omega)}(\kinparvec) \, ,
\end{align}
where $Q$ is the dimensionless expansion parameter of the EFT and $\genobsref$
is a reference scale for the observable $\genobs_k$.  For the EFT power
counting to hold, the (dimensionless) observable coefficients $c_n$ should be
approximately of order unity.  (Note: we leave implicit the dependence of
$c_n$ on $\lecs$ in subsequent equations.)  equation
(\ref{eq:observable_expansion_compton}) differs from
refs.~\cite{Furnstahl:2015rha,Melendez:2017phj,Melendez:2019izc} by the
inclusion of $\nu_n(\omega)$ rather than a simple $n$ as the exponent of the
expansion parameter.  The use of $\nu_n(\omega)$ reflects the fact that the
power counting changes as one moves from $\omega\simeq m_\pi$ to
$\omega \simeq \Delta$, resulting in a re-ordering of contributions to the
amplitude~\cite{Pascalutsa:2002pi}.

Because the EFT amplitudes are squared to compute observables (and in the case
of spin observables, further divided by the differential cross section), the
$c_n$ are not linearly related to the polarizabilities; instead they appear
naturally sized and randomly distributed. We will exploit this fact in the
convergence model.

For given choices of $\genobsref$, $Q$ and $\nu_n(\omega)$, the coefficients
$c_n$ are in 1-to-1 correspondence with the results $\genobs_n$ for orders
$n \leq k$.  We will begin by discussing these choices and how they lead to a
physically motivated distribution for $\delta\genobs_k$.  The choice of
$\nu_n(\omega)$ is more technical, and is described in
sec.~\ref{sec:compton_pc_rearrangement}.

\subsubsection{The Distribution of \texorpdfstring{$\delta\genobs_k$}{dyk}}
\label{sec:truncation_distribution_compton}

The reference scale $\genobsref$ should capture the overall size of
$\genobs_k$ in the appropriate units.  The spin observables $\Sigma_i$ are
dimensionless and bounded in $[-1, 1]$, hence the natural choice is
$\genobsref = 1$.  In contradistinction, the cross section varies over orders
of magnitude, and can contain cusp-like behavior near the pion-production
threshold.  We capture its overall trend, without cusps, by using a
$\genobsref$ comprised of the basic Born, pion pole, and Delta-pole parts of
the proton (and neutron) cross section.  With this choice, the proton cross
section still shows some growth of the $c_n$ near $\omega_\pi$ and at forward
angles, so we multiply this reference by a shifted 2-dimensional Lorentzian
\begin{align}
    \left[\left(\frac{\omegalab-\omega_\pi}{50\MeV}\right)^2 + \left(\frac{\thetalab}{150^\circ}\right)^2 + \frac{1}{3}\right]^{-1} + 1 \, .
\end{align}
There is no particular physics in this function. It serves only to produce $c_n$ that look similar across kinematic space. 

EFTs exploit a separation of scales, from which one can construct a small
expansion parameter $Q$.  Here we choose the expansion parameter
\begin{align} \label{eq:expansion_parameter_compton}
    Q(\kinparvec) = \sqrt{\frac{\omega_{\text{cm}} + m_\pi}{2\Lambda_b}} \, ,
\end{align}
where the low-momentum scale is the average of $m_\pi$ and
$\omega_{\text{cm}}$, the photon momentum in the center-of-momentum frame.
This extends the expansion parameter $Q = \sqrt{m_\pi/\Lambda_b}$ proposed for
$\omega \lesssim m_\pi$ in ref.~\cite{Pascalutsa:2002pi}. The $Q$ of
eq.~\eqref{eq:expansion_parameter_compton} explicitly builds in our
expectation that the EFT's accuracy degrades at large $\omega$.  The breakdown
scale of this \chiEFT\ in the single-nucleon sector was identified in
ref.~\cite{Pascalutsa:2002pi} as $\Lambda_b = 650\MeV$.  We also ran the
analysis using the expansion parameter $Q\to\sqrt{\mpi/\Lambda_b}$ and found
the results were essentially unchanged.

The crux of the EFT truncation error model is induction on the $c_n$: the
coefficients for $n > k$, which we have not yet seen, are assumed to have
approximately the same size and dependence on $(\omega,\theta)$ as the
lower-order $c_n$ that we already have from our EFT calculation.  To formalize
this inductive step, we model the $c_n$ as independent and identically
distributed (\iid) curves and assign them a Gau\3ian process (GP) prior. This
is notated as:
\begin{align} \label{eq:coefficient_prior_compton}
    c_n(\kinparvec) \given \sdth^2, \ell_\omega, \ell_\theta \overset{\text{\tiny \iid}}{\sim} \GP[0, \sdth^2 r(\kinparvec, \kinparvec';\ell_\omega, \ell_\theta)] \, .
\end{align}
Here and below $z\given g$ is read as ``$z$ given $g$'', while $z \sim \dots$
is statistical shorthand for ``$z$ is distributed as $\dots$.'' In this case
the $c_n$'s are distributed as $\GP[m(x), \kernel(x,x';\param)]$, which
denotes a GP with mean function $m(x)$ and covariance function
$\kernel(x,x';\param)$.  GPs are popular machine-learning algorithms that have
been employed in a wide variety of disciplines to perform nonparametric
regression and classification~\cite{sacks1989design,
  cressie1992statistics,rasmussen2006gaussian}.  The samples from a GP are
\emph{functions}, as opposed to numbers or vectors.  A brief introduction to
GPs in this context is given in ref.~\cite{Melendez:2019izc}; see also
refs.~\cite{rasmussen2006gaussian, Mackay:1998introduction,
  Mackay:2003information} for more in-depth discussions.  We adopt a mean
function of 0 since, \emph{a priori}, the \chiEFT\ corrections $c_n$ are just
as likely to be positive as they are to be negative.

\begin{figure}[tb]
    \centering
    \includegraphics[width=\linewidth]{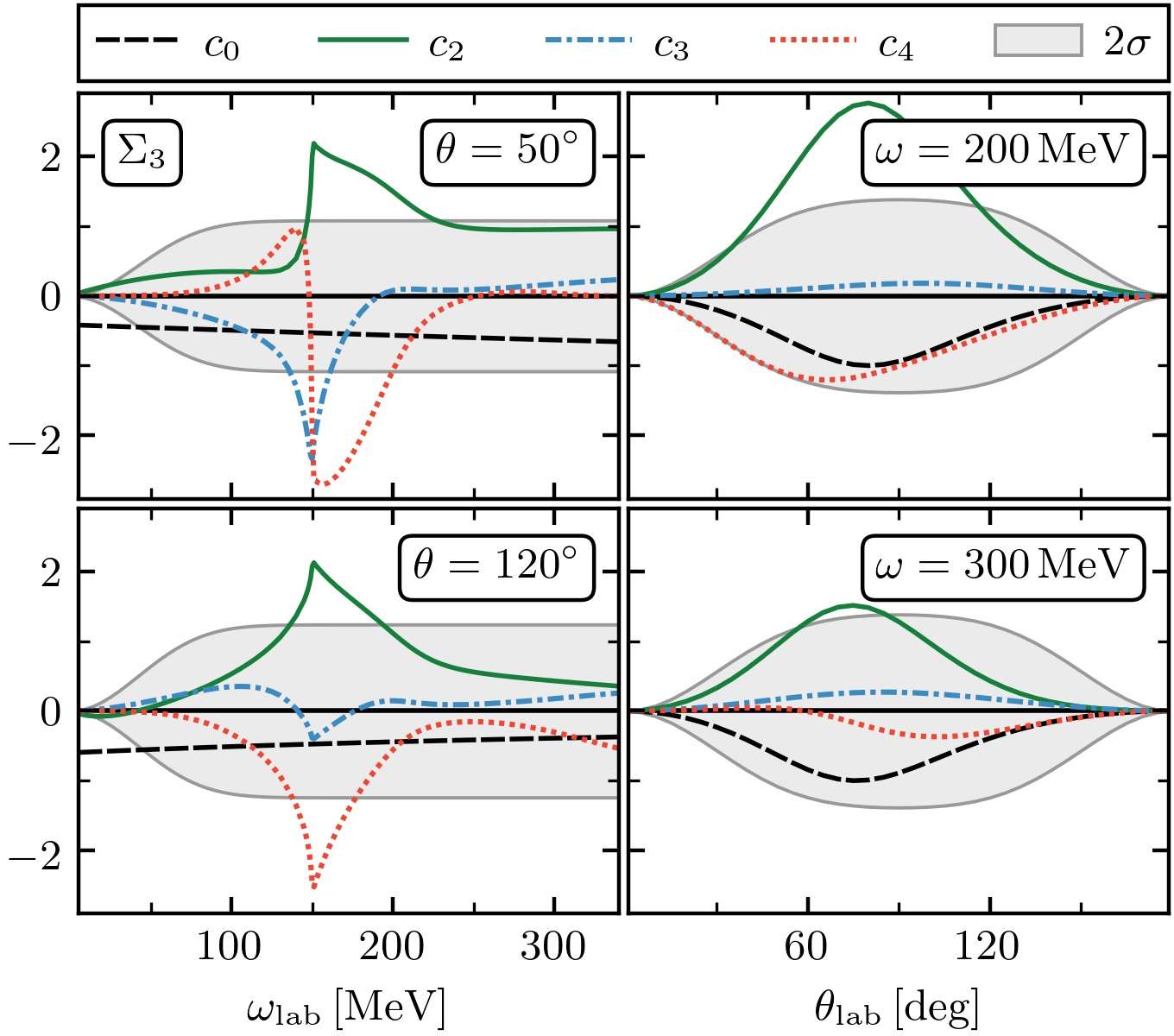}
    \caption{(Colour online) Observable coefficients for $\Sigma_3$.  The gray $2\sigma$
      bands indicate the expected 95\% credible interval for all higher order
      coefficients.  Note that, although $c_0 \neq 0$ at $\omega = 0$, all of
      the corrections $c_n$, along with their derivatives, do vanish.  A
      similar situation occurs at forwards and backwards angles.  These
      constraints, and the corresponding ones for other observables, are built
      into the EFT truncation error model.  }
    \label{fig:eft_coefficients}
\end{figure}

The values of our GP hyperparameters $\sdth^2$ and $\ell_i$, whose meaning are
discussed below, are tuned to the known $c_n$ with $n \leq k$ (at the best
known values of the polarizabilities $\lecs$ for each EFT order).  An example
of how these hyperparameters, combined with symmetry constraints on the
observables, lead to a distribution for higher-order $c_n$ is shown in
fig.~\ref{fig:eft_coefficients}. We observe that there is no obvious
systematic increase or decrease of the coefficients with $n$. This supports
our adoption of a breakdown scale $\Lambda_b=650$~MeV. But we also observe
that cusps in the observables around the pion-production threshold,
$\omega\approx\omega_\pi$, can grow rather large.  These are expected, and are
not a problem in and of themselves (as discussed when tuning $\ell_\omega$
below).  But, for spin observables $\Sigma_i$, their large size is
uncharacteristic compared to the $c_n$ away from $\omega_\pi$; hence we choose
to exclude $125< \omegalab < 200\MeV$ when training $\sdth^2$ and
$\ell_i$.\footnote{ Interestingly, the spikes disappear when considering the
  $c_n$ for rate-differences $\diffcs \times \Sigma_i$, but a Gau\3ian
  uncertainty in these and in $\diffcs$ does not lead to a simple Gau\3ian
  uncertainty in $\Sigma_i$.  Further study of the convergence patterns of
  $\Sigma_i$ near $\omega_\pi$ is needed.  } Despite providing the most
rigorous accounting of uncertainties to date, we are thus less confident in
the estimate of the \chiEFT\ uncertainty $\delta\genobs_k$ for $\Sigma_i$ very
close to $\omega_\pi$.  [The convergence pattern of $\diffcs$, in contrast,
remains regular at $\omega_\pi$, giving us confidence in our design results
there.]  The details of the fitting procedure, the symmetry constraints on the
observables, and figures for the remaining observables, are reserved for
app.~\ref{sec:truncation_model_details_compton}.  The results of the fits are
shown in table~\ref{tab:truncation_details_observables}.  By the inductive
step [eq.~\eqref{eq:coefficient_prior_compton}], these hyperparameters tell us
about the unknown higher-order $c_n$.

\begin{table}[tb]
\renewcommand{\tabcolsep}{3pt}
\caption{
  \label{tab:truncation_details_observables} The optimized truncation error hyperparameters for each observable listed in
  ref.~\cite{Griesshammer:2017txw}, which also discusses the definition and properties of the observables.
  The length scale $\ell_\omega$ is given in units of MeV, whereas $\ell_\theta$ is in units of degrees.
  The $c_n$ are dimensionless quantities, so $\cbar$ is dimensionless as well. 
}
\begin{ruledtabular}
\begin{tabular}{Sl d{2.2}*{2}{d{2.0}} d{2.2}*{2}{d{2.0}}}
 & \multicolumn{3}{Sc}{Proton} & \multicolumn{3}{Sc}{Neutron} \\
\cline{2-4}\cline{5-7}
 & \multicolumn{1}{Sc}{$\sdth$} & \multicolumn{1}{Sc}{$\ell_\omega$} & \multicolumn{1}{Sc}{$\ell_\theta$} & \multicolumn{1}{Sc}{$\sdth$} & \multicolumn{1}{Sc}{$\ell_\omega$} & \multicolumn{1}{Sc}{$\ell_\theta$} \\
\colrule
$\diffcs$ &    0.9 &       56 &       64 &     3.2 &       40 &       78 \\
$\Sigma_{1x}$      &   0.76 &       35 &       46 &    0.68 &       58 &       49 \\
$\Sigma_{1z}$      &   0.48 &       35 &       56 &     0.4 &       56 &       42 \\
$\Sigma_{2x}$      &   0.59 &       42 &       37 &     0.7 &       50 &       39 \\
$\Sigma_{2z}$      &    1.5 &       50 &       45 &     2.1 &       46 &       52 \\
$\Sigma_{3}$       &    0.7 &       49 &       35 &     0.5 &       70 &       44 \\
$\Sigma_{y}$       &   0.63 &       41 &       52 &    0.57 &       61 &       44 \\
$\Sigma_{3y}$      &   0.83 &       36 &       45 &    0.66 &       60 &       45 \\
$\Sigma_{3y'}$     &   0.64 &       41 &       46 &    0.66 &       49 &       46 \\
$\Sigma_{1x'}$     &   0.66 &       37 &       47 &    0.49 &       55 &       43 \\
$\Sigma_{1z'}$     &   0.28 &       33 &       44 &    0.29 &       55 &       43 \\
$\Sigma_{2x'}$     &    1.1 &       32 &       58 &     1.3 &       40 &       53 \\
$\Sigma_{2z'}$     &   0.75 &       34 &       52 &     1.8 &       48 &       57 \\

\end{tabular}
\end{ruledtabular}
\end{table}

The marginal variance $\sdth^2$ in eq.~\eqref{eq:coefficient_prior_compton}
controls the size of the $c_n$.  If the $c_n$ are naturally sized, then
$\sdth$ should be of order unity.  We place an inverse chi-squared prior on
$\sdth^2$:
\begin{align}
    \sdth^2 \sim \chi^{-2}(\nu_0, \tau_0^2) \, ,
\end{align}
where $\nu_0$ and $\tau_0$ are the prior degrees of freedom and scale
parameters, respectively.  This is a conjugate prior and allows the posterior
for $\sdth^2$ to be found analytically; see ref.~\cite{Melendez:2019izc} for
details.  We choose $\nu_0 = \verifyvalue{1}$ and $\tau_0 = \verifyvalue{1}$,
which is weakly informative~\cite{BDA3}.  We take the posterior mean as an
estimate for $\sdth^2$ in eq.~\eqref{eq:coefficient_prior_compton}.

The smoothness of the $c_n$ is dictated by the correlation function $r$.  We
take the correlation of $c_n$ between two kinematic points $x=(\omega,\theta)$
and $x^\prime=(\omega^\prime,\theta^\prime)$ to be given by a radial basis
function (RBF)
\begin{align} \label{eq:rbf_kernel_compton}
    r(x, x'; \ell_\omega, \ell_\theta) = \exp{-\frac{(\omega - \omega')^2}{2\ell_\omega^2} - \frac{(\theta - \theta')^2}{2\ell_\theta^2}} \, ,
\end{align}
where the correlation lengths $\ell_i$ control how quickly the $c_n$ vary as a
function of $\omega$ and $\theta$.  There is no conjugate prior for $\ell_i$;
rather, we use a uniform prior and find the best fits by maximizing the log
likelihood.  Choosing the RBF as a correlation function for the $c_n$ implies
that they are quite smooth. This assumption is validated empirically, except,
as already noted, at the pion-production threshold, which occurs at photon
energy $\omega_\pi\approx150\MeV$ in the lab frame.  Tuning $\ell_\omega$ to
data with cusps will bias it towards very small values.  To fix this bias, we
set the correlations between $c_n(x)$ and $c_n(x')$ to zero if they are on
opposite sides of the pion-production threshold, but still use the same
correlation lengths $\ell_\omega$ and $\ell_\theta$ below and above the cusps.
This procedure is consistent with the behavior of the coefficients excluding
the threshold region and with tests using GP toy models.  The details for
$\ell_\omega$ are not critical for the present analysis as when we look at
multiple design points they are all at the same energy: while correlations in
angle matter for the design assessment, correlations in energy do not.

Assuming we have estimates of $\sdth^2$ and $\ell_i$, we can construct the
distribution for $\delta\genobs_k$.  It follows from extending
eq.~\eqref{eq:observable_expansion_compton} that
\begin{align} \label{eq:discrepancy_sum_compton}
    \delta\genobs_k(\kinparvec) = \genobsref(\kinparvec) \sum_{n=0}^\infty c_{n+k+1}(\kinparvec) Q^{\nu_{\delta k}(\omega) + n}(\kinparvec) \, ,
\end{align}
where $\nu_{\delta k}(\omega)$ captures the first incomplete order of the
EFT\@, and we assert for simplicity that powers of $Q$ increment in integer
steps afterwards.  [For an EFT with a single power counting, one might expect
$\nu_{\delta k}(\omega) = k+1$.]  equation~\eqref{eq:discrepancy_sum_compton}
is a geometric sum of Gau\3ian random variables, from which it follows that
\begin{align} \label{eq:discrepancy_gp_compton}
    \delta\genobs_k(\kinparvec) \given \sdth^2, \ell_\omega, \ell_\theta \sim \GP[0, \sdth^2 \discrcorr{k}(\kinparvec, \kinparvec';\ell_\omega, \ell_\theta)] \, ,
\end{align}
where
\begin{align}
    \discrcorr{k}(\kinparvec, \kinparvec';\ell_\omega, \ell_\theta) & \equiv \genobsref(\kinparvec)\genobsref(\kinparvec')\frac{Q^{\nu_{\delta k}(\omega)}(\kinparvec)Q^{\nu_{\delta k}(\omega')}(\kinparvec')}{1 - Q(\kinparvec)Q(\kinparvec')} \notag \\
    & \times r(\kinparvec, \kinparvec'; \ell_\omega, \ell_\theta) \, . \label{eq:compton_truncation_corrfunc}
\end{align}
Given choices of $\genobsref$, $Q$, $\nu_{\delta k}$, and
$r(\kinparvec, \kinparvec')$, along with estimates of $\sdth^2$ and $\ell_i$,
the above equations completely define a physics-based uncertainty due to
truncation.

\subsubsection{The Power-Counting and Its Rearrangement}
\label{sec:compton_pc_rearrangement}

An EFT begins with an infinite set of operators that one orders via a power
counting in a ratio of a small to a large scale, denoted $Q$.  There is then a
finite number of parameters, and a finite set of diagrams, that contribute to
the process of interest at any given order in the EFT expansion.

Here, we briefly describe the power counting employed in our calculations of
Compton scattering; details and amplitudes are given in
refs.~\cite{McGovern:2012ew, Griesshammer:2015ahu, Griesshammer:2017txw}; see
also references therein.  Because of a different hierarchy of physical
mechanisms, the power counting is different at different energies, and so is
the first order at which the EFT amplitude is incomplete. We thus also detail
the model used to interpolate between the two regimes in which the EFT power
counting is well understood: $\omega\sim\mpi$ and $\omega\sim M_\Delta$.  The
EFT expansion breaks down entirely as $\omega\to\Lambda_b\approx650\MeV$.

The first regime concerns low photon energies up to around the pion mass,%
\footnote{In this regime, the relation between the expansion used here and the
  notation of refs.~\cite{Pascalutsa:2002pi,McGovern:2012ew,
    Griesshammer:2017txw, Griesshammer:2015ahu} is $Q^n=e^2\delta^n$; the
  simpler symbol leads to more compact formulae later. The LO defined in this
  presentation corresponds to performing the \chiEFT\ power counting on the
  structure part of the nucleon Compton amplitude, \ie, what remains after the
  (relativistic) nucleon and pion Born terms are subtracted, and thus differs
  in that detail from the power counting described in
  refs.~\cite{McGovern:2012ew, Griesshammer:2015ahu, Griesshammer:2017txw}.}
$\omega\lesssim\mpi\approx140\MeV$. Here, the power-counting mirrors that of
standard heavy-baryon $\chi$PT for graphs involving nucleons and pions; all
pion and nucleon Born graphs are counted as LO, or $\calO(Q^0)$, where
$Q\sim\sqrt{(\omega,m_\pi)/\Lambda_b}$. There is no NLO [$\calO(Q^1)$]
correction.  The first corrections come at \NkLO{2} [$\calO(Q^2)$] from the
pion cloud around the nucleon. It is at this order that the polarizabilities
enter first. At \NkLO{3} [$\calO(Q^3)$], effects from the lowest-lying
nucleonic resonance, the $\Delta(1232)$, and its pion cloud are added; the
counting exploits the numerical fact that $M_\Delta/\Lambda_b\sim Q$ to avoid
handling multiple expansion parameters. At \NkLO{4} [$\calO(Q^4)$],
corrections to pion-cloud effects are accounted for, and no new effects from
the Delta enter.  Our implementation of \chiEFT\ for Compton scattering from
the nucleon contains all contributions up to and including \NkLO{4} for photon
energies $\omega \simeq m_\pi$ as well as some terms that are \NkLO{5} there.
Therefore, the theory error $\delta\genobs_k$ in this regime follows
eq.~\eqref{eq:discrepancy_sum_compton} with $\nu_{\delta k}=k+1$, and proceeds
indeed in integer steps.

However, the contributions are re-ordered in the regime where the photon
energy approaches the excitation energy of the Delta resonance,
$\omega\approx300\MeV$. (Hereafter, we use the symbol $\Delta$ for this
energy, $\Delta=M_\Delta-M_N$.)  As this resonance is very strong, it is
physically intuitive that the most dramatic reordering involves Delta-pole
diagrams.  On the formal level, this change stems from the propagator in the
Delta pole graph which becomes large near the resonance.  The onset of the
Delta-dominated region can be estimated from the Delta resonance width to
occur at $\omega\sim 230\MeV$.

In order to describe the relative order of contributions in the
Delta-resonance region, and the transition, we first define $\nu_n(\Delta)$ as
the lowest order at which those diagrams which are of order $n$ at
$\omega\lesssim m_\pi$ contribute when $\omega \simeq \Delta$.  The leading
Delta pole diagram is promoted from $\mathcal{O}(Q^3)$ (\NkLO{3}) for
$\omega \simeq m_\pi$, to $\calO(Q^{-1})$ (LO) for $\omega \simeq \Delta$.
Hence $\nu_3(\Delta)=-1$. This is now the leading contribution in the
resonance region. (The Delta-$\pi$ loops are less strongly promoted, to
\NkLO{2}, so it is the pole graph that determines $\nu_3(\Delta)$.)
Subleading Delta pole diagrams (with dressed vertices) transition from
$\mathcal{O}(Q^5)$ (\NkLO{5}) to $\calO(Q^{0})$ (NLO). (Indeed, we include
them in our amplitudes even though they are not a complete set of low-energy
\NkLO{5} diagrams, because we want to catch all NLO contributions in the
resonance region.)
Born effects and corrections to resonance parameters also enter at NLO (now
$\calO(Q^0)$), so $\nu_0(\Delta)=0$. This makes the NLO contributions be
complete in the resonance region.
The contributions from pion loops round the nucleon change from $\calO(Q^2)$
(\NkLO{2}) to $\calO(Q^{1})$ so $\nu_2(\Delta)=1$, and from $\calO(Q^4)$
(\NkLO{4}) to $\calO(Q^{2})$ so $\nu_4(\Delta)=2$. These therefore enter in
the resonance region at \NkLO{2} and \NkLO{3}, respectively. However, in the
resonance region our amplitude is not complete at \NkLO{2} and \NkLO{3}: there
are further contributions at these orders that we have not accounted for. For
$\omega\sim\Delta$ the amplitude is complete only up to and including NLO.

Summarizing the previous paragraph: the relevant reorderings turn out to
follow the rule that, for diagrams of order $n$ at $\omega\simeq\mpi$,
$\nu_n(\Delta)=n/2$ for even orders but $\nu_n(\Delta)=(n-5)/2$ for odd
orders. The first omitted order for $\omega \sim \Delta$,
$\nu_{\delta k}(\Delta)$, is then given by the smaller of $\nu_{k+1}(\Delta)$
or $\nu_{k+2}(\Delta)$.

Now, we wish to be able to handle data in the transition region lying
somewhere in between $\omega\simeq\mpi$ and $\omega\simeq\Delta$, by defining
an interpolator $\nu_{n}(\omega)$ that is a function of $\omega$.  If we
define a suitable monotonic function $f(\omega)$ satisfying
 \begin{equation}
    \label{eq:fofomega}
    f(\omega\approx\mpi)\approx0\,, \quad f(\omega\approx\Delta)\approx1\,,
\end{equation}
the reordering is smoothly captured by
 \begin{equation}
    \nu_n(\omega) =
    \begin{cases}
        [1 - f(\omega)/2] n \, , & n \text{ even} \\
        [1 - f(\omega)/2] n - 5 f(\omega)/2 \, , & n \text{ odd}
    \end{cases}
    \label{eq:order_transition}
\end{equation}
which is tabulated in table~\ref{tab:order_and_truncation_functions}.  For
definiteness we use a logistic form inspired by the Fermi function
\begin{equation}
    f(\omega) = \left[1 + \exp(-4\ln 3 \cdot \frac{\omega - \omega_m}{\omega_2 - \omega_1})\right]^{-1},
    \label{eq:fermi_interp}
\end{equation}
where $\omega_1 = 180\MeV$ and $\omega_2 = 240\MeV$ are the locations where
$f(\omega_1) = 1/10$ and $f(\omega_2) = 9/10$, and
$\omega_m = (\omega_1 + \omega_2)/2\approx210\MeV$ is the midpoint
$f(\omega_m) = 1/2$.  The same form was already used in the plots of
ref.~\cite{Griesshammer:2017txw} to parametrize the ``gray mist'' at high
energies, but our framework puts this ``mist'' on a quantitative footing via
the EFT-inspired theory error,
eqs.~\eqref{eq:discrepancy_sum_compton}--\eqref{eq:compton_truncation_corrfunc}.
These choices are consistent with the estimate above that the Delta resonance
region starts around $230\MeV$.  This form for $f(\omega)$ is only one of
several possibilities. Other sensible models for $f(\omega)$ lead to results
which are compatible with those presented below.

\begin{table}[!tb]
\renewcommand{\tabcolsep}{3pt}
\caption{The rearrangement of orders of diagrams between the $m_\pi$ regime and the regime of the Delta resonance, and the approximate power of the first omitted term in the expansion.
The most general transition function $f(\omega)$ is defined in  eq.~\eqref{eq:fofomega}, with a particular choice (logistic function) given in  eq.~\eqref{eq:fermi_interp}.}
\label{tab:order_and_truncation_functions}
\begin{ruledtabular}
\begin{tabular}{SlSlSl}
    Order & Transition with $\omega$ & Leading \\[-1ex]
     at $\omega \simeq m_\pi$& &Truncation Error \\
    \colrule
    0 (LO) & $\nu_0(\omega) = 0$ & $\nu_{\delta 0}(\omega) = 2 - 3 f(\omega)$ \\
    2 (\NkLO{2}) & $\nu_2(\omega) = 2 - f(\omega)$ & $\nu_{\delta 2}(\omega) = 3 - 4f(\omega)$ \\
    3 (\NkLO{3}) & $\nu_3(\omega) = 3 - 4f(\omega)$ & $\nu_{\delta 3}(\omega) = 4 - 4f(\omega)$ \\
    4 (\NkLO{4}) & $\nu_{4}(\omega) = 4 - 2f(\omega)$ & $\nu_{\delta 4}(\omega) = 5 - 5f(\omega)$ \\
    5 (\NkLO{5}) & $\nu_5(\omega) = 5 - 5f(\omega)$ & $\nu_{\delta 5}(\omega) = 6 - 4f(\omega)$
\end{tabular}
\end{ruledtabular}
\end{table}

\begin{figure}[tb]
    \centering
    \includegraphics[width=\linewidth]{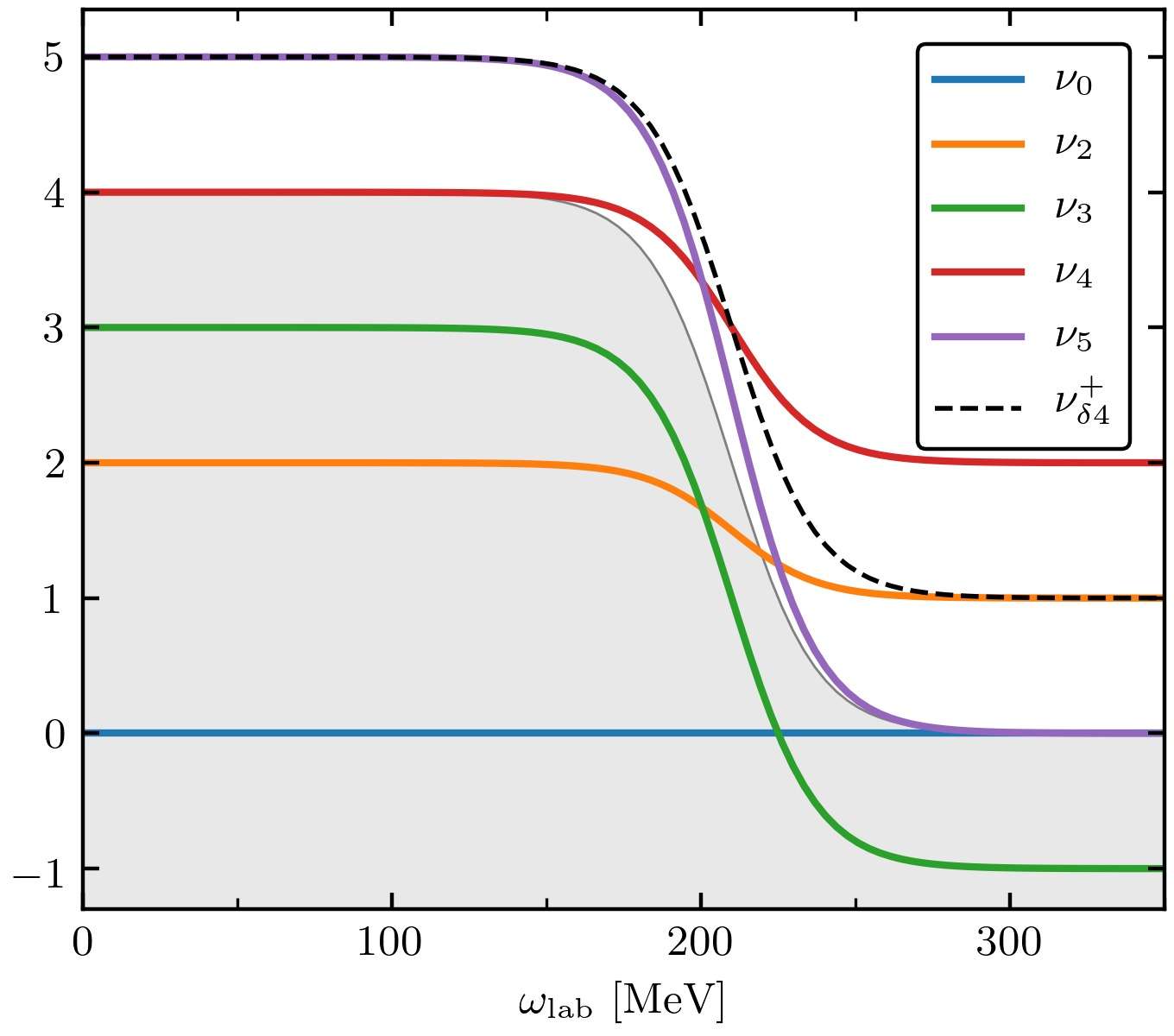}
    \caption{(Colour online) The power counting transitions from the $m_\pi$
      regime to the regime around the Delta resonance.  Solid lines
      corresponding to $\nu_n$ capture the most relevant reordering of
      diagrams, as described in the text.  The shaded region is the
      approximate order up to which the $\NkLO{4}^+$ EFT is complete.  The
      dashed line $\nu_{\delta 4}$ is one unit above the shaded boundary and
      represents the first order to be included in the EFT truncation error.
    }
    \label{fig:eft_order_transition}
\end{figure}

In fig.~\ref{fig:eft_order_transition}, we translate the first column of
table~\ref{tab:order_and_truncation_functions} and the logistic
function~\eqref{eq:fermi_interp} to a graphical representation of the
re-ordering of contributions. It is straightforward to read off the dominant
theory uncertainty that an amplitude which is complete up to $\calO(Q^n)$ in
the $\omega \simeq m_\pi$ regime has in the $\omega \simeq \Delta$
regime. This defines $\nu_{\delta n}(\Delta)$. The resulting form of the
leading truncation error $\nu_{\delta n}(\omega)$ as function of $\omega$ is
given in the third column of table~\ref{tab:order_and_truncation_functions}.
 
We observe that starting with the full amplitude up to and including
$\mathcal{O}(Q^4)$ (\NkLO{4}) for $\omega \simeq m_\pi$ only yields an
amplitude that is complete at $\calO(Q^{-1})$ (LO) for $\omega \simeq \Delta$.
However, there are only a small number of diagrams that are missing at
$\calO(Q^0)$ for $\omega \simeq \Delta$. These were identified, computed and
added to the amplitude in both regimes in ref.~\cite{McGovern:2012ew}. This
produces an amplitude that is complete up to $\calO(Q^0)$ for
$\omega \simeq \Delta$ and is ``$\NkLO{4}^+$'', \ie, more than \NkLO{4} but
not fully \NkLO{5}, for $\omega \simeq m_\pi$.  Since the truncation error
must include all orders that do not contain a complete set of diagrams, we
therefore identify
\begin{align} \label{eq:n3loplus_truncation_power}
    \nu_{\delta 4}^+(\omega) = 5 - 4 f(\omega)
\end{align}
as the first omitted power of our $\NkLO{4}^+$ EFT.

\subsection{Experimental Design}

The process of designing an experiment must begin with defining a goal.  For
example, this goal could be to make an accurate prediction of some future
measurement, to discriminate between competing models, or to precisely
constrain parameters of the theory.  The goal could even be designed with a
compromise between several different experimental aims in mind. It could also
incorporate time and cost constraints. But in this work, we simply take
constraining the nucleon polarizabilities as the goal of the experiments we
are designing---although time and cost constraints will be assessed indirectly
when we define different scenarios for the experimental accuracy.

\begin{figure*}
    \centering
    \includegraphics[width=\textwidth]{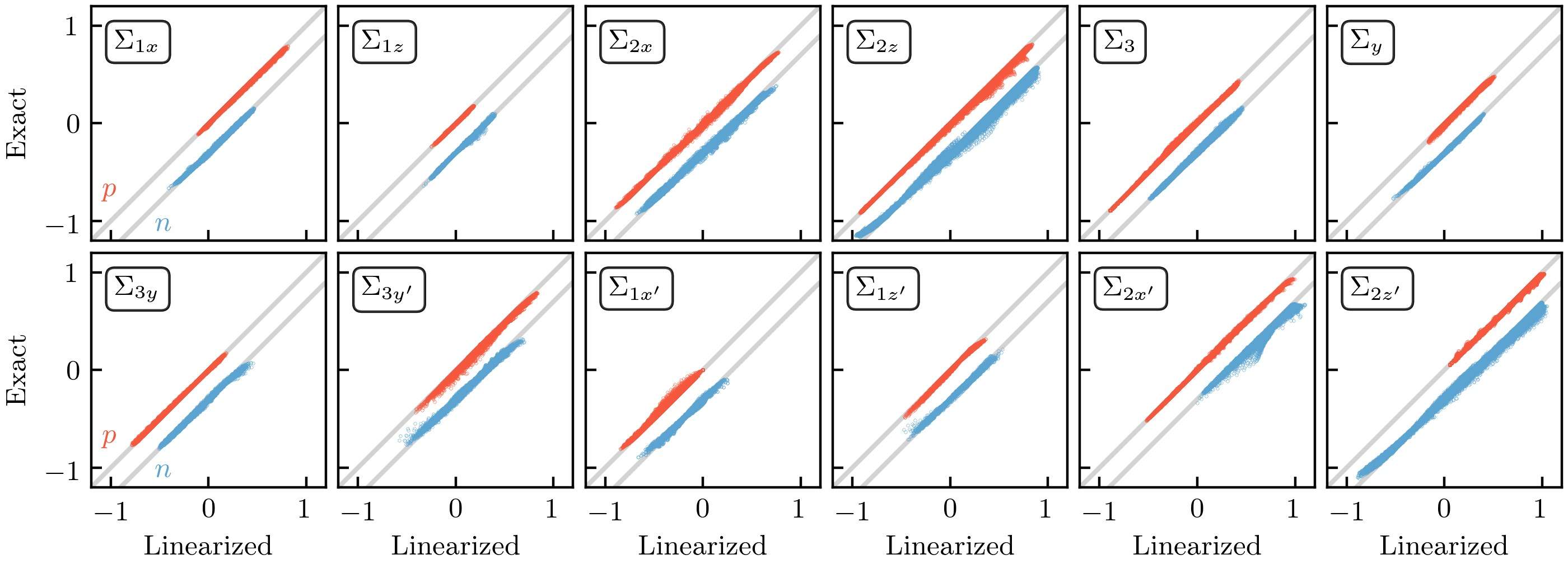}
    \caption{(Colour online) Exact vs linearized predictions of proton (`p',
      red) and neutron (`n', blue) observables from the $\NkLO{4}^+$ EFT\@.
      The neutron predictions have been vertically offset for clarity.  The
      points shown use all predictions at $\omega = 60,70,\dots,340\MeV$ and
      $\theta=40,50,\dots,140^\circ$, with 1000 sets of polarizability values
      sampled from the prior.  The markers are so tightly clustered around the
      gray diagonals that they are difficult to distinguish.  Here we have
      used the linear combinations of polarizabilities listed in
      table~\ref{tab:polarizability_info}. This significantly improves the
      linear approximation.  }
    \label{fig:true_vs_linearized_predictions}
\end{figure*}

\begin{figure}
    \centering
    \includegraphics[width=\linewidth]{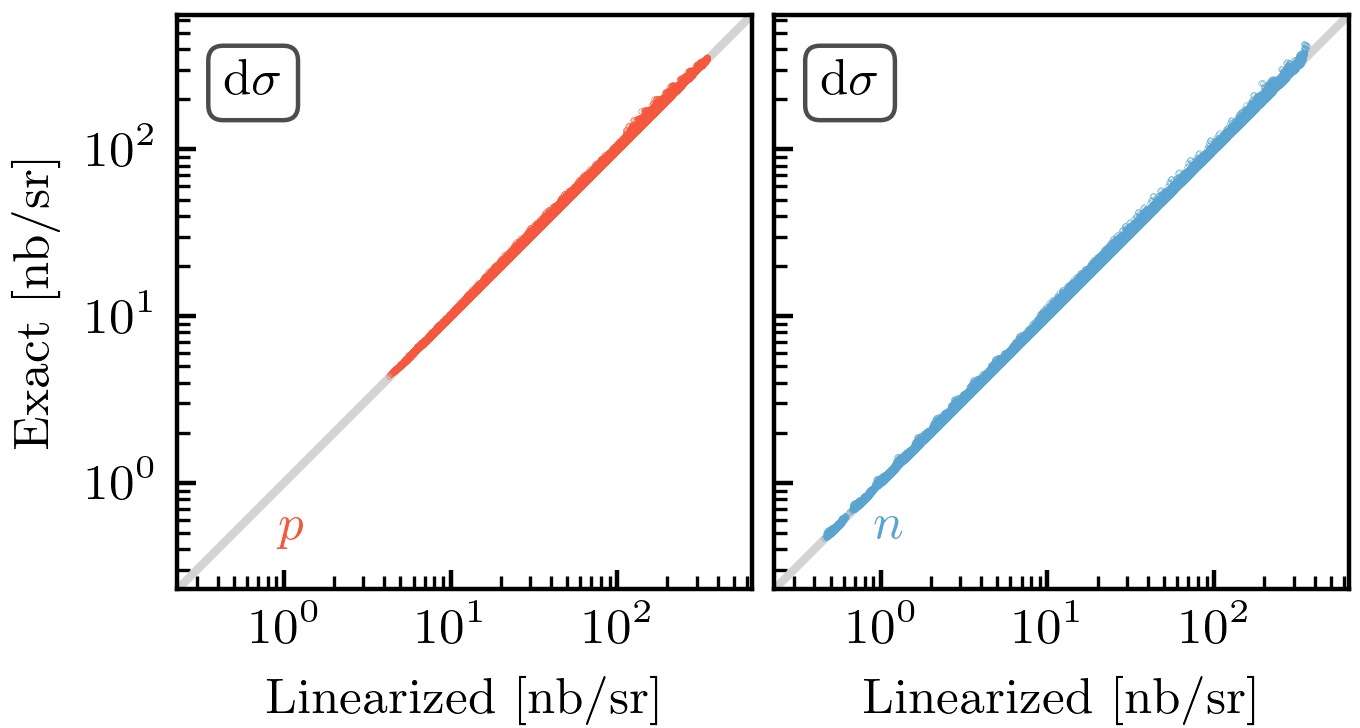}
    \caption{(Colour online) Exact vs linearized predictions of the
      differential cross section $\diffcs$ for proton (`p', red) and neutron
      (`n', blue) from the $\NkLO{4}^+$ EFT\@, with points as in
      fig.~\ref{fig:true_vs_linearized_predictions}.  }
    \label{fig:true_vs_linearized_dsg}
\end{figure}

The next step is to encode as mathematical objects the experimental goal and
all uncertainties.  Once encoded, our goal is known as a utility function, or
design criterion, $U(\design, \lecs, \genobsset)$, that depends on the design
points\footnote{ A single \emph{design} $\design$ in this work is specified by
  an observable and a set of kinematic points at which to measure it, and
  possibly the experimental noise levels.  The space $D$ is the set of all
  considered experiments over which the utility is optimized, \eg, all
  possible 5-angle measurements at a given energy.  } $\design$ in the design
space $D$ from which experimental data $\genobsset$ is then measured, and the
theory parameters $\lecs$.  But, of course, $\genobsset$ will not be known
until the experiment is conducted, and $\lecs$ is exactly the quantity we have
constructed our experiment to find.  Hence the optimal design $\design^\star$
is that which maximizes the \emph{expected} utility
$U(\design) = \E[U(\design, \lecs, \genobsset)]$.  That is,
\begin{align}
    \design^\star & = \argmax_{\design \in D} U(\design) \notag \\
    & = \argmax_{\design \in D} \int U(\design, \lecs, \genobsset) \pr(\lecs, \genobsset \given \design) \dd{\lecs} \dd{\genobsset} \label{eq:expected_utility} \\
    & = \argmax_{\design \in D} \int \Big\{ U(\design, \lecs, \genobsset) \pr(\lecs \given \genobsset, \design) \dd{\lecs} \Big\} \pr(\genobsset \given \design) \dd{\genobsset}. \notag
\end{align}
These integrals are usually intractable for nonlinear theories such as the
observable predictions from \chiEFT, but we show in
figs.~\ref{fig:true_vs_linearized_predictions}
and~\ref{fig:true_vs_linearized_dsg} that linearizing \chiEFT\ predictions
around the best known $\lecs$ is a very good approximation, and we employ it
from here on.

Equation~\eqref{eq:expected_utility} says that the process of experimental
design requires a theory $\genobs(\kinparvec; \lecs)$ and a probabilistic
model relating data to theory parameters,
$\pr(\lecs, \genobsset \given \design)$. To calculate that pdf we employ a
different (but equivalent) form of the product rule to that used below
eq.~\eqref{eq:expected_utility} and write
$\pr(\lecs, \genobsset \given \design) = \pr(\genobsset \given \lecs, \design)
\pr(\lecs)$.
Our truncation error model from eq.~\eqref{eq:errormodel_compton} now comes
into play.  If a Gau\3ian prior is placed on the polarizabilities,%
\footnote{ The notation $\normal(\vec{\mu}_0, V_0)$ denotes a Gau\3ian with
  mean $\vec{\mu}_0$ and covariance $V_0$. See also discussion of notation
  below eq.~\eqref{eq:coefficient_prior_compton}.  }
\begin{align} \label{eq:polarizability_prior}
    \lecs \sim \normal(\vec{\mu}_0, V_0) \, ,
\end{align}
then under the assumption that $\genobs(\kinparvec;\lecs)$ is linear in $\lecs$, one can show that the posterior is given by
\begin{align} \label{eq:polarizability_posterior}
    \lecs \given \genobsset, \design \sim \normal(\vec{\mu}, V) \, ,
\end{align}
where $\vec{\mu}(\genobsset, \design)$ and $V(\design)$ take into account both
the truncation error and the experimental errors, and both depend on the
values of the GP hyperparameters that have already been tuned to the EFT
convergence pattern (see sec.~\ref{sec:truncation_distribution_compton} and
app.~\ref{sec:experimental_design_details}).  The linearization point is
chosen to be $\vec{\mu}_0$ and the prior for each nucleon is given in
table~\ref{tab:polarizability_info}.  We will discuss these priors
momentarily.

Our goal is to constrain polarizabilities, so the optimal design is that which
is likely to provide the most information about $\lecs$.  It is reasonable
then to choose the utility to be the gain in Shannon information for $\lecs$
based on the experiment $(\design, \genobsset)$.  This is equivalent to the
Kullback-Leibler (KL) divergence between the prior and posterior for $\lecs$,
followed by marginalizing over $\genobsset$:
\begin{align}
    U_{\text{KL}}(\design)
    & = \int \!\!\left\{\! \ln\!\!\left[\frac{\pr(\lecs \given \genobsset, \design)}{\pr(\lecs)}\right]\! \pr(\lecs \given \genobsset, \design)  \dd{\lecs}\right\}\! \pr(\genobsset \given \design) \dd{\genobsset}\!. \label{eq:expected_utility_kl}
\end{align}
The assumptions of eqs.~\eqref{eq:polarizability_prior}
and~\eqref{eq:polarizability_posterior} allow \eqref{eq:expected_utility_kl}
to be computed exactly (see app.~\ref{sec:experimental_design_details}) with
the result
\begin{align} \label{eq:utility_kl_analytic}
    U_{\text{KL}}(\design) = \frac{1}{2} \ln \frac{|V_0|}{|V(\design)|} \equiv \ln\shrinkage(\design) \geq 0 \, ,
\end{align}
where we have defined the posterior shrinkage factor $\shrinkage \geq 1$.
Consider the hyperellipsoids defined by given confidence levels for the
$\lecs$ prior and posterior, \eqref{eq:polarizability_prior} and
\eqref{eq:polarizability_posterior}.  Then $\shrinkage$ is the factor by which
the volume of the prior ellipsoid shrinks as it is updated to the posterior,
with larger values of $\shrinkage$ (or $U_{KL}$) being more informative than
smaller values.  An experiment yielding $\shrinkage = 1$ (or $U_{KL} = 0$) is
then completely uninformative.  The utility of an experiment designed to
constrain any subset of $\lecs$, without regard to the others, can be assessed
by simply computing eq.~\eqref{eq:utility_kl_analytic} with the corresponding
submatrices of $V_0$ and $V$.

\begin{table}[tb]
\centering
\renewcommand{\arraystretch}{1.6}
\caption{Priors for the linear combinations of polarizabilities defined in eqs.~(12) and~(13) of ref.~\cite{Griesshammer:2017txw}.
  The prior covariance matrix $V_0$ for the neutron is constructed as a diagonal matrix with $[V_0]_{ii} = [\vec{\sigma}_0]_i^2$. This formula also defines the diagonal elements of the proton covariance matrix. But for $\alpha_{E1}-\beta_{M1}$ and $\gamma_{M-}$ there are correlations in the prior, as described in the text. 
  Unless otherwise referenced, values are from table 1 in ref.~\cite{Griesshammer:2015ahu}.
}
\label{tab:polarizability_info}
\begin{ruledtabular}
\begin{tabular}{l@{}d{4.2}d{0.3}@{}l@{}d{2.1}@{}d{1.1}@{}l}
    & \multicolumn{3}{c}{Proton} & \multicolumn{3}{c}{Neutron} \\
    \cline{2-4}\cline{5-7}
    $\lecs$ &
    \multicolumn{1}{c}{$\vec{\mu}_0$} & \multicolumn{1}{c}{$\vec{\sigma}_0$} &ref.&
    \multicolumn{1}{c}{$\vec{\mu}_0$} & \multicolumn{1}{c}{$\vec{\sigma}_0$}& ref.\\
    \colrule
$\alpha_{E1}+\beta_{M1}$ &
14.0 &       0.2&\cite{Gryniuk:2015eza} &
15.2 &       0.4&\cite{Levchuk:1999zy}\\
$\alpha_{E1}-\beta_{M1}$ &
7.5 &       0.9&\cite{McGovern:2012ew}&
7.9 &       3.0&\cite{Myers:2015aba, Myers:2014ace}\\
$\gamma_{0}$             &
-0.90 &     0.14&\cite{Pasquini:2010zr}&
0.4 &       2.2& \\
$\gamma_{\pi}$           &
7.7 &       1.8& \cite{Schumacher:2005an}&
7.8 &       2.2& \\
$\gamma_{E-}$            &
-0.7 &       2.0& &
-3.9 &       2.0& \\
$\gamma_{M-}$            &
0.3 &       0.9& &
-1.1 &       0.9& \\
\end{tabular}
\end{ruledtabular}
\end{table}

Although the posterior shrinkage has the benefit of being strictly
non-negative and increasing with increasing information, it is unbounded,
making it difficult to compare plots on different scales.  Thus, we choose to
show the percent decrease in uncertainty (sometimes referred to as
``information gain" below)
\begin{align} \label{eq:percent_decrease}
    \text{\% Decrease} & = \frac{|V_0|^{\frac{1}{2}} - |V|^{\frac{1}{2}}}{|V_0|^{\frac{1}{2}}} \times 100\% \notag \\
    & = {\left(1 - \frac{1}{\shrinkage}\right)} \times 100 \% \, .
\end{align}
This shares the beneficial aspects of $\shrinkage$, but is bounded in the range of 0--100\%.

Our assumptions lead to a form of the expected utility that is analytic, easy
to understand, and quick to compute.  This makes
eq.~\eqref{eq:utility_kl_analytic} very attractive. It allows quick assessment
of both:
\begin{itemize}
\item Optimal designs for various assumptions, such as experimental noise
  levels and truncation error forms.
    
\item Which polarizability subsets will have their constraints improved by a
  particular experiment---and by how much.
\end{itemize}
Constraints from previous experiments are built in naturally via the prior on
the polarizabilities. For example, a large utility in a previously
well-measured observable or region of kinematic space means that there is
still valuable constraining information to be gained there.  One might be
concerned that this choice of prior has undue influence on the final results,
so here we note two ways in which it does {\it not} influence them. First,
eq.~\eqref{eq:utility_kl_analytic} is invariant under any linear
transformation of $\lecs$, meaning, \eg, that the choice of units for $\lecs$
is irrelevant, and that this analysis would be consistent if we had instead
used
$\lecs = \{\alpha_{E1}, \beta_{M1}, \gamma_{E1E1}, \gamma_{M1M1},
\gamma_{E1M2}, \gamma_{M1E2}\}$,
so long as $V_0$ were transformed accordingly (see
sec.~\ref{sec:basic-compton-facts}).  Second, the final covariance matrix $V$
is independent of the data $\genobs$: it is determined by the accuracy of the
experiment, the sensitivity of observables to the polarizabilities at the
design points, and the prior covariance matrix $V_0$.  Because we have
linearized the problem, the central values shown in
table~\ref{tab:polarizability_info} only enter our experimental-design
assessment insofar as they affect the value of the vector $B$ that encodes the
sensitivity of the observable $\genobs$ to the six polarizabilities (see
eq.~\ref{eq:pol_posterior_cov_linear}).  The {\it width} of the prior---the
choice of $V_0$---does affect the design analysis though.

\subsection{Choice of Priors \label{sec:priors}}

The priors summarized in table~\ref{tab:polarizability_info} are the
uncertainties to which the polarizabilities are known at present.
As we base our design on the results of the \chiEFT\ variant of
refs.~\cite{Griesshammer:2017txw, Griesshammer:2015ahu, McGovern:2012ew,
  Griesshammer:2012we}, it is natural to resort to table~1 of
ref.~\cite{Griesshammer:2015ahu} for the central values and uncertainties for
all polarizabilities which are not well-determined by other means.
Uncertainties (theory and, as applicable, statistical) were combined in
quadrature in table~1 of ref.~\cite{Griesshammer:2015ahu}. Theory
uncertainties were derived as advocated in ref.~\cite{Furnstahl:2015rha}:
based on the order-by-order convergence of the series, one uses a Bayesian
analysis to predict the size of the first omitted order in the EFT expansion.

Four values in table~\ref{tab:polarizability_info} are not taken from
ref.~\cite{Griesshammer:2015ahu}. The best known is $\alpha_{E1}+\beta_{M1}$
which is most accurately determined not from Compton scattering experiments,
but from evaluations of the Baldin Sum Rule for the
proton~\cite{Gryniuk:2016gnm, Gryniuk:2015eza} and
neutron~\cite{Levchuk:1999zy, Levchuk:1999zy}. This recasts it as an
energy-weighted integral over photoproduction cross sections.  Likewise, the
GDH Sum Rule provides a highly precise value for the proton's
$\gamma_0$~\cite{Pasquini:2010zr}.

Meanwhile, for the spin-polarizability $\gamma_\pi$ of the proton and neutron,
there is information available from back-scattering Compton experiments. For
the proton we take the result $\gamma_\pi=7.7 \pm 1.8$ from
ref.~\cite{Schumacher:2005an}.  For the neutron the spin-polarizability
situation is less clear, since ``experimental" results that do exist are
obtained from experiments on a deuteron target after nuclear effects have been
removed. Therefore in the neutron case we employ the predictions inferred
from the \chiEFT\ results for the spin polarizabilities in table 1 of
ref.~\cite{Griesshammer:2015ahu}. We also employ the \chiEFT\ result from that
table for the proton spin polarizability, $\gamma_{E-}$. The last proton spin
polarizability, $\gamma_{M-}$, is derived from the the \chiEFT\ result for
$\gamma_{M1E2}$ and a fit value for $\gamma_{M1M1}$~\cite{McGovern:2012ew},
see below for further comments.

Alternative values for the spin polarizabilities with overall similar
uncertainty estimates are available~\cite{Gryniuk:2016gnm,
  Gryniuk:2015eza,Lensky:2015awa, Babusci:1998ww, Hildebrandt:2003fm,
  Holstein:1999uu}, as well as some from recent data
analyses~\cite{Martel:2014pba,Sokhoyan:2016yrc, Paudyal:2019mee, MartelPhD}.
Other recent extractions of polarizabilities from unpolarized data should also
be mentioned~\cite{Pasquini:2017ehj,Krupina:2017pgr,Pasquini:2019nnx}.

The values of $\alpha_{E1}-\beta_{M1}$ in table~\ref{tab:polarizability_info}
were determined in the \chiEFT\ variant we employ here from Compton scattering
data on the proton~\cite{McGovern:2012ew} and, for the neutron values, on the
deuteron~\cite{Myers:2014ace, Myers:2015aba}.  The fit of proton Compton
cross-section data in ref.~\cite{McGovern:2012ew} revealed a correlation
between $\alpha_{E1}-\beta_{M1}$ and $\gamma_{M-}$. Since our knowledge of
these two parameters comes from these data, that correlation should be
included in the prior. Translating the correlation matrix from fig.~12 of
ref.~\cite{McGovern:2012ew} into the polarizability basis used in this paper
and inflating the errors to account for theory uncertainty in the extraction
per the procedure of ref.~\cite{Griesshammer:2015ahu} we find a covariance
matrix of:
\begin{equation}
    \left(\begin{array}{cc}
\sigma^2_{\alpha_{E1}-\beta_{M1}} &  \sigma^2_{\alpha_{E1}-\beta_{M1} \; \gamma_{M-}}\\
    \sigma^2_{\alpha_{E1}-\beta_{M1} \; \gamma_{M-}} & \sigma^2_{\gamma_{M-}}     \end{array}
    \right)
    =\left( \begin{array}{cc}
    0.82 & 0.52 \\
    0.52& 0.80 \\ 
    \end{array} \right).
    \label{eq:corrmatrix}
\end{equation}

Ultimately the prior represents how extant experimental data and \chiEFT\
analysis constrain the polarizabilities. Therefore it should be obtained from
a \chiEFT\ analysis of Compton data that uses the statistical model for
theoretical uncertainties in observables that we formulated here. While the
analysis of ref.~\cite{McGovern:2012ew} included theoretical uncertainties in
its final results, it did so in a less sophisticated manner. Redoing the
extraction of polarizabilities in a way that is completely consistent with the
experimental-design analysis of this work is an important project for the
future. Until that is done though, we use a prior based on
ref.~\cite{McGovern:2012ew} supplemented by other information, namely the
prior we have described in this section.  We reiterate that the conclusions of
the design analysis are insensitive to the central values adopted for the
polarizabilites in the prior, $\mu_0$. Changes in the correlation structure
could affect them. But adopting different, reasonable correlations in the
prior does not produce marked changes in expected utilities or posterior
shrinkages---as the results we now present will demonstrate.

\section{Results}
\label{sec:results_experimental_design}

It is useful to start the presentation of results with the customary word of
caution in mathematical statistics.  The predictions which form the output of
this formalism should be understood as likely outcomes, not as guarantees.
They carry ``errors on the errors''.  Details depend on our input choices
(priors) and model assumptions, and it is an advantage of the Bayesian
approach that these must be discussed explicitly.  We found our results to be
robust against other reasonable choices, though reasonable people can make
other reasonable choices, which then leads to scientific progress by
discussion.  Overall, the choices we explored led to different details, but
not to substantially different outcomes.

We would therefore not label one design's superiority as significant if its
decrease in uncertainty [see eq.~\eqref{eq:percent_decrease}] is within a few
percentage points of others. But we are confident that a difference of, say,
ten percent indicates a clear preference of one design over others.

The guidance we provide for observables and kinematic locations is documented
in a publicly available Jupyter notebook~\cite{BUQEYEgithub}.  We hope this
will facilitate improvements on this analysis, which is meant to be the first,
not the last, word in the ongoing conversation regarding the best way to
improve the constraints on the nucleon polarizabilities.

\subsection{Precision Levels and Constraints of Compton Experiments}
\label{sec:experiments}

We attempt to choose estimates of experimental input which are realistic for
modern accelerators and detectors, but also realize that the specifics depend
on experimental details. For a first take, we focus on a scenario which is not
optimized to a \emph{particular} facility but should be at least of some use
for planning and design at \emph{any} facility. We also do not assess the
impact of common-mode errors on the accuracy with which polarisabilities can
be extracted, but instead consider only point-to-point (``statistical")
errors.

We consider three levels of point-to-point detector precision, see
table~\ref{tab:experimental_precision_levels}. We believe these provide a
range of plausibly achievable experimental uncertainties for measurements on
the proton.  The ``standard'' scenario assumes uncertainties in the cross
section of $\pm5\%$ (point-to-point systematic and statistical uncertainties
combined in quadrature), and an absolute uncertainty in spin observables of
$\pm0.10$.  This is state-of-the-art for proton Compton experiments for the
cross section and those spin observables that have already been
measured~\cite{Martel:2014pba, Sokhoyan:2016yrc, Paudyal:2019mee,
  Martel:2019tgp,Martel:2017pln, Ahmed:2020hux, privcomm}.  A second scenario
lists experimental error bars $\Delta\!\diffcs\approx\pm4.0\%$ and an absolute
$\Sigma_i$ error of $\pm0.06$ which are deemed ``doable'' nowadays without
excessive improvements.  The ``aspirational'' scenario assumes considerable
but realistic new resources and possibly new equipment.  Our choices were
informed by discussions with our experimental colleagues who work on Compton
scattering at MAMI and HI$\gamma$S, for whose input we are very
grateful~\cite{privcomm}. Unless otherwise stated, all results in figures
assume the ``doable'' level of experimental precision, with the remaining
levels reserved for \verifyvalue{the Supplemental Material.}

\begin{figure*}[!tbp]
    \centering
    \includegraphics[width=\textwidth]{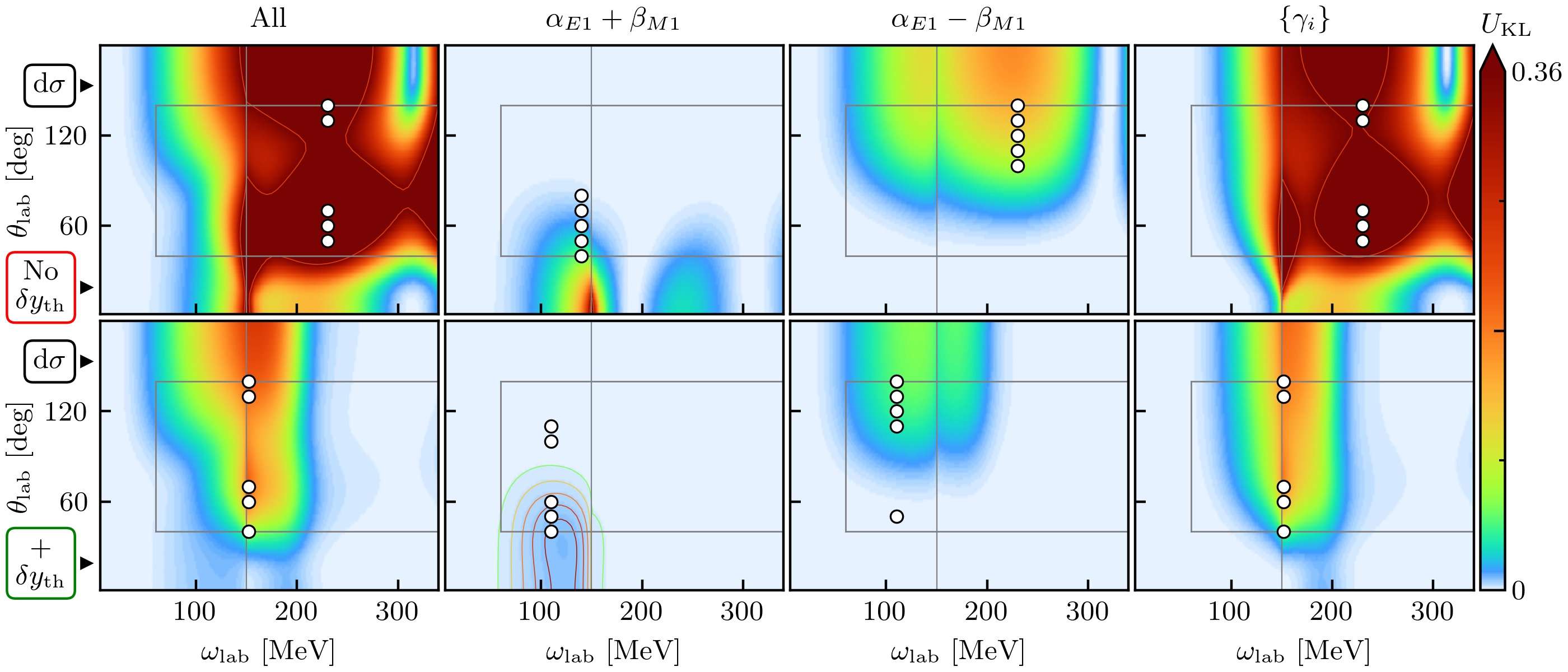}
    \label{fig:utility_grid_compare_subsets_and_truncation_proton}
    \caption{(Colour online) The expected utility
      eq.~$\eqref{eq:utility_kl_analytic}$ of proton differential cross
      section ($\diffcs$) measurements.  Colors indicate the utility of one
      measurement conducted at each kinematic point $(\omegalab, \thetalab)$,
      with the point of largest utility $U_{KL}$ being by definition the
      optimal 1-point design. (The color bar is on a linear scale, though the
      hue varies \emph{much} more quickly for small $U_{KL}$.)  The top row
      (with the red, ``No $\delta\genobsth$'' box) does not include EFT
      truncation estimates, whereas the bottom row does include the EFT
      uncertainty.  Each column shows the utility one could expect to achieve
      for a subset of the proton polarizabilities: the first column considers
      all polarizabilities together, while the second and third show the gain
      for $\alpha_{E1}\pm\beta_{M1}$ individually, and the final column
      reflects the collective information to be learned about the four spin
      polarizabilities $\{\gamma_i\}$.  The interior box that excludes
      $\omega < 60\MeV$ along with forward and backward angles marks the
      experimentally accessible regime (see text for further discussion).  The
      vertical line marks the cusp at the pion-production threshold.  To
      visualize the full range of variation in the subplots, the color ranges
      from zero to a saturation point calculated by averaging the optimal
      utilities across all subplots.  Contour lines are added for utilities
      above the saturation point, with the first contour at saturation and
      subsequent contour lines at intervals of 50\% of the color range, see
      the top left subplot.  To help visualize the utilities from
      less-constraining measurements, additional contours are added to any
      subplot whose maximum utility is less than 10\% of the color range, see,
      \eg,
      fig.~\ref{fig:utility_grid_subset_polarizabilities_observable_set_1}.
      Unless otherwise stated, the color ranges are common to all subplots
      within a figure, but different between figures.  The white circles with
      black borders show the optimal five-point design kinematics, as
      described in the text.  The effect of including EFT truncation errors is
      striking: it shifts the region of optimal utility to lower energies and
      moderates the expected utility.  }
    \label{fig:utility_grid_compare_subsets_and_truncation}
\end{figure*}

\begin{table}[tb]
\renewcommand{\arraystretch}{1.6}
\caption{The experimental noise levels for the differential cross section $\diffcs$ and spin observables $\Sigma_i$. The standard deviation for $\diffcs$ is given in a percentage of its predicted value (using the prior mean for the polarizabilities), whereas the standard deviation for the spin observables is on an absolute scale.
}
\label{tab:experimental_precision_levels}
\begin{ruledtabular}
\begin{tabular}{ld{2.2}d{2.2}}
Level & \multicolumn{1}{c}{$\diffcs$ (\%)} &  \multicolumn{1}{c}{$\Sigma_i$ (abs.)}\\
\colrule
Standard     & \pm 5.00 & \pm 0.10 \\
Doable       & \pm 4.00 & \pm 0.06 \\
Aspirational & \pm 3.00 & \pm 0.03 \\
\end{tabular}
\end{ruledtabular}
\end{table}

It should be noted that achieving even ``standard'' errors of $\pm0.10$ for
some of the hitherto-unmeasured spin observables is not simple. Especially for
the spin-polarization transfer observables, the experimental challenges of
detecting recoil spin polarizations are considerable.  In that case, the
estimate can serve as benchmark, with the ``standard'' scenario already an
``aspirational goal.''  Due to the absence of quasi-stable free-neutron
targets, a ``standard'' uncertainty for neutron Compton scattering is of
course well beyond ``aspirational.'' We nonetheless chose to use the same
error bars for the neutron, to ease comparison.

We search for the optimal one-point design and the optimal five-point design,
\ie,~a search over all accessible combinations of five unique angles at a
given $\omegalab$ (``5-point design''). In line with experimentalists'
constraints on the placement of bulky detectors, we require that the angles be
at least $\verifyvalue{10^\circ}$ apart.  The focus on one photon energy and
multiple angles mirrors the capabilities of ``monochromatic-beam'' facilities
like HI$\gamma$S which measure several angles at one energy simultaneously,
but many other choices could be made. The assessment can easily be extended to
``bremsstrahlung facilities,'' where a number of both angles and energies can
be measured simultaneously.  In that case, a typical spacing between the
central energy of each energy bin of about $10$ to $20\MeV$ appears realistic,
given that a sufficient number of events must be collected in each ``bin'' for
meaningful statistics~\cite{Martel:2014pba, Sokhoyan:2016yrc, Paudyal:2019mee,
  Martel:2019tgp,Martel:2017pln, Ahmed:2020hux, privcomm}.  Hence, our results
attempt to be as realistic as possible given the choices above, and are a
proof of principle for further, more specific research.

While the plots show a full range of energies and angles, we also indicate on
them regions defined by $\omega\le\verifyvalue{60}\MeV$ or
$\theta\le\verifyvalue{40}^\circ$ or $\theta\ge\verifyvalue{150}^\circ$ in
which experiments are unlikely to be conducted, because forward and backward
angles are physically hard to access, or because sensitivity to
polarizabilities at very low energies is negligible. Therefore, we do not
elaborate on designs that involve these kinematic regions.

Only the cross section and $\Sigma_{3}$ are non-vanishing as $\omega\to0$, the
physics of both being governed by the Thomson limit, with polarizability
corrections very small. Our LO result provides the correct Thomson limit for
each observable automatically, and we constrained the unknown higher-order
corrections so that they do not change this; see
app.~\ref{sec:truncation_model_details_compton}.  Indeed, we find a typical
energy correlation length of $\ell_\omega\approx\verifyvalue{50}\MeV$ for the
proton; see table~\ref{tab:truncation_details_observables}, so that such a
constraint becomes less important around and above $60\MeV$.

In addition, due to the coordinate singularity at $\theta=0^\circ$ and
$180^\circ$, observables or their derivatives with respect to $\theta$ must be
zero there. We implemented these constraints as described in
app.~\ref{sec:truncation_model_details_compton}.  As the angular correlation
lengths from table~\ref{tab:truncation_details_observables} are all smaller
than $\ell_\theta\approx\verifyvalue{60^\circ}$, this is not a strong
constraint on observables at intermediate angles where experiments are most
feasible.

Our design model does not include the constraint that spin observables
$\Sigma_i$ can only have values between $-1$ and $1$.  This is a reasonable
omission because the mean value of most $\Sigma_i$ and their \chiEFT\
uncertainties are mostly well contained within these bounds (except maybe at
the largest $\omega$), see app.~\ref{sec:truncation_model_details_compton} for
details.

Finally, we reiterate that around the pion-production threshold
($\omega_\pi\,\pm\,20\MeV$ or so, with $\omega_\pi\approx150\MeV$ marked by a
vertical line in plots), the \chiEFT\ truncation error estimates for the spin
observables $\Sigma_i$ are less understood, and experimental conditions are
difficult as well; see sec.~\ref{sec:truncation_distribution_compton} and
app.~\ref{sec:truncation_model_details_compton}.

\subsection{First Discussion and Impact of Accounting for Theory Uncertainties: The Cross Section}
\label{sec:crosssection}

The unpolarized differential cross section is the most extensively studied
nuclear Compton scattering observable.  Therefore we begin by showing the
expected utility of further measurements, with the goal of constraining
various subsets of the polarizabilities.  Our results focus on proton
observables unless otherwise stated, due to the difficulty of performing
experiments on neutrons; see \verifyvalue{the Supplemental Material} for the
corresponding neutron design results.  We start by considering the following
subsets: all polarizabilities simultaneously, only $\alpha_{E1}+\beta_{M1}$,
only $\alpha_{E1}-\beta_{M1}$, and only the spin polarizabilities
$\{\gamma_i\} \equiv \{\gamma_{0}, \gamma_{\pi}, \gamma_{E-}, \gamma_{M-} \}$,
or each of them separately.

Figure~\ref{fig:utility_grid_compare_subsets_and_truncation} shows the
expected utility of future proton experiments, with and without an estimate of
the truncation error.  Without truncation, the utility of an experiment to
measure $\alpha_{E1}\pm\beta_{M1}$ mirrors the sensitivity analysis of fig.~8
in ref.~\cite{Griesshammer:2017txw}.  There, the derivative of the observable
with respect to a particular polarizability was plotted, and the truncation
error of the EFT was only accounted for indirectly by casting a ``gray mist''
over the plot which thickens into the Delta resonance region, starting at
$\omega\gtrsim210\MeV$, \ie,~where our transition region starts.

In fact, the constraining power of truncation-free 1-point measurements (as
measured by $U_{\text{KL}}$) on each individual polarizability follow exactly
the patterns of the local sensitivities for all observables and
polarizabilities.  This can be verified by comparing the remainder of our
zero-truncation-error results in \verifyvalue{the Supplemental Material} with
the appropriate subplots of figs.~9--20 in ref.~\cite{Griesshammer:2017txw}.

However, when truncation-error estimates are included, the optimal designs are
pushed to lower $\omegalab$, with a particularly dramatic shift for
$\alpha_{E1} - \beta_{M1}$.  This is expected because the \chiEFT\ uncertainty
$\delta\genobs_k$ increases with energy.  Still, the optimal locations for
constraining the spin polarizabilities often remain at or above the
pion-production threshold.

One of the benefits of our Bayesian analysis over a purely derivative-based
approach---like that of ref.~\cite{Griesshammer:2017txw}---is that we can
examine the collective gain in information for multiple polarizabilities.  The
second, third, and fourth panels in the lower row of
fig.~\ref{fig:utility_grid_compare_subsets_and_truncation} provide the optimal
kinematics at which to constrain either of the scalar polarizabilities, or the
spin-polarizabilities collectively. In these second and third panels (fourth
panel) the design choice is computed unconditional on the other five (two)
polarizabilities, \ie we marginalize over the polarizabilities we are not
interested in for that panel.  Looking across them reveals that the collective
utility in the first panel is approximately the sum of the utilities of each
subset.  In this study, we found that the correlations between these linear
combinations of polarizabilities that are induced by fitting are rather small;
see the extended discussion of fig.~\ref{fig:shrinkage_per_subset}.
Equation~\eqref{eq:utility_kl_analytic} then says that, to the extent that the
covariance matrices $V_0$ and $V$ are diagonal, the total utility is the sum
of the individual utilities.  The feature seen here is thus generic in the
absence of correlations: the amount of benefit derived from collectively
constraining $\lecs$ is related to how much the utilities for individual
components of $\lecs$ overlap in kinematic space.

\subsection{All Observables: Discussion}
\label{sec:observables}

\begin{figure*}[!t]
    \centering
    \includegraphics[width=\textwidth]{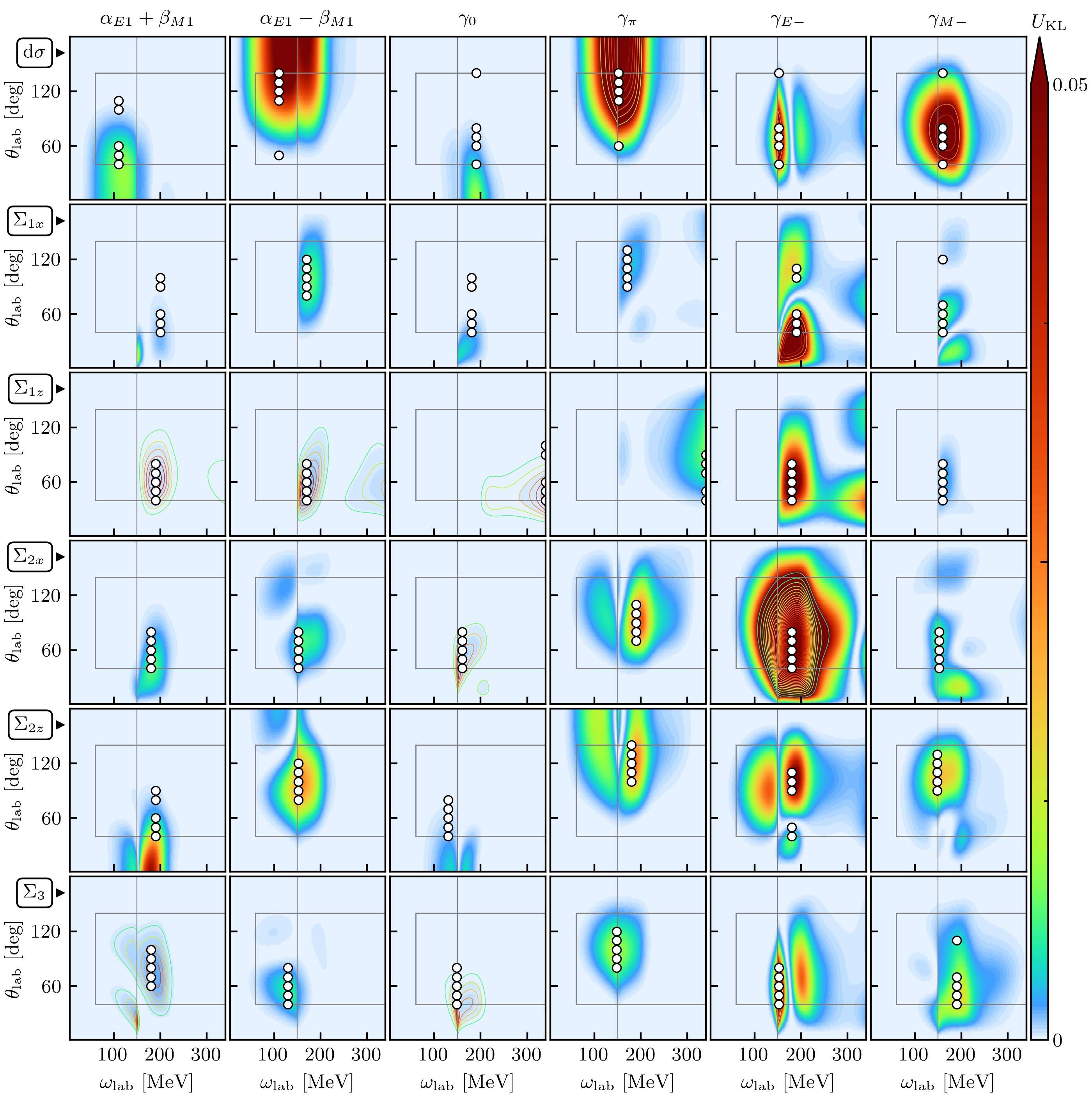}
    \caption{(Colour online) Expected utilities of conducting various
      experiments, with the goal of constraining individual proton
      polarizabilities.  Truncation errors are included throughout.  See
      fig.~\ref{fig:utility_grid_compare_subsets_and_truncation} for a
      detailed description of the figure notation.  Again, the color scale is
      common among all subplots so that both the location and relative
      magnitudes of utilities can be uncovered.  To facilitate comparisons
      between observables, the color range is identical to that in
      fig.~\ref{fig:utility_grid_subset_polarizabilities_observable_set_2}. However
      it must be noted that the color scale is very, very different from
      fig.~\ref{fig:utility_grid_compare_subsets_and_truncation}, and that
      dark red in the current plot reflects only a modest increase in
      information.  }
    \label{fig:utility_grid_subset_polarizabilities_observable_set_1}
\end{figure*}

\begin{figure*}[!t]
    \centering
    \includegraphics[width=\textwidth]{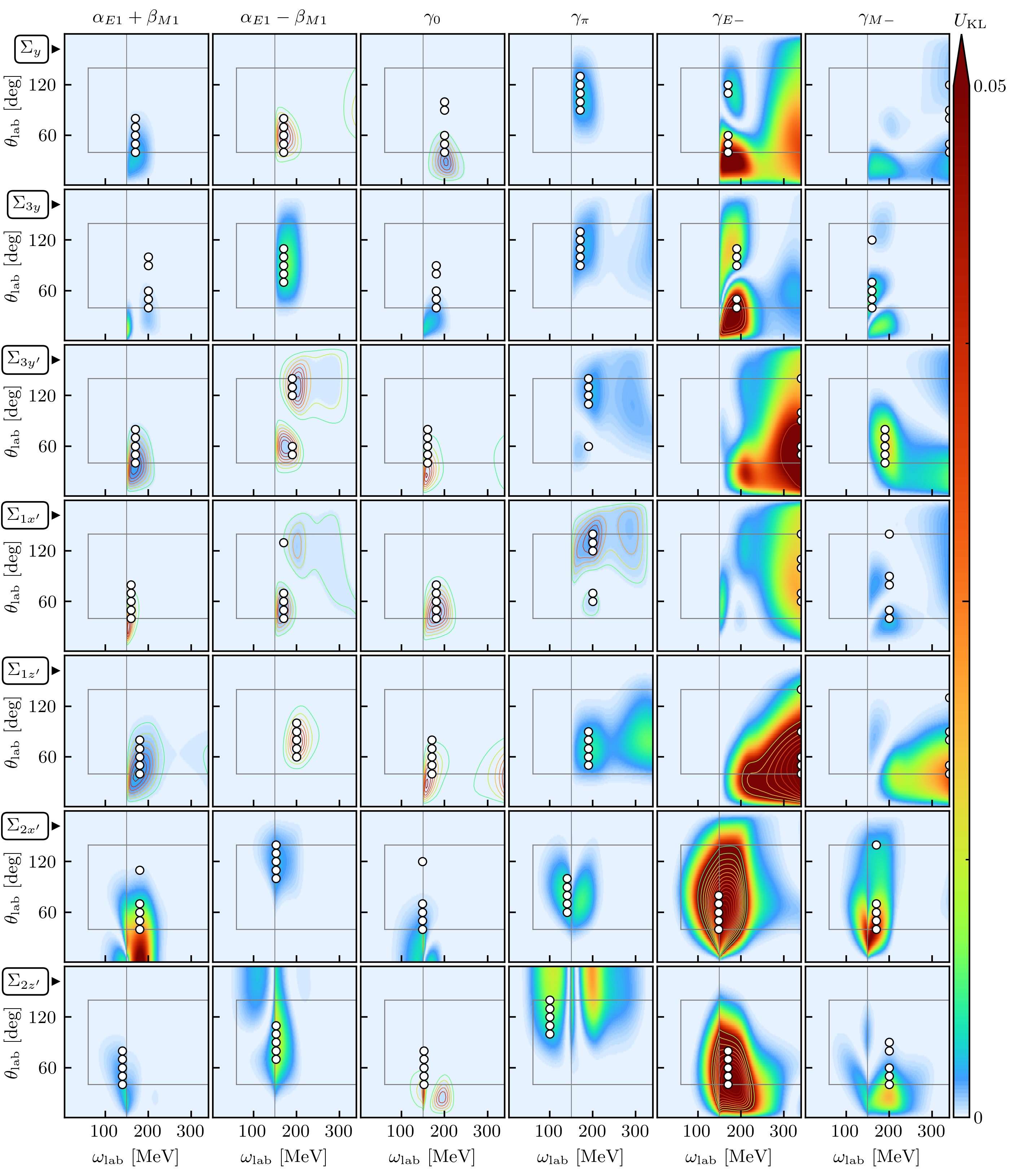}
    \caption{(Colour online) The remaining observables, see
      fig.~\ref{fig:utility_grid_subset_polarizabilities_observable_set_1}.
      Even with a saturated color scale that greatly exaggerates small values,
      the utilities from different measurements cover a vast range; see
      comments in
      fig.~\ref{fig:utility_grid_subset_polarizabilities_observable_set_1}.  }
    \label{fig:utility_grid_subset_polarizabilities_observable_set_2}
    \vspace*{1in}
\end{figure*}

We now extend our analysis of the differential cross section to the spin
observables $\Sigma_i$.  For the remainder of this work, we include truncation
error estimates, because otherwise the constraining power of any measurement
would be overstated; see \verifyvalue{the Supplemental Material} for
corresponding results without truncation errors included.
Figures~\ref{fig:utility_grid_subset_polarizabilities_observable_set_1}
and~\ref{fig:utility_grid_subset_polarizabilities_observable_set_2} show heat
maps of the expected utility of all proton observables, with truncation-error
estimates included.  Note that the scale has changed dramatically, as can be
seen by comparing the results in the bottom row of
fig.~\ref{fig:utility_grid_compare_subsets_and_truncation} to the same results
repeated in the top row of
fig.~\ref{fig:utility_grid_subset_polarizabilities_observable_set_1}.

\begin{figure*}[!bth]
    \centering
    \includegraphics[width=\textwidth]{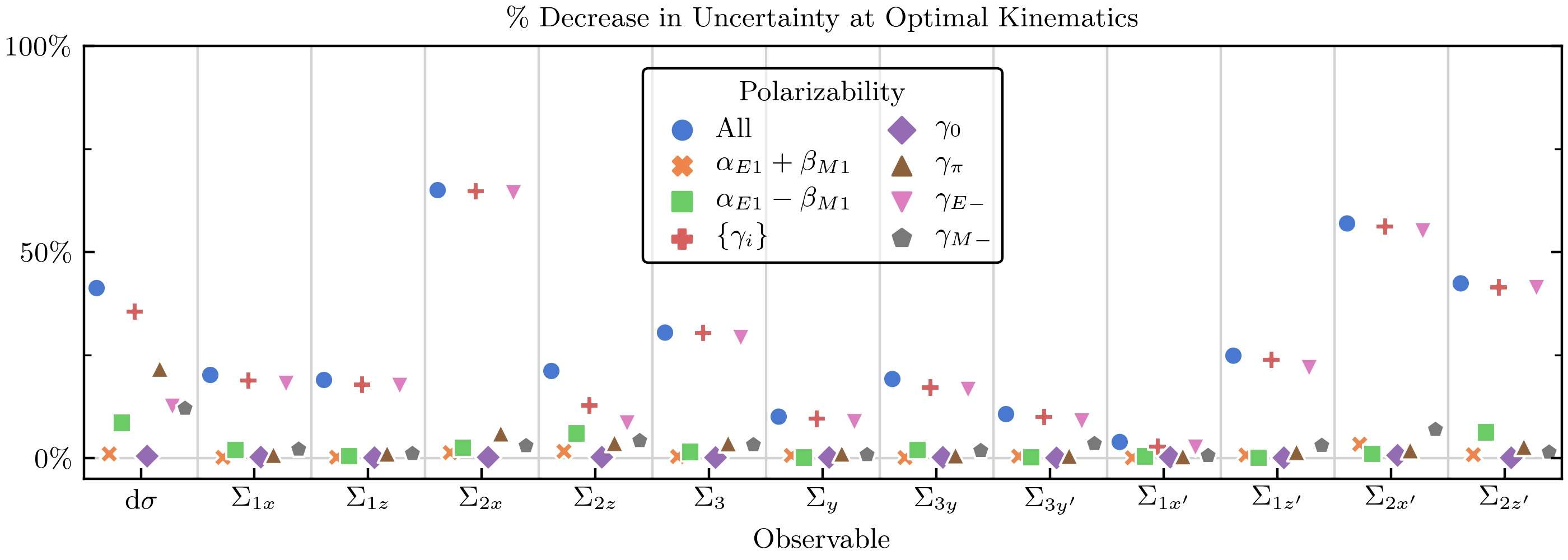}
    \caption{(Colour online) The percent decrease in uncertainty or
      ``information gain" [eq.~\eqref{eq:percent_decrease}] computed while
      including truncation errors for the optimal 5-point proton Compton
      experiments.  The maximum for each polarizability and observable is
      likely to be obtained at its \emph{own} unique kinematic points, see
    figs.~\ref{fig:utility_grid_compare_subsets_and_truncation},~\ref{fig:utility_grid_subset_polarizabilities_observable_set_1},
      and~\ref{fig:utility_grid_subset_polarizabilities_observable_set_2}.
      The spin observables $\Sigma_{2x}$ and $\Sigma_{2x'}$ show powerful
      constraining power of $\gamma_{E-}$, with the potential to shrink its
      uncertainty by 50\%.  In contrast, measurements of the differential
      cross section $\diffcs$ can reduce the uncertainty in $\gamma_{\pi}$,
      $\gamma_{E-}$ and $\gamma_{M-}$ by about 30\% (collectively), and also
      can shrink the error on $\alpha_{E1}-\beta_{M1}$ by about 25\%.  Other
      observables show slight constraining power. Note that the (primed)
      spin polarization transfer observables are notoriously difficult to
      measure.  }
    \label{fig:shrinkage_per_subset}
\end{figure*}

Figures~\ref{fig:utility_grid_subset_polarizabilities_observable_set_1} and
\ref{fig:utility_grid_subset_polarizabilities_observable_set_2} contain a
wealth of information about the relative utility between observables at
various kinematics, but for more readily interpretable statements about
potential constraining power, we turn first to
fig.~\ref{fig:shrinkage_per_subset}.  In what follows, we will discuss the
observables in turn based on the promise shown in that figure, at the same
time looking at the relevant row of
Figures~\ref{fig:utility_grid_subset_polarizabilities_observable_set_1} and
\ref{fig:utility_grid_subset_polarizabilities_observable_set_2} for more
detailed information.

From fig.~\ref{fig:shrinkage_per_subset} we can see the largest percent
decrease in uncertainty or ``information gain"
[eq.~\eqref{eq:percent_decrease}] of all optimal 5-point designs for each
observable, with utilities split up based on the polarizabilities one might be
interested in measuring. Thus, given a decision about which polarizability is
of most interest---a decision which we do not encode mathematically---our
approach provides a quantitative method for evaluating the worth of future
experiments.

Any \emph{set} of utilities, such as ``All'' or ``$\{\gamma_i\}$'' are
guaranteed to be greater than or equal to the optimal utility of any
individual polarizability that contributes to it.  How much more is learned by
considering multiple polarizabilities, depends on how much their optimal
designs overlap in kinematic space---because we have found that in this case
only small correlations are induced in the covariance matrix $V$ by fitting.

For the proton, the combinations $\alpha_{E1}+\beta_{M1}$ and $\gamma_0$ are
well-constrained by sum rules~\cite{Gryniuk:2015eza, Gryniuk:2016gnm}; see the
small error bars in table~\ref{tab:polarizability_info}. In Compton
scattering, these are the only two linear combinations of polarizabilities
which enter the cross section as $\theta\to0$.
Figure~\ref{fig:shrinkage_per_subset} reveals that indeed little information
on them can be gained from direct Compton experiments. The other four
combinations, $\alpha_{E1}-\beta_{M1}$, $\gamma_\pi$, $\gamma_{E-}$ and
$\gamma_{M-}$, will therefore dominate our discussion.

One observable that stands out in fig.~\ref{fig:shrinkage_per_subset} is
$\Sigma_{2x}$ (circularly polarized photons on a transversely polarized
target). Its overall utility of over 60\% stems near-exclusively from the gain
in $\gamma_{E-}$.  Combining this with
fig.~\ref{fig:utility_grid_subset_polarizabilities_observable_set_1}, we see
that the gain occurs in a quite robust and large region which extends from
about $\omega_\pi$ to above $200\MeV$, at angles between
$\theta \approx 30^\circ$ and $90^\circ$.
The other polarizabilities are optimally constrained in a similar kinematic
region, but according to fig.~\ref{fig:shrinkage_per_subset}, their
contribution to the overall utility from such an experiment is negligible
($\lesssim5\%$) compared to what would be learned about $\gamma_{E-}$.  This
means that measurements of $\Sigma_{2x}$ in that region allow for an
extraction of $\gamma_{E-}$ which is highly insensitive to the particular
values of $\alpha_{E1}$, $\beta_{M1}$ and the other spin-polarizabilities
used.  This observable was already explored in a pioneering experiment at MAMI
for $\omega\approx290\dots330\MeV$ \cite{Martel:2014pba}, where unfortunately
the information content of an EFT interpretation is not very high.  (Remember
that the color scale in
fig.~\ref{fig:utility_grid_subset_polarizabilities_observable_set_1} changes
\emph{rapidly} with decreasing utility.)  This analysis implies that much more
information can be gained from an experiment at $\omega\lesssim200\MeV$.

A similarly large constraint on the polarizabilities comes from the analogous
polarization-transfer observable $\Sigma_{2x'}$ (incident circularly polarized
photon on unpolarized target, transverse spin polarization of recoil proton
detected).  In this observable $\gamma_{E-}$ is a little less constrained than
it is in $\Sigma_{2x}$, but the information gain for $\Sigma_{2x'}$ is still
more than $50\%$. Meanwhile, the shrinkage for $\gamma_{M-}$ increases
slightly compared to $\Sigma_{2x}$: it is now a bit less than $10\%$. Some
limited information about $\gamma_{M-}$ can perhaps be gained here, but such a
small information gain should not be over-interpreted.  To achieve these
shrinkages a $\Sigma_{2x'}$ experiment would need to be made near---or a few
dozen MeV above---$\omega_\pi$ and towards forward angles.

The polarization transfer $\Sigma_{2z^\prime}$ (incident circularly polarized
photon on unpolarized target with detection of longitudinal recoil
polarization) provides a gain of about $40\%$ overall, at similar energies but
slightly smaller angles.  Most of the gain is again in $\gamma_{E-}$, followed
by a small gain in $\alpha_{E1}-\beta_{M1}$.

Decent information gain on $\gamma_{E-}$ (about $30\%$) can also be found from
measuring the beam asymmetry $\Sigma_3$ (linearly polarized beam on
unpolarized target) at intermediate angles in two narrow corridors, namely
close to the pion-production threshold and slightly higher,
$\omega\approx200\MeV$.  Some data are actually available
there~\cite{Blanpied:2001ae, Sokhoyan:2016yrc, Martel:2017pln, CollicottPhD}
but these have not yet been analyzed in EFT\@ and thus do not enter the prior
in table~\ref{tab:polarizability_info}. These results for $\Sigma_3$ suggest
that such an analysis could be valuable.

Measurements of $\Sigma_3$ at lower $\omega$ have been used in attempts to
constrain the scalar polarizability $\beta_{M1}$~\cite{Sokhoyan:2016yrc}. But
we see that even in the most sensitive kinematics the impact of this
observable on $\beta_{M1}$ amounts to just a few percent.

Instead, the combination $\alpha_{E1}-\beta_{M1}$ can best be measured from
the cross section in a region somewhat above $100\MeV$ at back-angles. The
overall information gain from a five-point measurement there is
$\approx 10\%$.  Qualitatively, this angle regime is not surprising since it
is well known that this particular linear combination enters the cross section
as $\theta\to180^\circ$, as does $\gamma_\pi$. Nevertheless, the conclusion
may at first be surprising since there is already an extensive data set from
ref.~\cite{OlmosdeLeon:2001zn} in this energy region, and contemporary
evaluations of $\alpha_{E1}-\beta_{M1}$, encoded in our prior, lean heavily on
this data set. In fact, the possibility seen in
fig.~\ref{fig:shrinkage_per_subset} to improve knowledge of
$\alpha_{E1}-\beta_{M1}$ by measuring the cross section is supported by simple
estimates, as follows. There are 20 Olmos de Le\'on {\it et al.} data points
corresponding to lab photon energy $> 135$~MeV. How then, can five additional
data points result in significant information gain for
$\alpha_{E1}-\beta_{M1}$? In fact, these Olmos de Le\'on {\it et al.} points
have precision of 7--15\%. When combined with the point-to-point systematic of
5\% their statistical power is slightly less than that of five additional
points with 5\% precision.

Interestingly, the next-largest information gain for $\alpha_{E1}-\beta_{M1}$
appears to be found in $\Sigma_{2z}$ (circularly polarized beam on
longitudinally polarized target) and the corresponding polarization transfer,
$\Sigma_{2z^\prime}$, but these amount to only 5--10\%.  In both these cases
the region of greatest sensitivity lies right at the pion-production
threshold, where experiments are particularly challenging and where our
\chiEFT\ uncertainties may be less accurate (see
sec.~\ref{sec:truncation_distribution_compton}).

According to fig.~\ref{fig:shrinkage_per_subset}, $\gamma_{M-}$ is quite
elusive.  Only the differential cross section shows appreciable information
gain (about $15\%$), while the next-largest gain, in $\Sigma_{2x^\prime}$, is
$\approx 10\%$.  In all three observables, the region of largest sensitivity
to $\gamma_{M-}$ is right at the $\omega_\pi$ cusp, where the $\diffcs$
convergence pattern is well behaved (see
fig.~\ref{fig:coefficients_dsg_slices}).  This makes us more confident in our
design predictions for $\diffcs$ than for the spin observables, which
fluctuate more strongly.  Taking all this into account, we find that a
measurement of the cross section in a broad band around $\omega_\pi$ and at
intermediate angles is the best chance to constrain $\gamma_{M-}$.  As a
bonus, such a measurement would concurrently constrain other polarizabilities
``for free.''

For $\gamma_\pi$, any information gain can only be found in the cross section
and is about 20\%; several other polarizabilities contribute similar amounts;
see discussion above.  A dedicated $\theta=180^\circ$ experiment, like that of
ref.~\cite{Zieger:1992jq}, may be able to resolve $\gamma_\pi$ and
$\alpha_{E1}-\beta_{M1}$, but needs a special design.  No other polarizability
combination enters at that angle.

Indeed, optimal 5-point measurements of the differential cross section
$\diffcs$ can decrease the collective uncertainty of all polarizability
combinations by about $40\%$, but the information gain is spread out amongst
individual polarizabilities: about 15--20\% for $\gamma_\pi$, $\gamma_{E-}$,
and $\gamma_{M-}$, $10\%$ for $\alpha_{E1}-\beta_{M1}$, and no perceptible
information gain for $\gamma_0$ and $\alpha_{E1} + \beta_{M1}$. In part,
different kinematic regions are sensitive to individual combinations, so
measurements across a wide array of energies and angles can be used to
disentangle individual contributions.

The correlation structure implemented in the prior can affect these
experimental-design conclusions.  To check this, we also ran an
``uncorrelated" design analysis in which the off-diagonal elements of the
correlation matrix for $\alpha_{E1}-\beta_{M1}$ and $\gamma_{M-}$ given in
eq.~(\ref{eq:corrmatrix}) were set to zero. Results for this alternative prior
choice are provided in the \verifyvalue{Supplemental Material}, see
figs.~\ref{fig:utility_grid_subset_polarizabilities_observable_set_1_uncorrelated}--\ref{fig:shrinkage_per_subset_uncorrelated}. The
results for these two priors are very similar. The most notable difference is
that the information gain expected in $\alpha_{E1}-\beta_{M1}$ from
differential cross section measurements is a little larger if an uncorrelated
prior is employed, cf.\ fig.~\ref{fig:shrinkage_per_subset} and
\ref{fig:shrinkage_per_subset_uncorrelated}. The amount that we expect to be
learnt from $\Sigma_{2z}$ and $\Sigma_{2x'}$ also increases slightly with use
of an uncorrelated prior. This, as well as the other alterations in shrinkage
and utilities that result from changing the prior's correlation structure, are
well within the impact of different ``reasonable prior choices" that we
anticipated at the beginning of this section.

We now highlight an important point regarding the fact that our experimental
design analysis is actually ``aware" of the experimental information that is
presently available. The utility function does not explicitly trace the
kinematics and quality of available data. But the significant amount of proton
Compton cross section data enters via the priors on $\alpha_{E1}$,
$\beta_{M1}$ and $\gamma_{M-}$ in table~\ref{tab:polarizability_info}, via the
experimental (statistical plus systematic) uncertainties in the fits of those
quantities.  The fact that additional information can be gained from more
high-quality data in specific kinematic regions implies that the quantity and,
most importantly, the quality of future Compton data in that region can
provide important information gains on the polarizabilities, even if that
region appears at first glance to be already saturated.  On the other hand,
the available data for both $\Sigma_{2x}$~\cite{Martel:2014pba, MartelPhD} and
$\Sigma_{2z}$~\cite{Martel:2017pln,Paudyal:2019mee} did not enter in the
priors of table~\ref{tab:polarizability_info}. That is because those data were
taken in the Delta resonance region, where the sensitivity of these two
observables to any polarizability is minuscule, according to
fig.~\ref{fig:utility_grid_subset_polarizabilities_observable_set_1}.  Adding
their information to the priors will therefore not change our conclusions or
improve polarizability error bars.

The relatively steep differences in information-gain reflect to a large extent
the fact that the uncertainties for the spin polarizabilities and for
$\alpha_{E1}-\beta_{M1}$ are substantially larger than for the sum-rule
constrained combinations $\alpha_{E1}+\beta_{M1}$ and $\gamma_0$.  In
particular, the very small error bar on $\gamma_0$ makes it very hard to gain
information via Compton scattering; fig.~\ref{fig:shrinkage_per_subset} shows
that the expected shrinkage is indeed close to zero for all observables.

We caution, however, that the prior size of a polarizability's error bar is
not by itself a reliable predictor of possible information gain. The error bar
of $\gamma_{M-}$ is about half of that of $\gamma_\pi$ or $\gamma_{E-}$, so
one might expect the information gain in measuring it to be about half of that
for $\gamma_\pi$ or $\gamma_{E-}$. Instead, a measurement with substantial
information gain is much more elusive than that, as explained above.

The biggest sensitivity of $\Sigma_{3y^\prime}$ to $\gamma_{E-}$ and of
$\Sigma_{1z^\prime}$ to both $\gamma_{E-}$ and $\gamma_{M-}$ is pushed to the
maximum considered energies $\omega\approx300\MeV$.  On the one hand, such
behavior might be interpreted as in apparent tension with the fact that
\chiEFT\ is significantly less reliable in the Delta resonance region than at
lower energies.  On the other hand, the uncertainty of \chiEFT\ is accounted
for in our experimental design; see discussion in sec.~\ref{sec:crosssection}.
Possibly, the 5-point design in that region probes a sensitivity of the
correlated angular dependence at high energies, rather than on individual
values/rates at a particular angle. If so, and if \chiEFT\ predicts these
correlations more robustly than overall sizes of an observable, then the
phenomenon would be explained and measurements of the functional dependence of
observables on angle at such high energies could provide determinations of
$\gamma_{E-}$ and $\gamma_{M-}$.  However, optimal 1-point designs would not
be sensitive to correlations and still appear sometimes at very high energies
in these same observables. Apparently, the sensitivity is so strong at such
kinematics as to win over the decreased theory uncertainties.  As we did not
find an intuitively obvious resolution, this merits further study.

Equally as notable as these substantial information gains on polarizabilities
are those observables that seem to provide almost no information about the
polarizabilities at this level of experimental and theoretical precision. The
most prominent such example is $\Sigma_{1x'}$ (total gain $<5\%$).  Given our
truncation error estimates for this quantity, there is little information on
the polarizabilities to be gained from any 5-point experiment. Measuring it,
or indeed any observable, in a region where the information gain for
polarizabilities is negligible, can still be useful though. It provides
information about how accurately \chiEFT\ describes the Compton process,
independent of the polarizabilities. This is an important cross-check of
\chiEFT, even though it is not part of the utility used in this work.

\begin{figure}[tb]
    \centering
    \includegraphics[width=\linewidth]
    {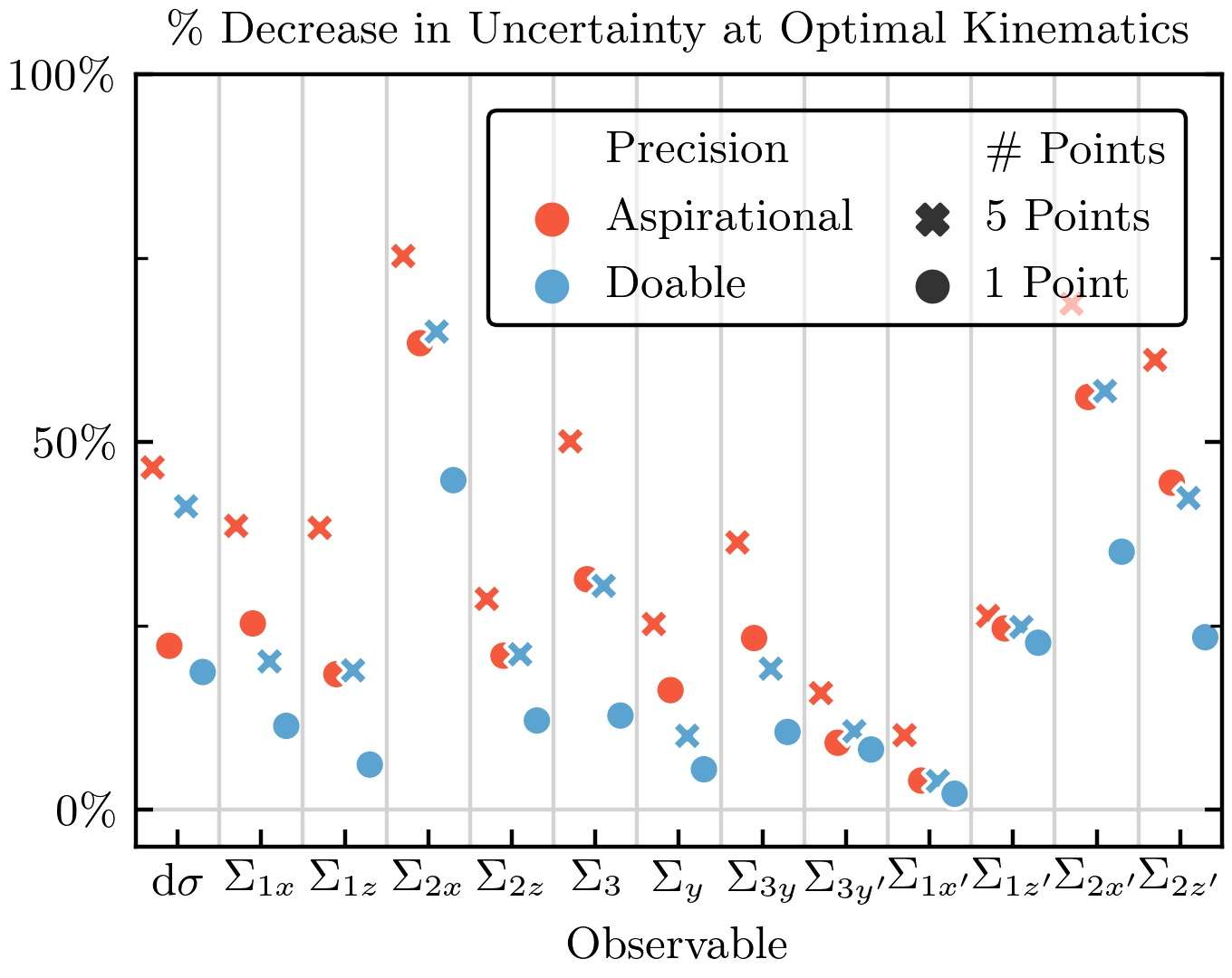}
    \caption{(Colour online) The percent decrease in $\lecs$ uncertainties, as
      in fig.~\ref{fig:shrinkage_per_subset}, applied to decide on the
      trade-off between different allocations of experimental resources
      (exploration vs.\ exploitation).  Larger values imply that the
      measurement is more informative.  The 1-point (5-point) optimal design
      is denoted by a circle (cross), and the experimental precision levels
      are given in table~\ref{tab:experimental_precision_levels}.  The
      decision to increase precision or measure at more kinematic points (or
      neither) can vary significantly by observable.  }
    \label{fig:shrinkage_fixed_photons}
\end{figure}

Such an analysis raises a further question: if experimental resources are
limited, does it make sense to measure 1 point very precisely or many points
less precisely?  Our framework can supply answers to this and many other such
questions.  By comparing the optimal designs of 1- and 5-point experiments at
both the ``doable'' and ``aspirational'' level of experimental precision (see
table~\ref{tab:experimental_precision_levels}) we get an idea of how to design
the most effective experiment.  The results are given in
fig.~\ref{fig:shrinkage_fixed_photons}.  Again, it is clear that the details
depend on the observable, which proves the usefulness of our approach: one
need not rely on heuristics when a quantitative scheme is readily available.

For example, the differential cross section does not appear to benefit as much
from an increase in precision (red circle) as it would from more data across
$\theta$ (blue cross).  In almost all other cases, such as $\Sigma_{2x}$,
$\Sigma_{2x'}$ or $\Sigma_{2z'}$, the gain in information an ``aspirational''
1-point experiment is about the same as 5 measurements from a ``doable''
experiment.  Surprisingly, some observables, such as $\Sigma_{1z'}$, benefit
very little from either increased precision or an increased number of data
points: one ``doable'' measurement in the right spot already realizes most of
the information gain to be had from them.

For completeness, we show the utility of performing neutron Compton scattering
experiments, with more plots for the neutron in the \verifyvalue{Supplemental
  Material.}  Figure~\ref{fig:dsg_neutron_utility} shows the 1-point profile
of the differential cross section with truncation error, which is similar to
the corresponding utility in
fig.~\ref{fig:utility_grid_compare_subsets_and_truncation}.  Such measurements
are notoriously difficult, so the interpretation of these results should
proceed with caution.  More realistically, our analysis should be applied to
\chiEFT\ predictions of light incident on the deuteron, $^3$He or other
few-nucleon targets for which calculations of Compton scattering are available
in the same \chiEFT\ formulation~\cite{Margaryan:2018opu, Hildebrandt:2005iw,
  Griesshammer:2013vga, Griesshammer:2012we}.

\begin{figure}[tb]
    \centering
    \includegraphics[width=\linewidth]
    {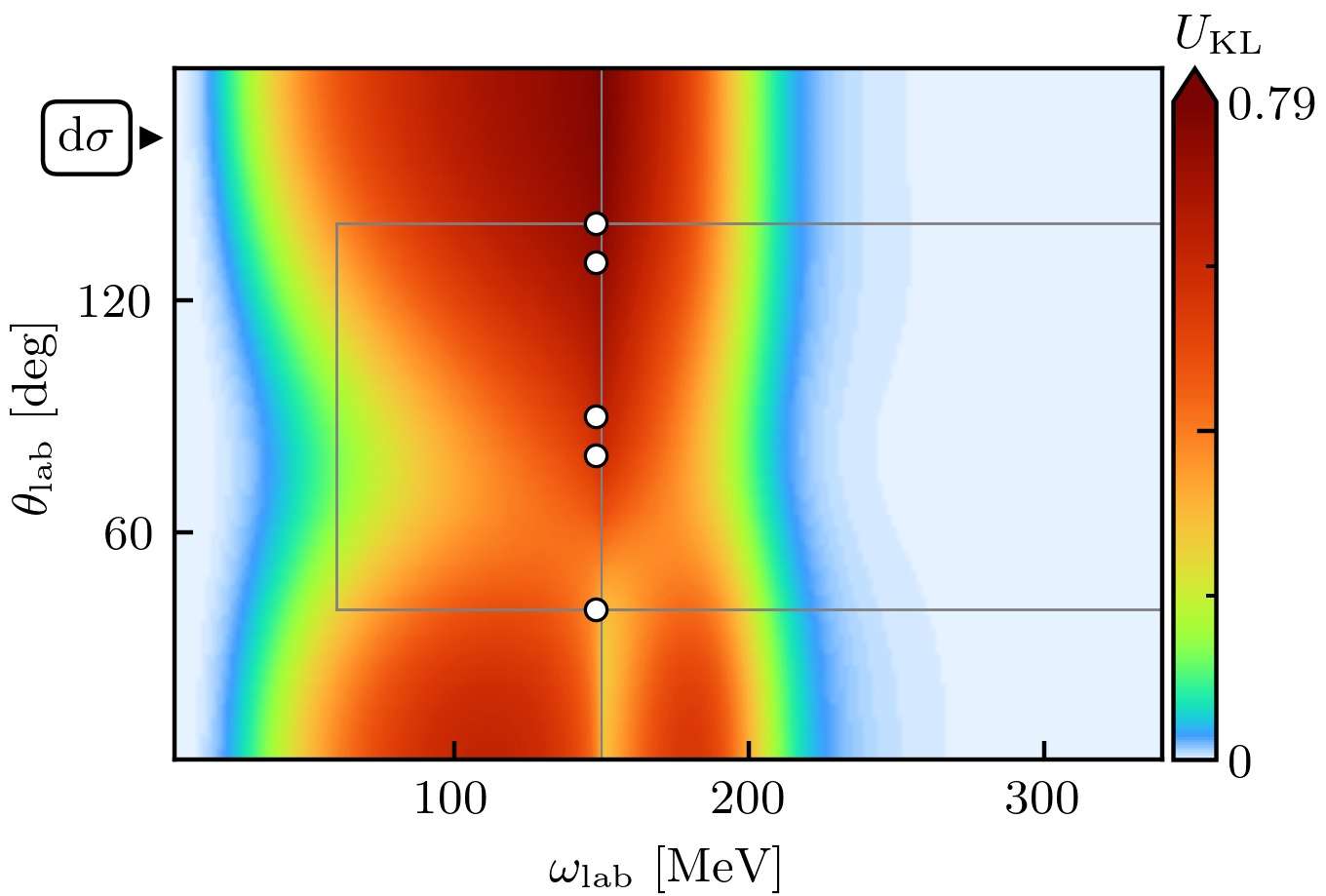}
    \caption{(Colour online) The expected utility from measuring the
      differential cross section for the neutron while including truncation
      error.  All neutron polarizabilities are included in this analysis.  Its
      profile resembles that of the analogous proton observable, see
      fig.~\ref{fig:utility_grid_compare_subsets_and_truncation}, though the
      color scales differ.  }
    \label{fig:dsg_neutron_utility}
\end{figure}

\section{Summary}
\label{sec:summary_design}

We have proposed a powerful and versatile framework to help plan experiments
which rely on EFT to extract or check parameters.  Using the example of
Compton experiments in order to constrain nucleon polarizabilities, this
method quantifies the expected gain in information from an experiment: it
maximizes shrinkage of the posterior.  The framework solves the problems of
theoretical errors conflicting with experimental considerations, and finds a
compromise between the two.  Under reasonable assumptions, we obtain an
algorithm that is analytic, easy to understand, and quick to compute.

Furthermore, we employed a Bayesian machine learning algorithm for estimating
EFT truncation errors whose power counting varies across the domain.  This is
a novel extension of the model introduced in ref.~\cite{Melendez:2019izc}.
Gau\3ian processes efficiently and accurately account for correlations in the
EFT truncation error, and impose the symmetry constraints on observables and
their derivatives that must vanish, \eg, at $\theta = 0^\circ$ or $180^\circ$.
This physically motivated model is crucial to the study of experimental design
with EFTs, as otherwise errors will be underestimated.

To facilitate reproduction and extension of our results, we provide all of the codes and data that generated our results~\cite{BUQEYEgithub}.

Our Bayesian experimental design framework has the following benefits:
\begin{enumerate}
\item It can incorporate the effects of both experimental and theoretical
  uncertainties.
\item Its output contains both the optimal design and an estimate of the gain
  in information for that design which can be understood quite easily.
\item It can include the effects from measuring multiple kinematic points and
  can assess the interaction of multiple polarizabilities at once.
\item It permits a quantitative analysis of competing choices, \eg, one can
  answer the question: should an experiment measure one point with high
  precision or many points with less precision?
\item Bayesian statistics mandates us to clearly specify our
  assumptions. Those who disagree with any assumption (size of error bars,
  priors on GP hyperparameters, the power counting in the transition region,
  design constraints, etc.) can readily modify our calculations, provided
  at~\cite{BUQEYEgithub}, to their own specifications, thereby facilitating an
  ongoing dialog regarding the robustness of our experimental design results.
\end{enumerate}

We also make the obvious point that while we have focused on Compton
scattering experiments here, our EFT-based Bayesian approach to experimental
design is easily adapted to other experiments informed by EFT calculations.

We tried to make a realistic assessment of experimental capabilities in this
work, but realize that experiments can differ greatly. We reiterate that our
analysis was meant to open the discussion on experimental design for Compton
scattering, not provide the last word. In the future, our framework could be
applied to a specific experiment at, \eg, MAMI or HI$\gamma$S, with additional
details accounted for. For example, the analysis carried out here did not
consider common-mode (correlated) errors, which can significantly affect the
accuracy of an absolute measurement of the differential cross section. In
fact, the correlation matrix of systematic errors across the space of data
points could in general be quite complex. Including a more sophisticated
treatment of such issues in the assessment of experimental designs would be
one place where collaboration with experimental colleagues could be very
fruitful. We emphasize that since we provide a Jupyter notebook for the
experimental design problem, colleagues who wish to make different assumptions
about the nature of correlations in errors between data taken in different
kinematics can rerun the analysis with a statistical model improved thusly.

Bayesian experimental design can then determine the expected amount or quality
of data required to reach a given level of polarizability precision within a
particular experimental setup.  However, we caution that such guidance is {\it
  always} in the light of particular prior information. Because experiments
are assessed by the information gain they are expected to provide relative to
a particular prior, that implies that a different choice of prior can produce
different conclusions about which experiments will be most enlightening.

In this context, we point out that a fully consistent design analysis would
begin with an extraction of scalar and spin polarizabilities from the present
proton Compton scattering database that uses the error model
\eqref{eq:errormodel_compton} + \eqref{eq:observable_expansion_compton},
\ie~the same error model as used to assess the information gain of future
experiments. Obtaining this six-dimensional probability distribution is an
important topic for future work, as it would accurately represent the current
state of knowledge of polarizabilities in \chiEFT. The prior we have used here
only approximately represents that state. But the results of the design
analysis are insensitive to the central values of the polarizabilities in that
prior.  And adopting different, reasonable correlations in the prior does not
produce marked changes.

This framework can be extended to sequential designs, where experimental
campaigns are split into a sequence of parts and the design of future
experiments depends on the results of the initial
experiments~\cite{Ryan:2015aaa}.  This ``sequential experimental design"
analysis can be computationally intensive and we have not attempted it here.
Although we have found the assumption of linearity to be good in this case,
one could perform a full Bayesian experimental design if this assumption no
longer holds~\cite{Ryan:2015aaa,jacksonDesignPhysicalSystem2018}.  Our
theoretical truncation estimates are the most comprehensive to date, but
further study of chiral EFT convergence for Compton observables should be
performed.  These are all tasks for future work.

\begin{acknowledgments}
  We thank Ian Vernon for useful discussions, and M.~Ahmed, E.~Downie,
  G.~Feldman, P.~P.~Martel, as well as the MAMI-A2/CB and Compton@HI$\gamma$S
  teams for their patience in discussing experimental constraints. We
  gratefully acknowledge the stimulating atmosphere created by organizers and
  participants of the workshops \textsc{Uncertainty Quantification at the
    Extremes (ISNET-6)} at T.U.~Darmstadt (Germany) and \textsc{Bayesian
    Inference in Subatomic Physics - A Wallenberg Symposium (ISNET-7)} at
  Chalmers U.~(G\"oteborg, Sweden), which triggered and expanded these
  investigations.
H.W.G.~gratefully acknowledges the warm hospitality and financial support of the A2/Crystall Ball Collaboration Meeting 2020 at MAMI (U.~Mainz, Germany), of both Ohio University and the Ohio State University, and of the University of Manchester, where part of this work was conducted. 
The work of R.J.F.~and J.A.M.~was supported in part by the National Science Foundation
under Grant Nos.~PHY--1614460 and PHY--1913069 and the NUCLEI SciDAC Collaboration under
US Department of Energy MSU subcontract RC107839-OSU\@.
The work of D.R.P.~was supported by the US Department of Energy under contract
DE-FG02-93ER-40756 and by the ExtreMe Matter Institute EMMI at the GSI
Helmholtzzentrum f\"ur Schwerionenphysik, Darmstadt, Germany.
The work of H.W.G.~was supported in part by the US Department of Energy under
contract DE-SC0015393, by the High Intensity Gamma-Ray Source \HIGS\ of the
Triangle Universities Nuclear Laboratory TUNL in concert with the Department
of Physics of Duke University, and by The George Washington University: by the
Dean's Research Chair programme and an Enhanced Faculty Travel Award of the
Columbian College of Arts and Sciences; and by the Office of the Vice
President for Research and the Dean of the Columbian College of Arts and
Sciences.  His work was conducted in part at GW's Campus in the Closet.  The
work of J.McG.~was supported by the UK Science and Technology Facilities
Council grant ST/P004423/1.  The work of M.T.P.~was supported in part by the
King Abdullah University of Science and Technology (KAUST) Office of Sponsored
Research (OSR) under Award No. OSR-2018-CRG7-3800.3.
\end{acknowledgments}

\appendix
\section{Experimental Design Details}
\label{sec:experimental_design_details}

Suppose that our theoretical model $y_k(\kinparvec;\lecs)$ is related to
measurements $\genobsexp(\kinparvec)$ via additive theoretical and
experimental noise, as in eq.~\eqref{eq:errormodel_compton}.  We can linearize
$y_k(\kinparvec;\lecs)$ about some point $\lecs_\star$ by keeping only the
first order terms in its Taylor expansion, \ie,
\begin{align}
    y_k(\kinparvec;\lecs) & \approx y_k(\kinparvec; \lecs_\star) + \sum_i b_i(\kinparvec) [\lecs_i - \lecs_\star] \notag \\
    & = c(\kinparvec; \lecs_\star) + \vec{b}(\kinparvec) \cdot \lecs \, ,
\end{align}
where
$\vec{b}(\kinparvec) \equiv \partial \genobs_k(\kinparvec;\lecs)/\partial
\lecs$
evaluated at $\lecs_\star$ are our basis functions and
$c(\kinparvec;\lecs_\star) \equiv y_k(\kinparvec;\lecs_\star) - \vec{b} \cdot
\lecs_\star$
is constant with respect to the polarizabilities $\lecs$ but depends on the
kinematic point $\kinparvec$.  Thus, the vector of $N$ measurements
$\genobsset$ is related to the polarizabilities via the likelihood
\begin{align} \label{eq:likelihood_linear}
    \genobsset \given \lecs \sim \normal[B\lecs + \mathbf{c}, \Sigma]
\end{align}
where $B \equiv \vec{b}(\kinparvecset)$ is an $N \times 6$ matrix,
$\mathbf{c} \equiv c(\kinparvecset; \lecs_\star)$ is a length $N$ vector, and
$\Sigma$ is the $N \times N$ covariance matrix due to theoretical and
experimental error.  That is, given some experimental covariance
$\Sigma_{\rm exp}$ and a theoretical covariance $\sdth^2 \discrcorr{k}$ from
eqs.~\eqref{eq:discrepancy_gp_compton}
and~\eqref{eq:compton_truncation_corrfunc}, then
\begin{align}
    \Sigma = \sdth^2 \discrcorr{k} + \Sigma_{\rm exp} \, .
\end{align}
Note that $\discrcorr{k}$ depends on the values of the tuned $\ell_\omega$ and
$\ell_\theta$, whose estimates from the order-by-order convergence pattern are
given in table~\ref{tab:truncation_details_observables}.

The linear model of eq.~\eqref{eq:likelihood_linear} is well known in the
statistics literature~\cite{gelman2013bayesian,o1994bayesian}, so here we will
simply state the relevant results.  If a Gau\3ian prior with mean
$\vec{\mu}_0$ and covariance $V_0$ is placed on $\lecs$ as in
eq.~\eqref{eq:polarizability_prior}, then the resulting posterior is also
Gau\3ian, with mean and covariance given by
\begin{align}
    \vec{\mu} & = V\left[V_0^{-1}\vec{\mu}_0 + B^\trans \Sigma^{-1} (\genobsset - \mathbf{c})\right] \, , \label{eq:pol_posterior_mean_linear} \\
    V & = (V_0^{-1} + B^\trans \Sigma^{-1} B)^{-1} \, . \label{eq:pol_posterior_cov_linear}
\end{align}
Importantly to our study of experimental design, the posterior covariance $V$
depends on the kinematic points $\kinparvecset$ where the experiment is
performed, and on the specifics of the observable through $\Sigma$, but not on
the exact results of the experiment $\genobsset$.

Given that we choose to maximize the expected information gain in the
polarizabilities, then the integrals of eq.~\eqref{eq:expected_utility_kl}
must still be performed.  The integral over $\lecs$ splits into the difference
of two terms: the differential entropy of the prior $\pr(\lecs)$ and of the
posterior $\pr(\lecs \given \genobsset, \design)$.  The differential entropy
of a Gau\3ian $\normal(\mu, \Sigma)$ is well known to be
$\frac{1}{2}\ln{|2\pi e\Sigma|}$.  Therefore
\begin{align}
    U_{\text{KL}}(\design)
    & = - \int \ln[\pr(\lecs)] \pr(\lecs) \dd{\lecs} \notag \\
    & ~~~ + \int \ln[\pr(\lecs\given \genobsset, \design)]\pr(\lecs\given \genobsset, \design) \dd{\lecs} \pr(\genobsset \given \design)\dd{\genobsset} \notag \\
    & = \frac{1}{2}\ln{|2\pi e V_0|} - \frac{1}{2}\ln{|2\pi e V|} \int \pr(\genobsset \given \design)\dd{\genobsset} \notag\\
    & = \frac{1}{2} \ln \frac{|V_0|}{|V|} \, ,
\end{align}
where we used the fact that $V$ does not depend on $\genobsset$ and then
performed the trivial integration over all possible measurements $\genobsset$.


\section{Observable Constraints and EFT Truncation Model Details}
\label{sec:truncation_model_details_compton}

Constraints on Compton observables are discussed in detail in
ref.~\cite{Griesshammer:2017txw}.  Some of this is reproduced here, with
particular attention paid to $n$th-order chiral \emph{corrections} to
observables $\Delta y_n$ rather than the value of the observable $y$ itself.
The $\Delta y_n$ impact the distribution for the \chiEFT\ uncertainty
$\delta \genobs_k$, but because we restrict the ``experimentally accessible
regime'' in this study from small $\omega$, and forwards/backwards angles,
these constraints are not as important as they otherwise would be.  These
constraints are summarized for particular $\omegalab$ and $\thetalab$ values
in table~\ref{tab:compton_observable_symmetry_constraints}.

All observables that are nonzero below $\omega_\pi$ approach the Thomson limit
as $\omega \to 0$~\cite{Griesshammer:2012we}.  Thus, higher-order corrections
must vanish there , and approach $\omega = 0$ as at least $\omega^2$.
Therefore, at least the first derivative of all corrections must vanish there
as well.

The remaining observables must vanish for $\omega \leq \omega_\pi$, but there
is no constraint on the the derivative of corrections at
$\omega = \omega_\pi$.  We have found that the corrections approach 0
\emph{very} quickly, so that imposing the constraint
$\Delta \genobs_n(\omega_\pi,\theta) = 0$ for all higher order terms is
actually a worse approximation than not imposing the constraint at all; see,
\eg, fig.~\ref{fig:coefficients_1x_and_1xp_slices} in the Supplemental
Material.  This comes back to the large cusps in the spin-observable $c_n$
found near $\omega_\pi$, discussed in
sec.~\ref{sec:truncation_distribution_compton}, which remain an unresolved
aspect of this model.

Due to the coordinate singularity at $\theta=0^\circ$ and $180^\circ$,
observables or their derivative with respect to $\theta$ must vanish
there~\cite{Griesshammer:2017txw}.  But this does not preclude \emph{both} the
value and their derivatives from vanishing there.  These constraints can be
deduced by symmetry arguments, and are summarized in
table~\ref{tab:compton_observable_symmetry_constraints}.

The hyperparameters $\sdth^2$ and $\ell_i$, shown in
table~\ref{tab:truncation_details_observables}, are tuned to coefficients
$c_n$ at the best known $\lecs$ (see table~\ref{tab:polarizability_info}) for
$\Lambda_b = 650\MeV$.  The training data are on a grid with
$\thetalab = \{30^\circ, 50^\circ, 70^\circ, 90^\circ, 110^\circ, 130^\circ\}$
and $\omegalab = \{200, 225, 250\}\MeV$ for observables which are zero below
$\omega_\pi$.  For observables that are non-zero below $\omega_\pi$, the
additional training points $\omegalab = \{50, 75, 100, 125\}\MeV$ are
included, and common $\ell_\omega$ and $\ell_\theta$ are used between the two
regions. The training region is well outside the kinematic endpoints where
additional constraints arise on observables or their derivatives, and excludes
the pion-production threshold region.

Because the first nonzero order often behaves differently than the
corrections, we do not use it for induction on the $c_n$; that is we only
train the hyperparameters on \emph{corrections}.  Hence, we train on
$c_2$--$c_4$ for $\diffcs$ and $\Sigma_3$, but otherwise we train on $c_3$ and
$c_4$.

The coefficients for various observable slices are shown in
figs.~\ref{fig:coefficients_dsg_slices} for the cross section, and
figs.~\ref{fig:coefficients_3_and_y_slices}--\ref{fig:coefficients_3y_and_3yp_slices}
of the Supplemental Material for the spin observables.  These plots also
include uncertainty bands for higher order coefficients, with the symmetry
constraints given in table~\ref{tab:compton_observable_symmetry_constraints}
included.  These constraints on both the coefficient functions and their
derivatives propagate directly to the truncation error $\delta\genobs_k$ by
replacing $r(x,x';\ell_\omega, \ell_\theta)$ in
eq.~\eqref{eq:compton_truncation_corrfunc} by its \emph{conditional} form
$\tilde r(x,x';\ell_\omega, \ell_\theta)$, see
refs.~\cite{rasmussen2006gaussian,Melendez:2019izc}.  For example, if the
value of $c_n$ is known at the set of points $\mathbf{x}$, then one can
compute its conditional GP, with covariance kernel given by
\begin{align}
    \tilde r(x, x') = r(x, x') - r(x, \mathbf{x}) r(\mathbf{x}, \mathbf{x})^{-1} r(\mathbf{x}, x') \, .
\end{align}
See refs~\cite{Rasmussen:2003gphmc, Solak:2003dgpds,Eriksson:2018scaling} for
details about adding derivative observations to GPs.  Because the RBF kernel
[eq.~\eqref{eq:rbf_kernel_compton}] is separable in $\omega$ and $\theta$,
these constraints can simply be applied to each one-dimensional kernel
separately, and multiplied to yield the total constrained kernel.  We employ
the \texttt{gptools} python package for easily implementing derivative
constraints~\cite{Chilenski_2015_gptools}.

\begin{figure}
    \centering
    \includegraphics[width=\linewidth]{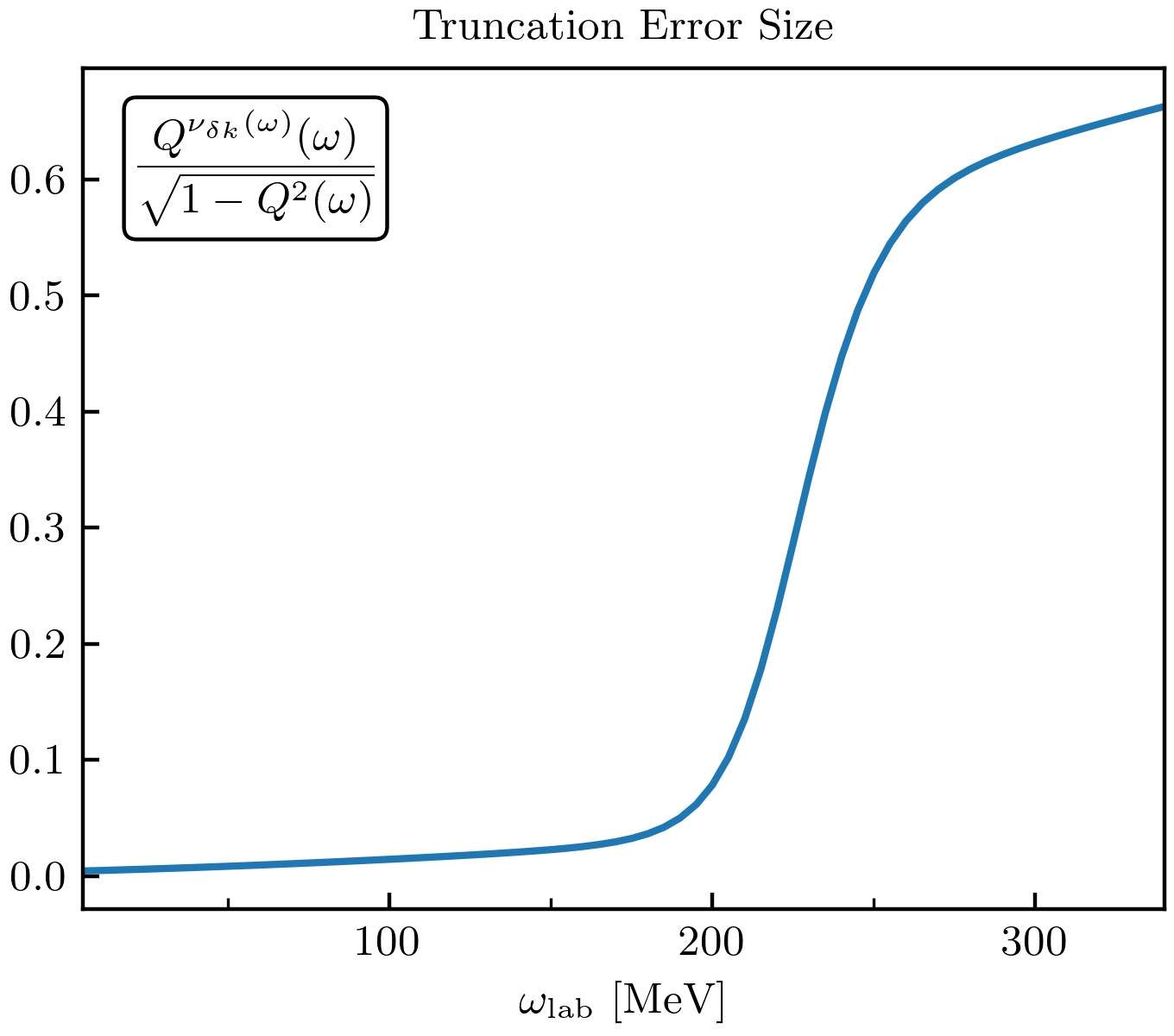}
    \caption{(Colour online) A component of the standard deviation due to \chiEFT\ uncertainty at $\NkLO{4}^+$, see  eq.~\eqref{eq:compton_truncation_corrfunc}.
    The factor of $\cbar$ is unique to each observable, and is not included.
    See table~\ref{tab:truncation_details_observables}.
    }
    \label{fig:truncation_error_stdv_compton}
\end{figure}

\begin{figure}[tb]
    \centering
    \includegraphics[width=0.495\textwidth]{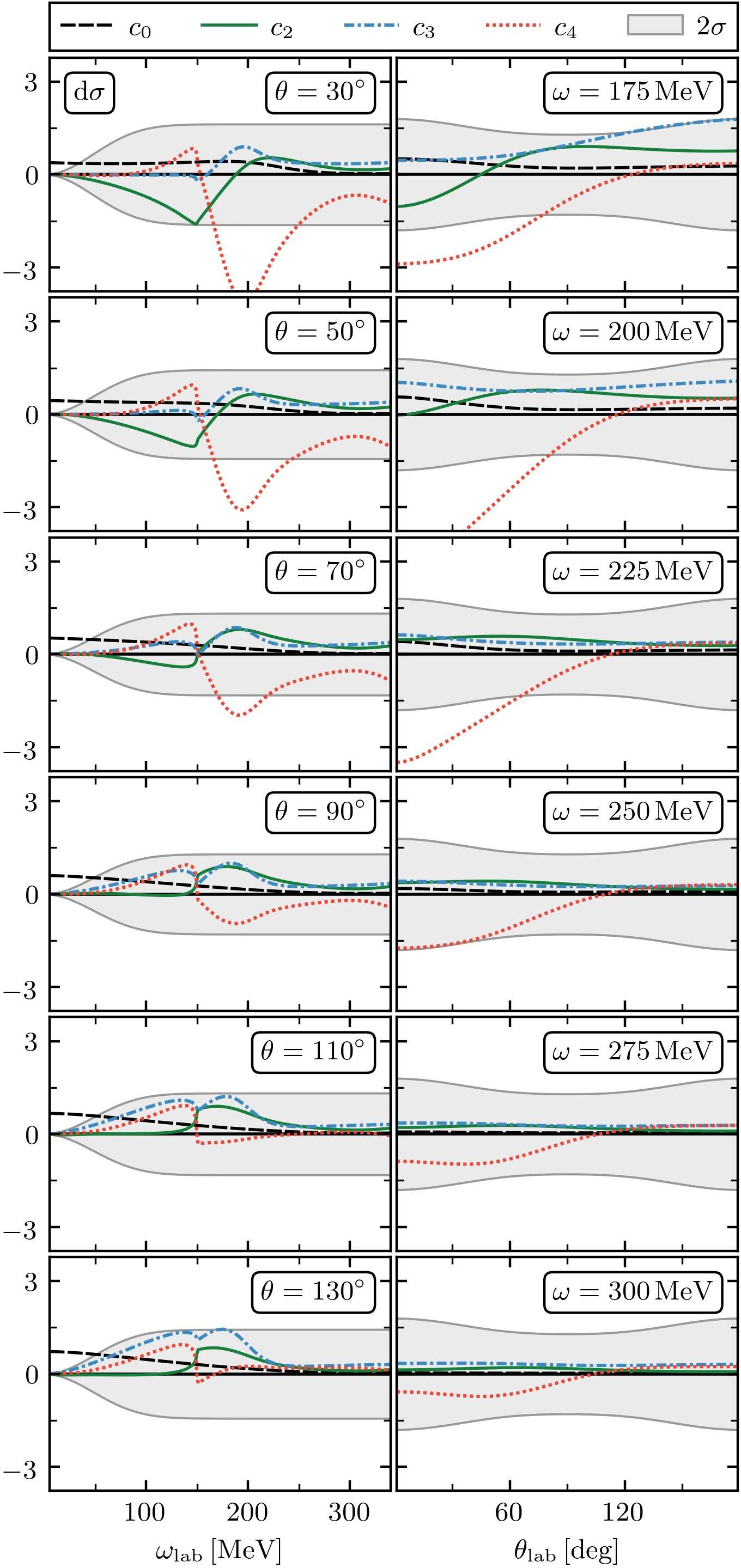}
    \caption{(Colour online) Coefficients for the differential cross section
      $\diffcs$.}
    \label{fig:coefficients_dsg_slices}
\end{figure}

For completeness, we also provide the profile for the truncation error
standard deviation (up to factors of $\cbar$, which vary by observable); see
fig.~\ref{fig:truncation_error_stdv_compton}.  It assumes the form of $Q$
provided in eq.~\eqref{eq:expansion_parameter_compton} along with the first
omitted \chiEFT\ order given in eq.~\eqref{eq:n3loplus_truncation_power}.

This allows us to return to the discussion of the omitted constraints
$\Sigma_i \in [-1, 1]$ on the spin observables in sec.~\ref{sec:experiments}.
Over the physically interesting kinematic range, the actual value of most spin
observables lies in the much more narrow interval $[-0.7,0.7]$; see fig.~5 in
ref.~\cite{Griesshammer:2017txw}.  So, then the question becomes: are the mean
prediction \emph{and} its theory uncertainty contained in $[-1,1]$ with a high
degree of probability?  From eq.~\eqref{eq:discrepancy_sum_compton}, one can
see that the $1\sigma$ interval for the truncation error $\delta\genobs_k$ is
$\genobsref\sdth$ times another factor
$Q^{\nu_{\delta k}(\omega)}/\sqrt{1-Q^2(\omega)}$.  Here $\genobsref=1$ and
$\sdth \lesssim 0.7$ for most spin observables (see
table~\ref{tab:truncation_details_observables}). The third factor is plotted
in fig.~\ref{fig:truncation_error_stdv_compton} and does not exceed
$\approx0.3$ at $\omega\lesssim230\MeV$, where our analysis shows the biggest
sensitivities.  Therefore, even for the spin observables with large
magnitudes, the $1\sigma$ upper range of a GP will only give values about
$0.2$ larger than the established maximum of $0.7$, namely about $0.9$ in
total. This is close but still below $|\Sigma_i|=1$.  Therefore, a majority of
our test functions in the GP will not probe, let alone exceed, the strict
bounds on those spin observables.  Furthermore, if observables and their
truncation errors vanish at $\theta = 0^\circ$ or $180^\circ$, this will make
the constraint even more trivially satisfied near these regions.  We are
therefore confident that implementing the constraint $\Sigma_i\in[-1, 1]$
would not impact our results for $\omega\lesssim220\MeV$, and cautiously
optimistic that the impact would be small even at higher energies.

Though we likewise do not constrain the cross section to be non-negative, we
are confident that within our constrained angle range, corrections are highly
unlikely to be large enough for this to be a worry.  According to fig.~4 in
ref.~\cite{Griesshammer:2017txw}, the \NkLO{4}cross section is small
($<10\,\mathrm{nb/sr}$) in a narrow region at forward angles around
$\omega_\pi$.  Figure~\ref{fig:truncation_error_stdv_compton} shows that the
expansion parameter is small, and fig.~\ref{fig:coefficients_dsg_slices} shows
that the coefficients $c_i$ are natural-sized.  Therefore, the GP corrections
are highly unlikely to exceed the size of the predicted cross section and
create negative (unphysical) values.

\begin{table}[t]
\renewcommand{\tabcolsep}{3pt}
\caption{\label{tab:compton_observable_symmetry_constraints}
  The constraints on \emph{corrections} to observables $\Delta y$ and their derivatives $\Delta y'$ at particular $\theta$ and $\omega$. The LO amplitude as well as all calculated higher orders fulfill them automatically, so these must only be enforced in the GP. 
  The observables marked by a dagger $\dagger$ are zero below the pion-production threshold, but we impose no constraint on them at $\omega=\omega_\pi$, as discussed in the text.}
\begin{ruledtabular}
\begin{tabular}{SlScScScSc}
& \multicolumn{2}{Sc}{$\theta$ [deg]} & \multicolumn{2}{Sc}{$\omega$ [MeV]} \\
\cline{2-3}\cline{4-5}
 & $\Delta y = 0$ & $\Delta y' = 0$ & $\Delta y = 0$ & $\Delta y' = 0$ \\
\hline
$\diffcs$ &           --- &              0, 180 &           0 &                 0 \\
$\Sigma_{1x}$      &        0, 180 &                 --- &         $\dagger$ &               $\dagger$ \\
$\Sigma_{1z}$      &        0, 180 &              0, 180 &         $\dagger$ &               $\dagger$ \\
$\Sigma_{2x}$      &        0, 180 &                 --- &           0 &                 0 \\
$\Sigma_{2z}$      &           --- &              0, 180 &           0 &                 0 \\
$\Sigma_{3}$       &        0, 180 &              0, 180 &           0 &                 0 \\
$\Sigma_{y}$       &        0, 180 &                 --- &         $\dagger$ &               $\dagger$ \\
$\Sigma_{3y}$      &        0, 180 &                 --- &         $\dagger$ &               $\dagger$ \\
$\Sigma_{3y'}$     &        0, 180 &                 --- &         $\dagger$ &               $\dagger$ \\
$\Sigma_{1x'}$     &        0, 180 &                   0 &         $\dagger$ &               $\dagger$ \\
$\Sigma_{1z'}$     &        0, 180 &                 180 &         $\dagger$ &               $\dagger$ \\
$\Sigma_{2x'}$     &           180 &                   0 &           0 &                 0 \\
$\Sigma_{2z'}$     &             0 &                 180 &           0 &                 0 \\
\end{tabular}
\end{ruledtabular}

\end{table}

\clearpage\newpage
\setcounter{section}{18} 
\setcounter{figure}{0}
\renewcommand\thefigure{\thesection.\arabic{figure}}  
\setcounter{equation}{0}
\renewcommand\theequation{\thesection.\arabic{equation}}  
\setcounter{table}{0}
\renewcommand\thetable{\thesection.\arabic{table}}  

\section{Supplemental Material}
\label{sec:extra_compton_figures}

\subsection{Coefficients and Uncertainties for the Proton Spin Observables}

\begin{figure*}[tb]
    \centering
    \includegraphics[width=0.495\textwidth]{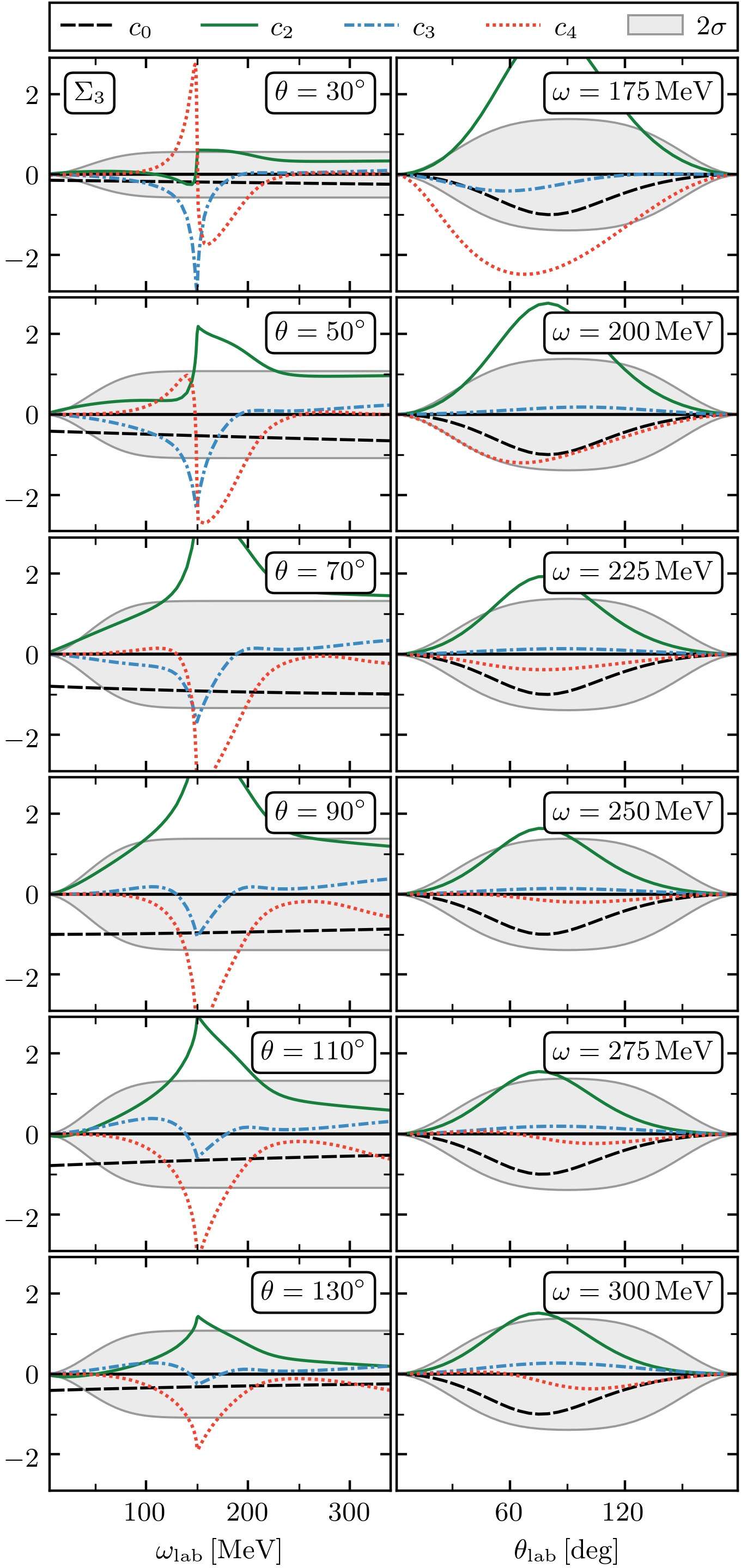}
    \includegraphics[width=0.495\textwidth]{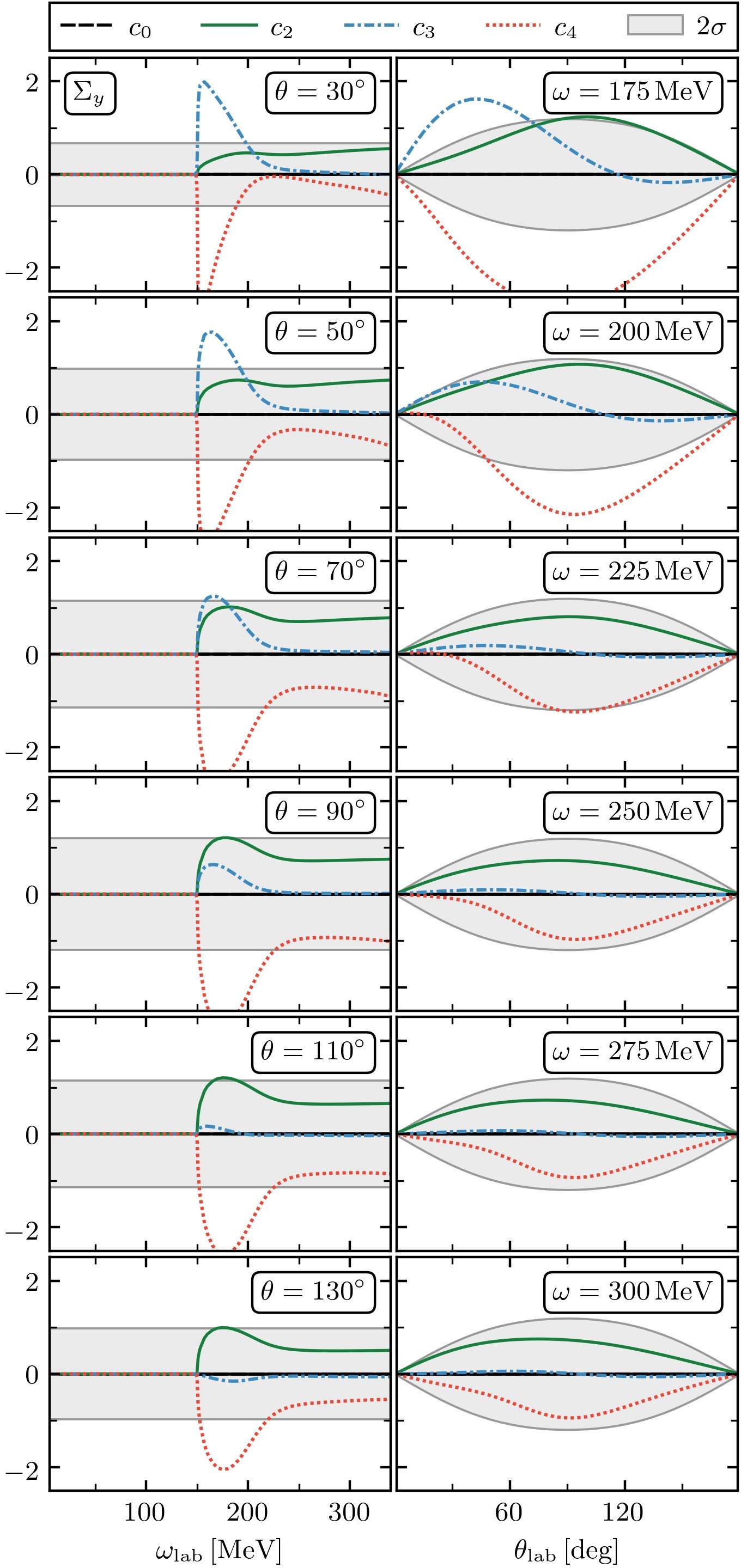}
    \caption{(Colour online) Coefficients for $\Sigma_3$ and $\Sigma_y$.}
    \label{fig:coefficients_3_and_y_slices}
\end{figure*}

\begin{figure*}[tb]
    \centering
    \includegraphics[width=0.495\textwidth]{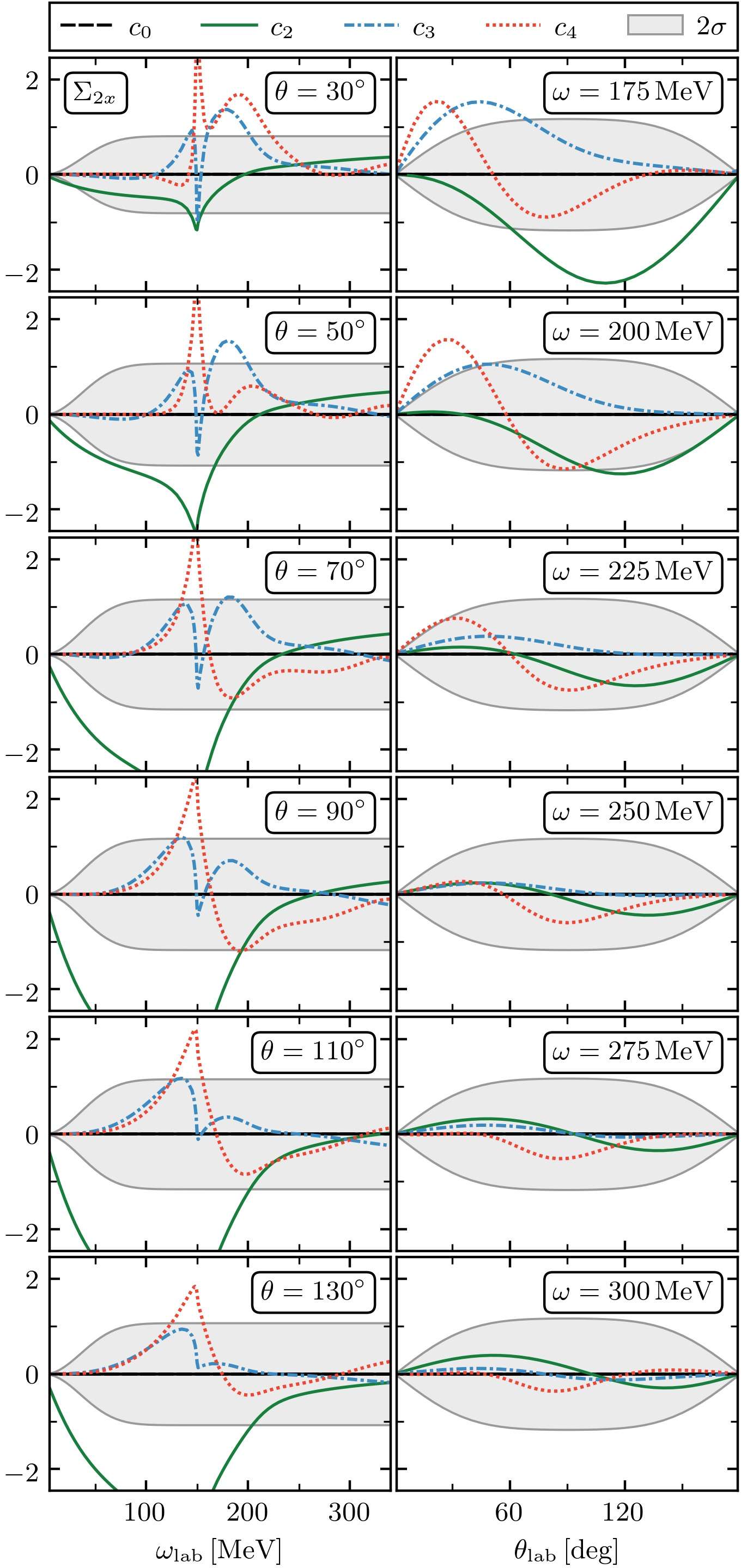}
    \includegraphics[width=0.495\textwidth]{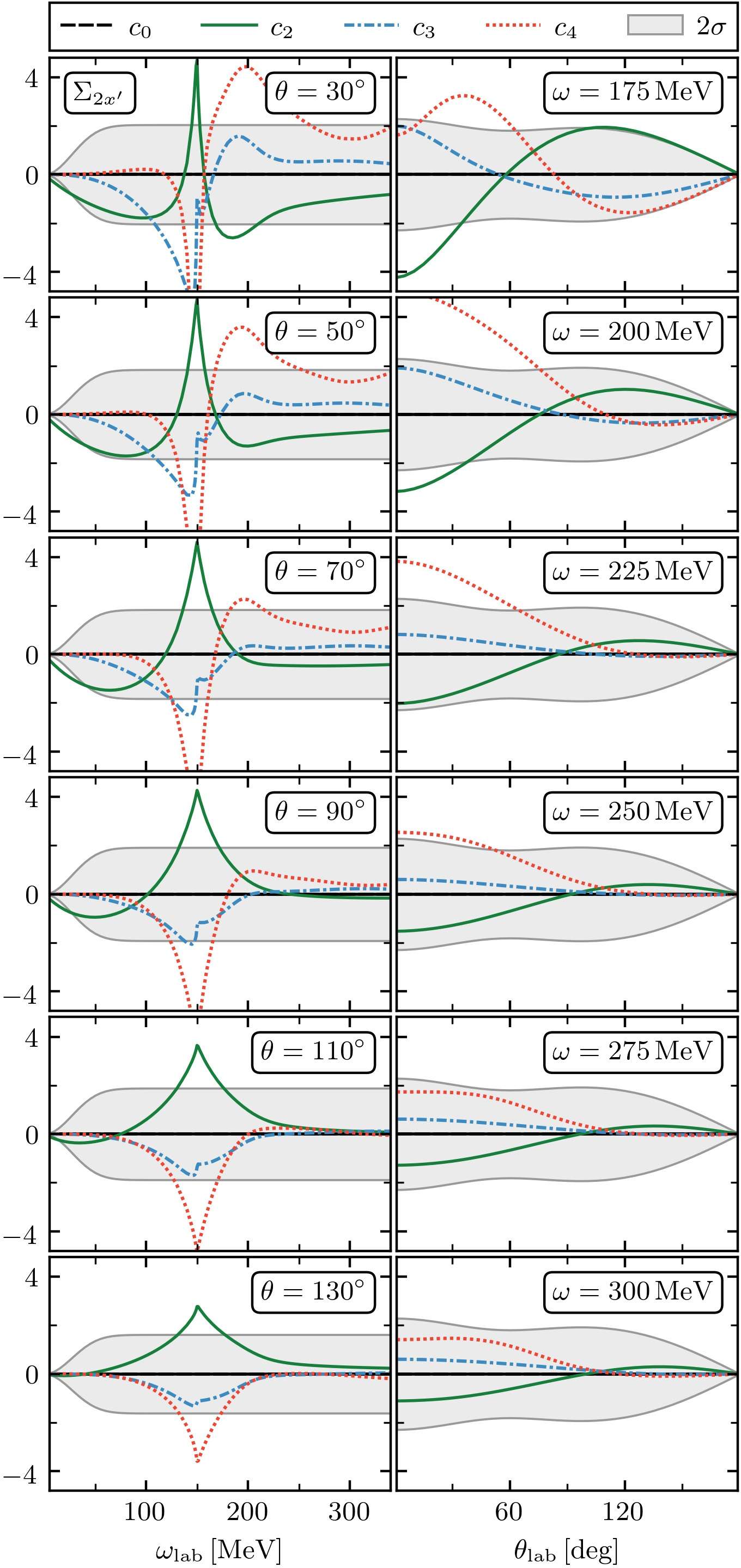}
    \caption{(Colour online) Coefficients for $\Sigma_{2x}$ and $\Sigma_{2x'}$.}
    \label{fig:coefficients_2x_and_2xp_slices}
\end{figure*}

\begin{figure*}[tb]
    \centering
    \includegraphics[width=0.495\textwidth]{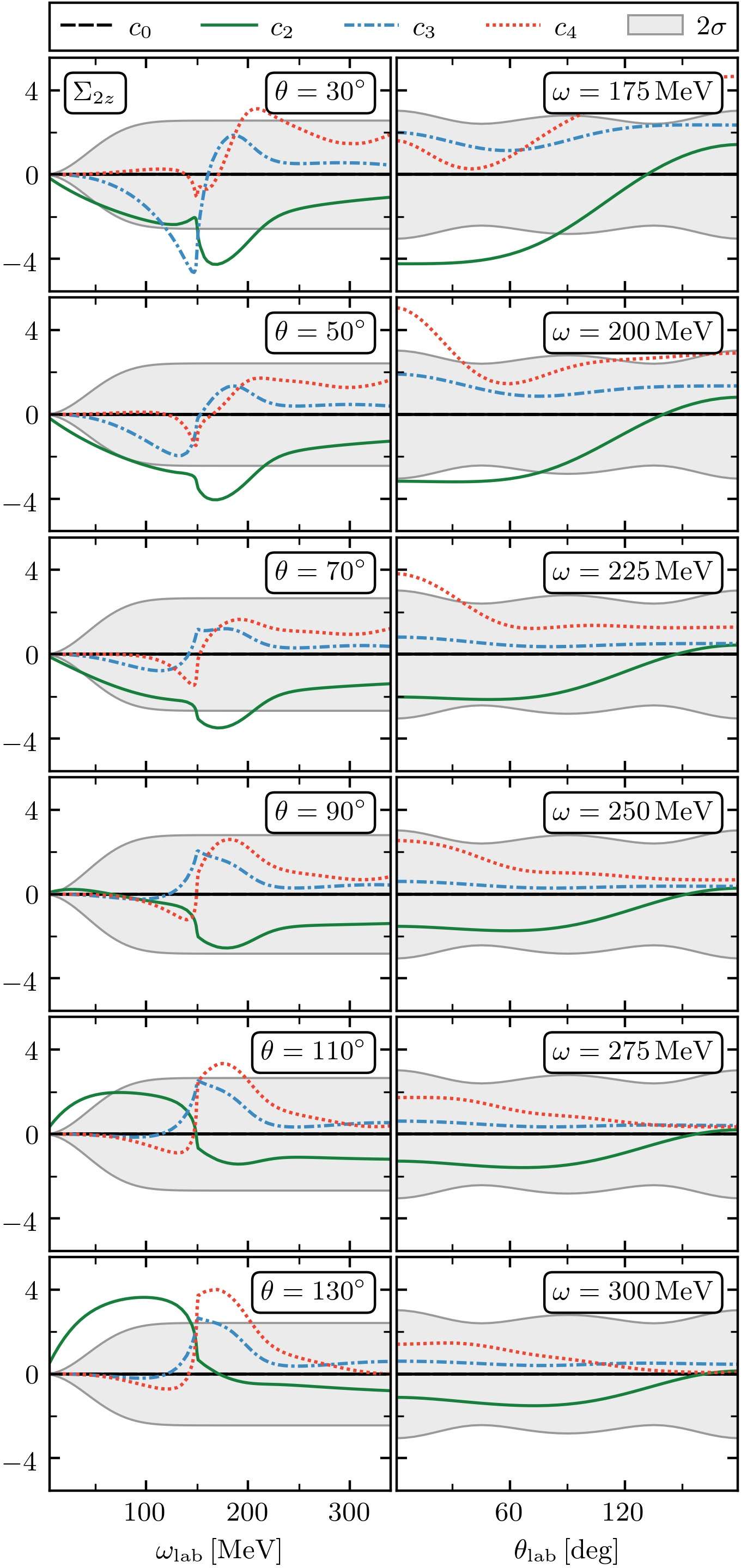}
    \includegraphics[width=0.495\textwidth]{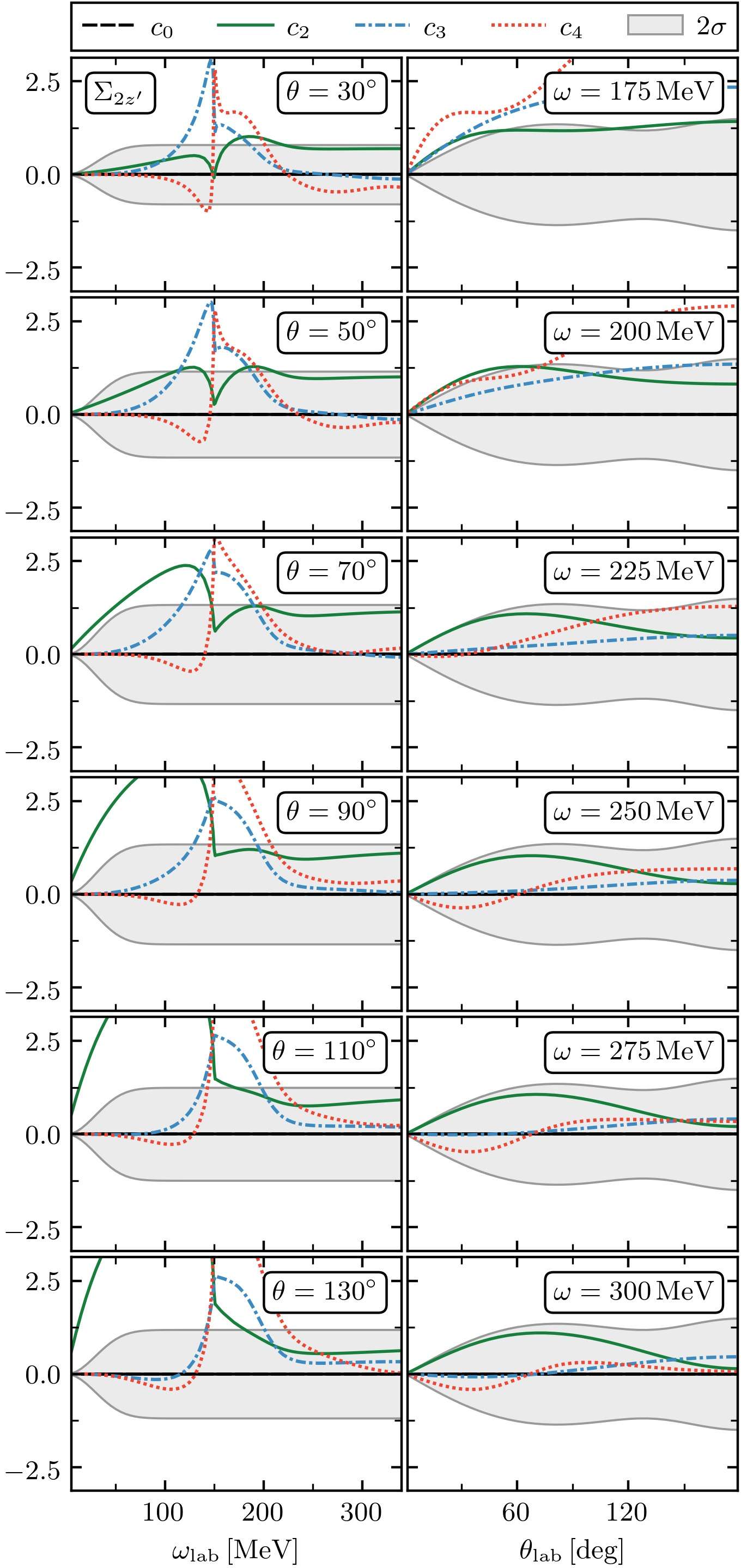}
    \caption{(Colour online) Coefficients for $\Sigma_{2z}$ and $\Sigma_{2z'}$.}
    \label{fig:coefficients_2z_and_2zp_slices}
\end{figure*}

\begin{figure*}[tb]
    \centering
    \includegraphics[width=0.495\textwidth]{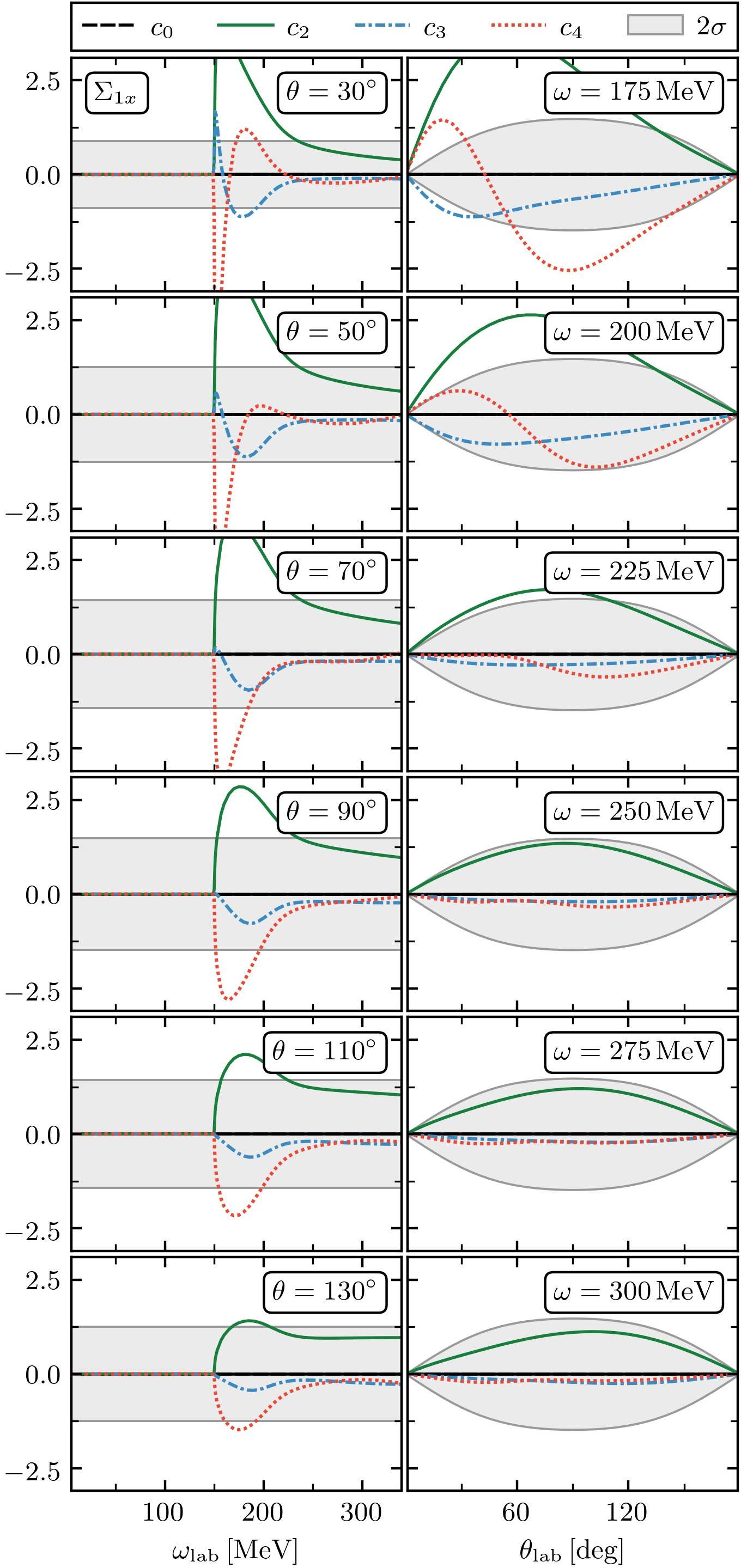}
    \includegraphics[width=0.495\textwidth]{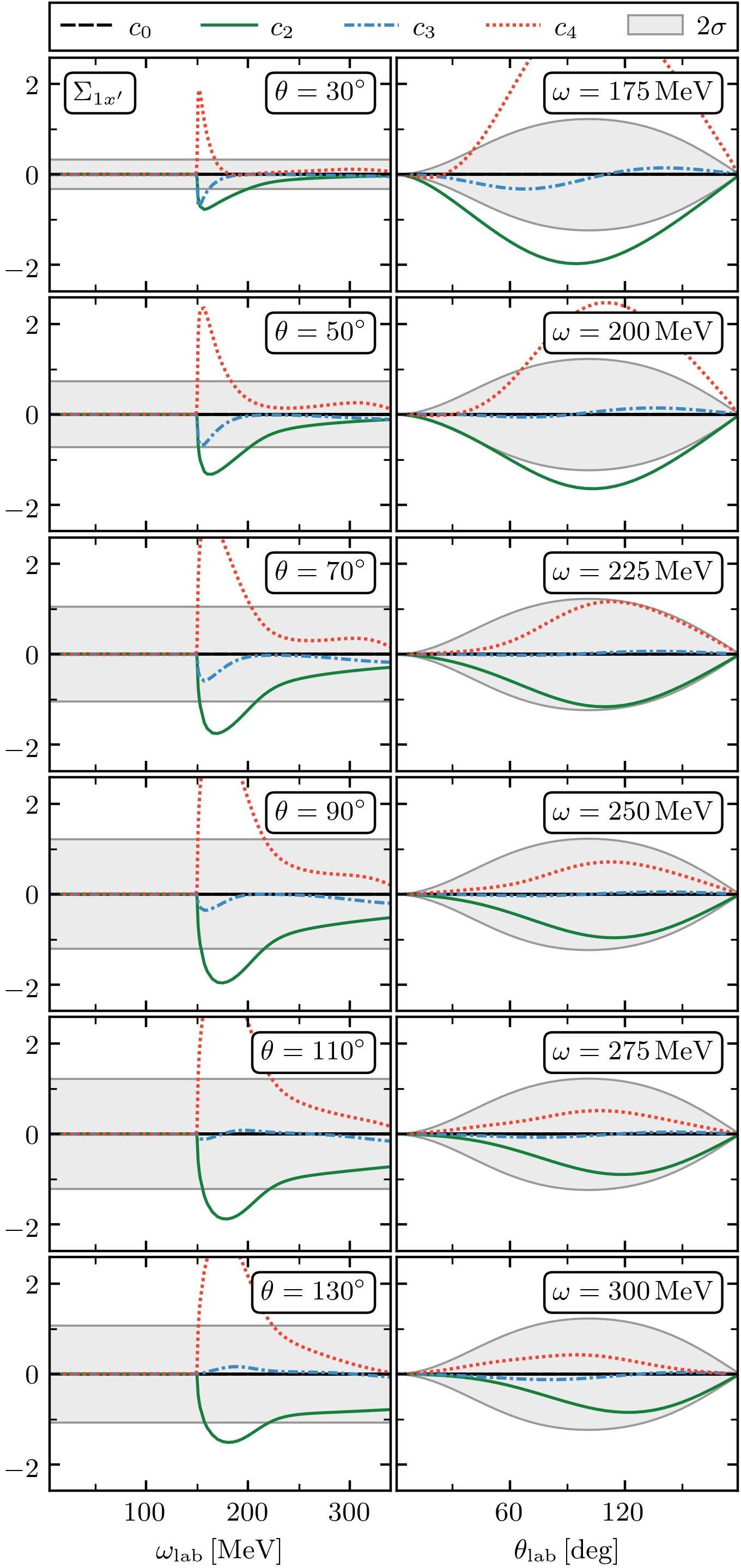}
    \caption{(Colour online) Coefficients for $\Sigma_{1x}$ and $\Sigma_{1x'}$.}
    \label{fig:coefficients_1x_and_1xp_slices}
\end{figure*}

\begin{figure*}[tb]
    \centering
    \includegraphics[width=0.495\textwidth]{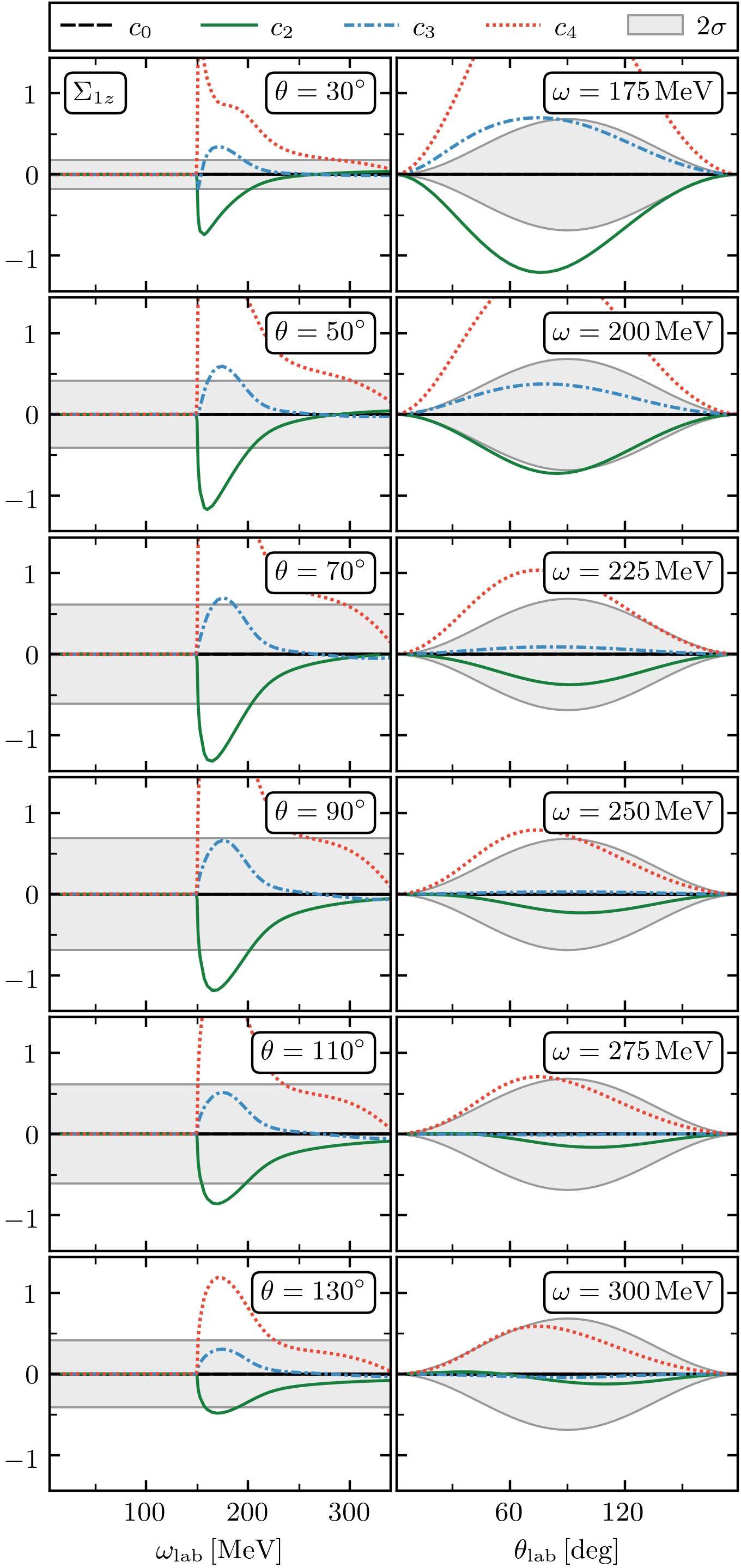}
    \includegraphics[width=0.495\textwidth]{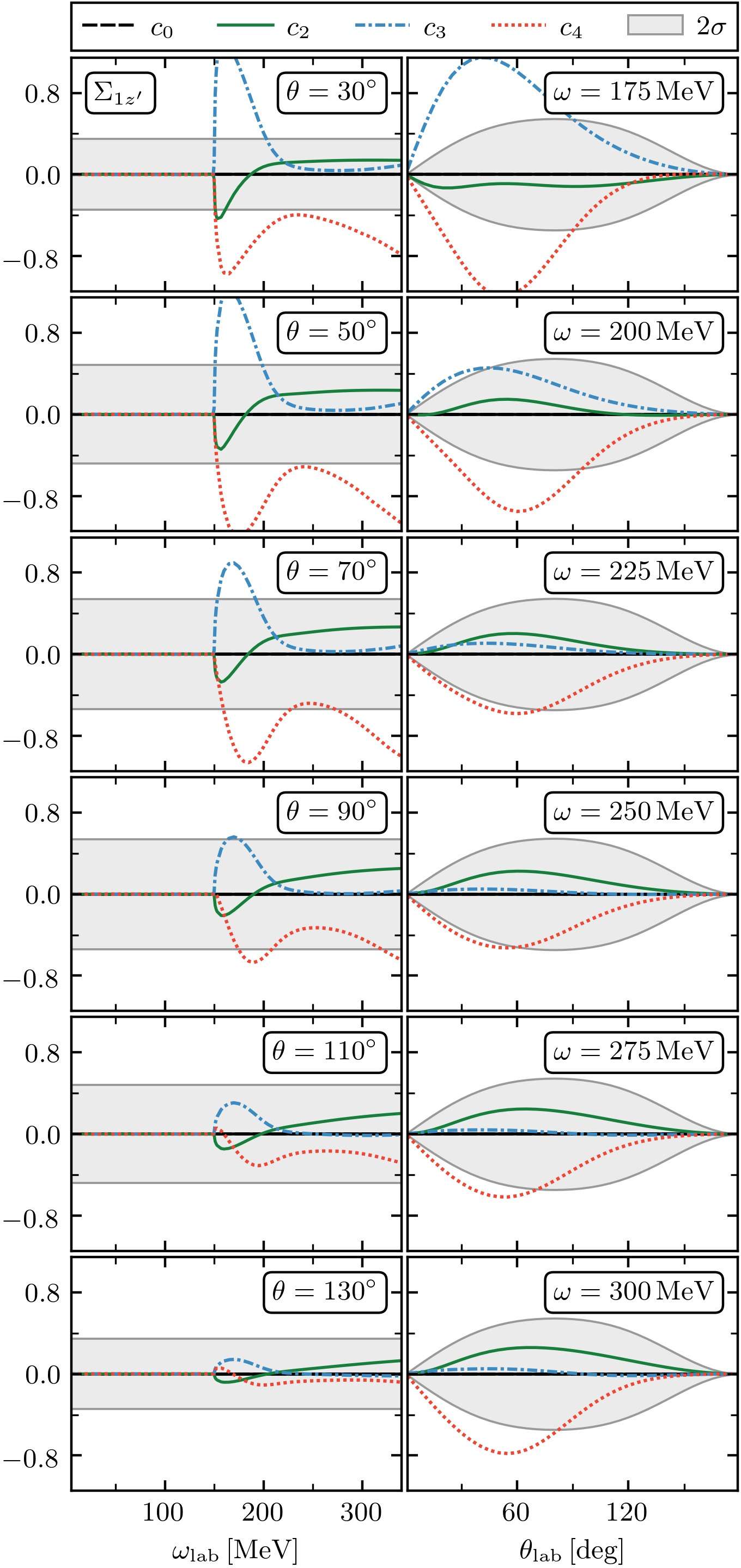}
    \caption{(Colour online) Coefficients for $\Sigma_{1z}$ and $\Sigma_{1z'}$.}
    \label{fig:coefficients_1z_and_1zp_slices}
\end{figure*}

\begin{figure*}[tb]
    \centering
    \includegraphics[width=0.495\textwidth]{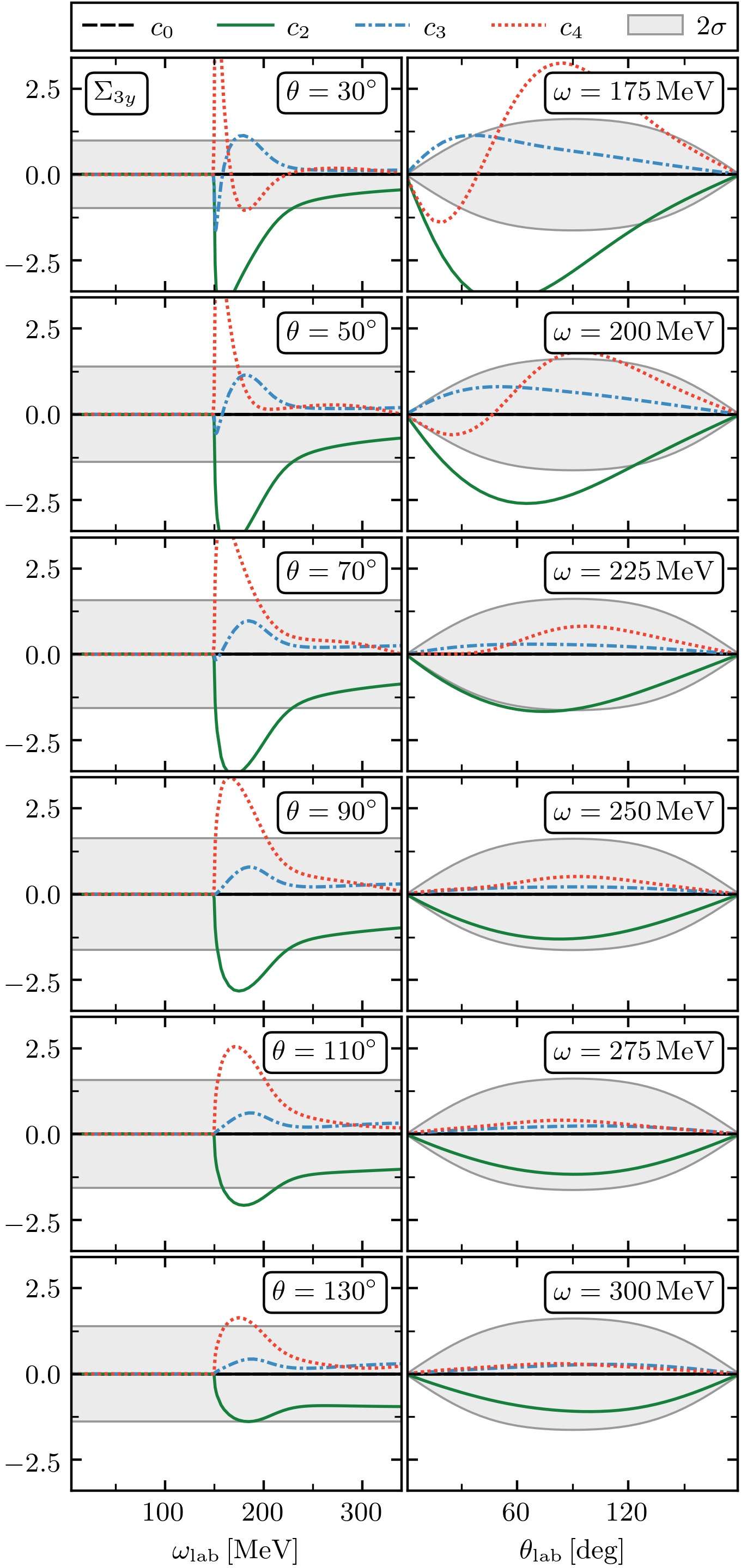}
    \includegraphics[width=0.495\textwidth]{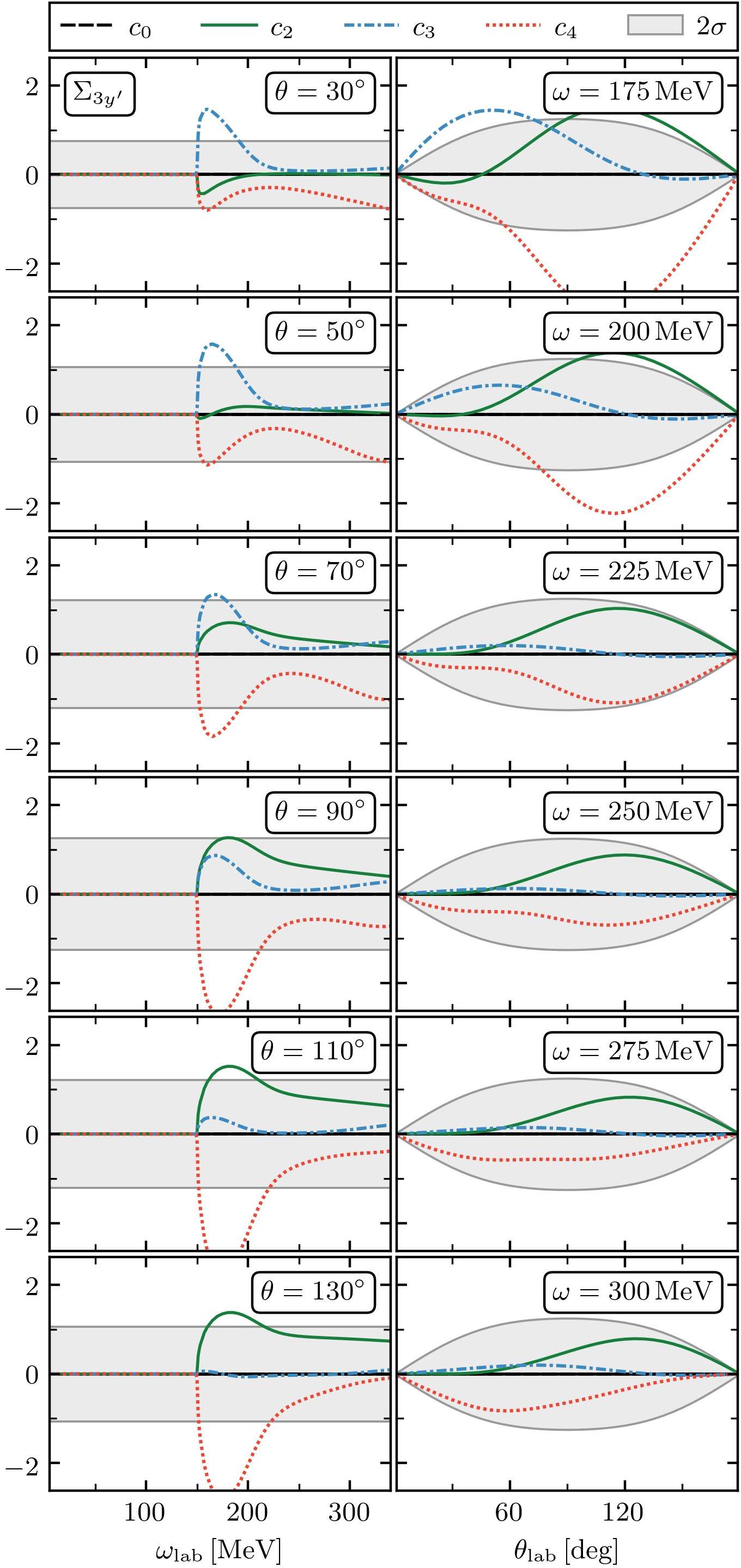}
    \caption{(Colour online) Coefficients for $\Sigma_{3y}$ and $\Sigma_{3y'}$.}
    \label{fig:coefficients_3y_and_3yp_slices}
\end{figure*}

In this section, we present plots of the sizes of the coefficients $c_i$ as
function of $\omega$ and slices of $\theta$, and vice versa, as
figs.~\ref{fig:coefficients_3_and_y_slices} to
\ref{fig:coefficients_3y_and_3yp_slices}. The coefficients for the cross
section are shown in fig.~\ref{fig:coefficients_dsg_slices} of
App.~\ref{sec:truncation_model_details_compton}.

\subsection{Proton Observables for an Uncorrelated Prior}

In this section, we present results for the analysis of proton observables
where we neglect the correlations between $\alpha_{E1}-\beta_{M1}$ and
$\gamma_{M-}$, \ie~we neglect the off-diagonal elements in
eq.~(\ref{eq:corrmatrix}).

\begin{figure*}[p]
    \centering
    \includegraphics[width=\textwidth]{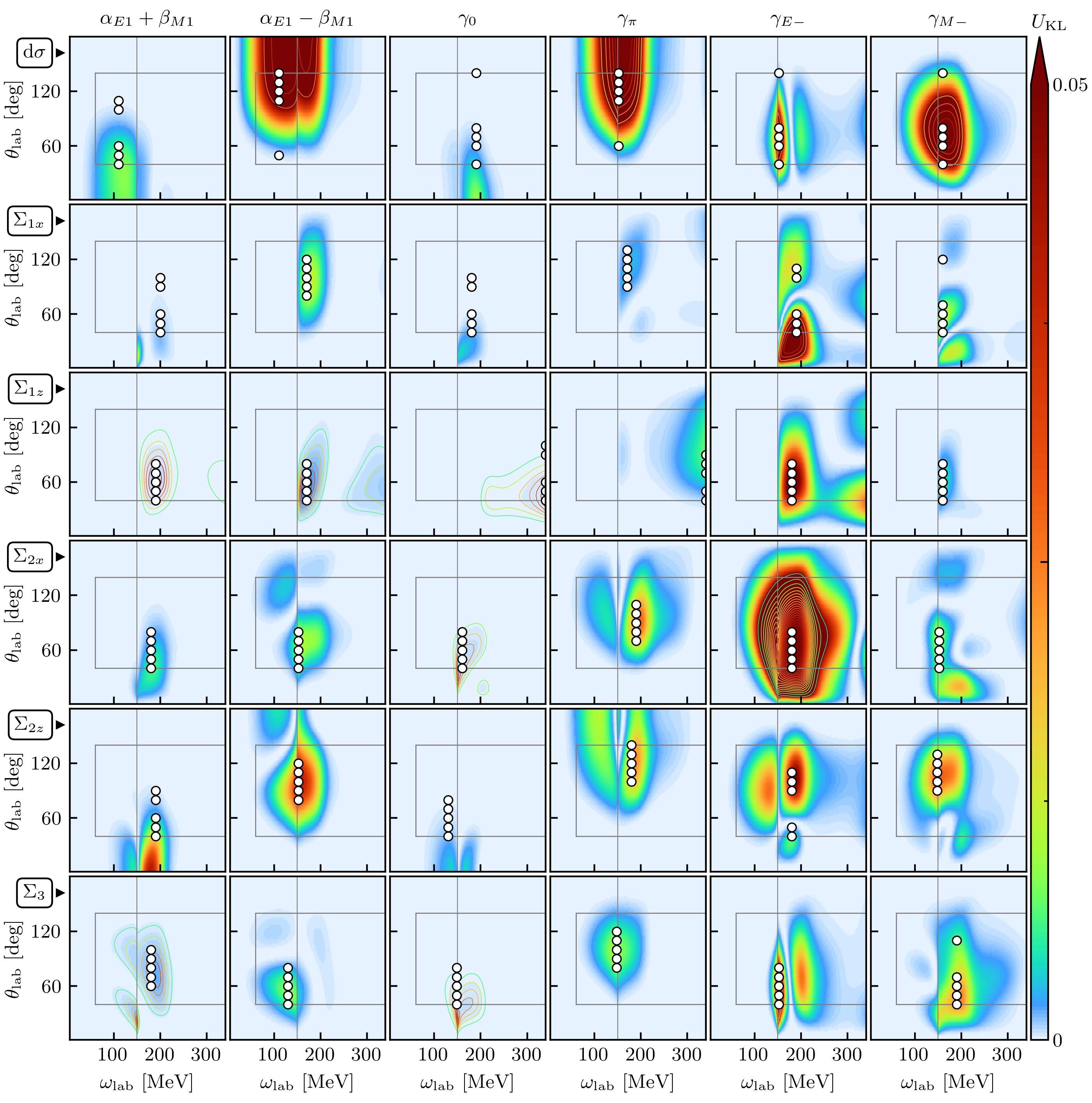}
    \caption{(Colour online) As in fig.~\ref{fig:utility_grid_subset_polarizabilities_observable_set_1}, with ``doable'' experimental precision, but for the case of an uncorrelated prior.
    }
    \label{fig:utility_grid_subset_polarizabilities_observable_set_1_uncorrelated}
\end{figure*}

\begin{figure*}[p]
    \centering
    \includegraphics[width=\textwidth]{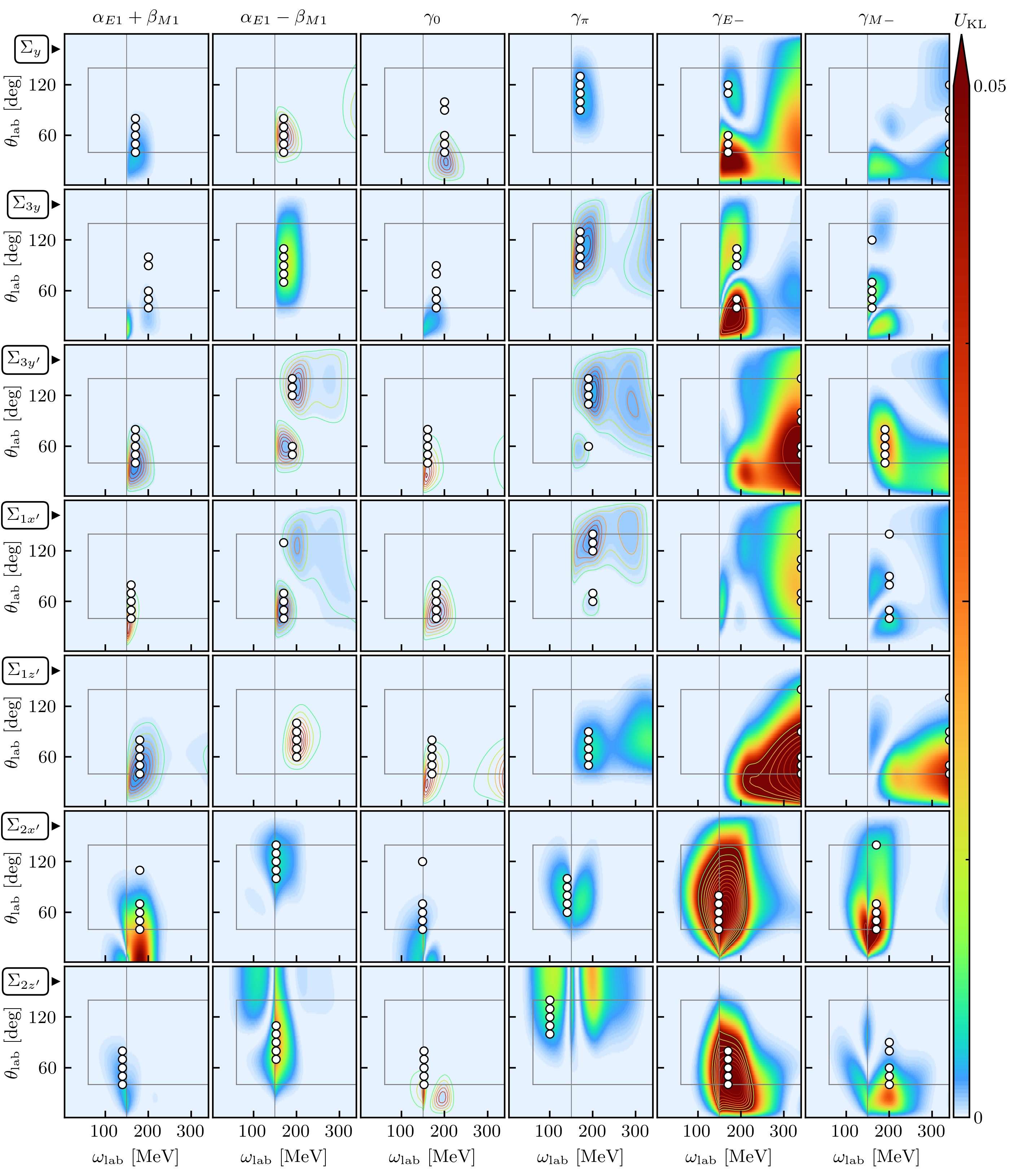}
    \caption{(Colour online) As in fig.~\ref{fig:utility_grid_subset_polarizabilities_observable_set_2}, with ``doable'' experimental precision, but for the case of an uncorrelated prior.
    }
    \label{fig:utility_grid_subset_polarizabilities_observable_set_2_uncorrelated}
\end{figure*}

\begin{figure*}[tbp]
    \centering
    \includegraphics[width=\textwidth]{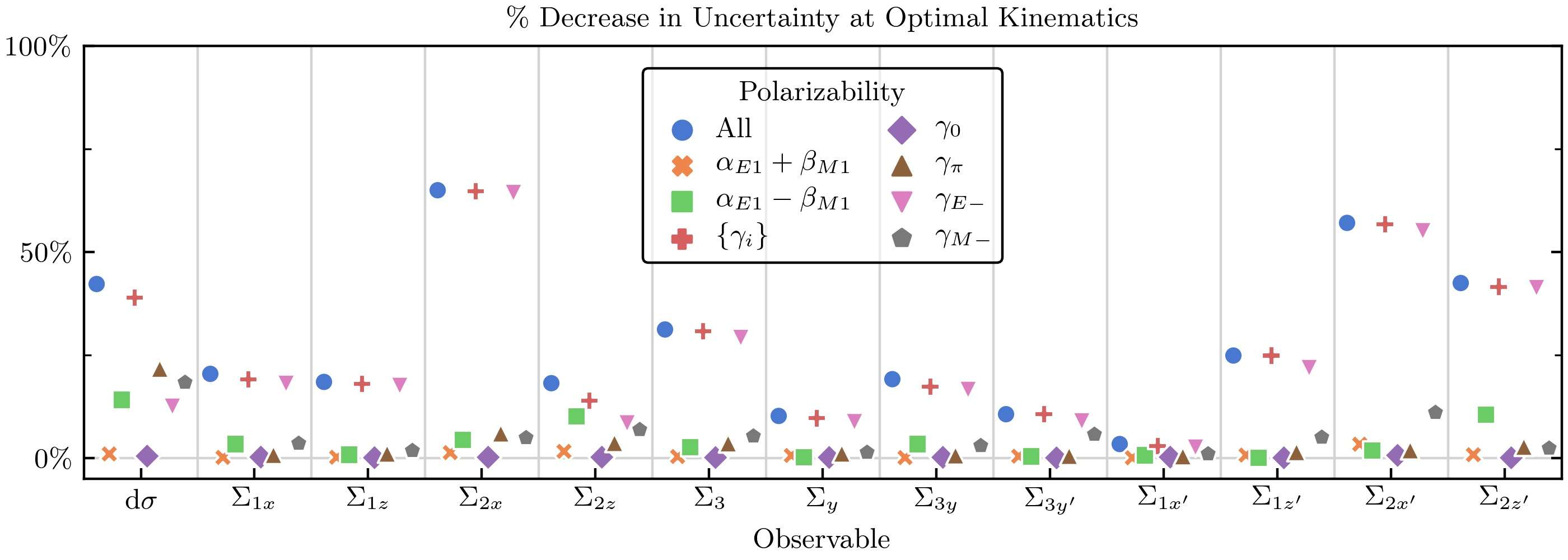}
    \caption{(Colour online) As in fig.~\ref{fig:shrinkage_per_subset}, but for the case of an uncorrelated prior.
    }
    \label{fig:shrinkage_per_subset_uncorrelated}
\end{figure*}

Figures~\ref{fig:utility_grid_subset_polarizabilities_observable_set_1_uncorrelated},~\ref{fig:utility_grid_subset_polarizabilities_observable_set_2_uncorrelated}, and~\ref{fig:shrinkage_per_subset_uncorrelated} are the analogs of figs.~\ref{fig:utility_grid_subset_polarizabilities_observable_set_1}, \ref{fig:utility_grid_subset_polarizabilities_observable_set_2}, and \ref{fig:shrinkage_per_subset} but for this uncorrelated prior.

\subsection{Proton Observables with Other Experimental Precision}

\begin{figure}[tbh]
    \centering
    \includegraphics[width=\linewidth]
    {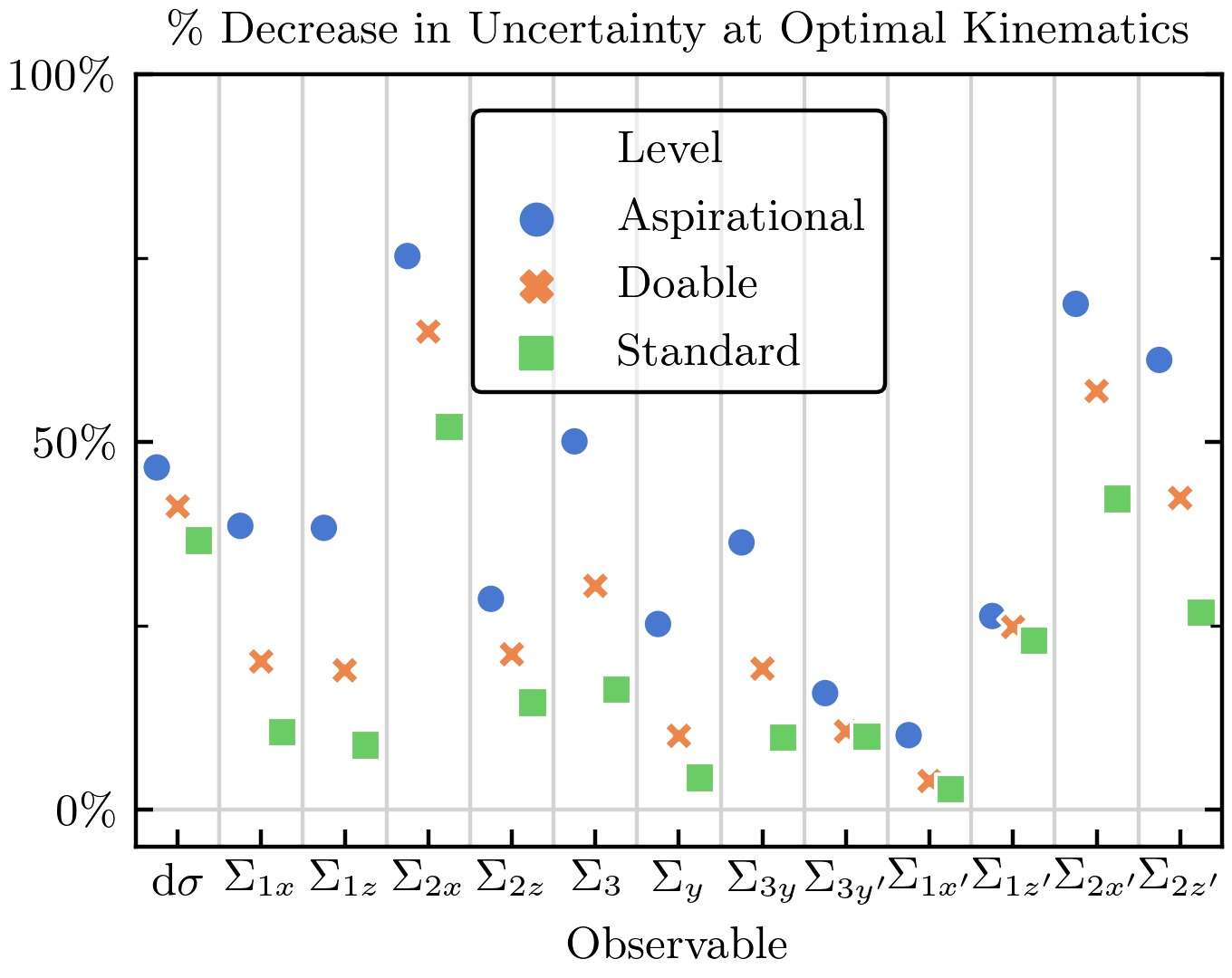}
    \caption{(Colour online) Comparison of the shrinkage power of the optimal designs for each precision level.
    Note that the optimal design likely differs between precision levels.
    }
    \label{fig:shrinkage_per_level}
\end{figure}

Figure \ref{fig:shrinkage_per_level} compares the maximal information gain in
each observable for ``standard'', ``doable'' and ``aspirational''
experiments. Not surprisingly, data with aspirational experimental error bars
are far superior to those with only standard ones. If theory errors were
absent, one would naively assume the information gain of the scenarios to
scale roughly like $1/\sqrt{\Delta\Sigma_i}$. This appears to be largely
fulfilled, except for $\Sigma_{2x^\prime}$ and, less noticeably,
$\Sigma_{3y^\prime}$.

\begin{figure*}[p]
    \centering
    \includegraphics[width=\textwidth]{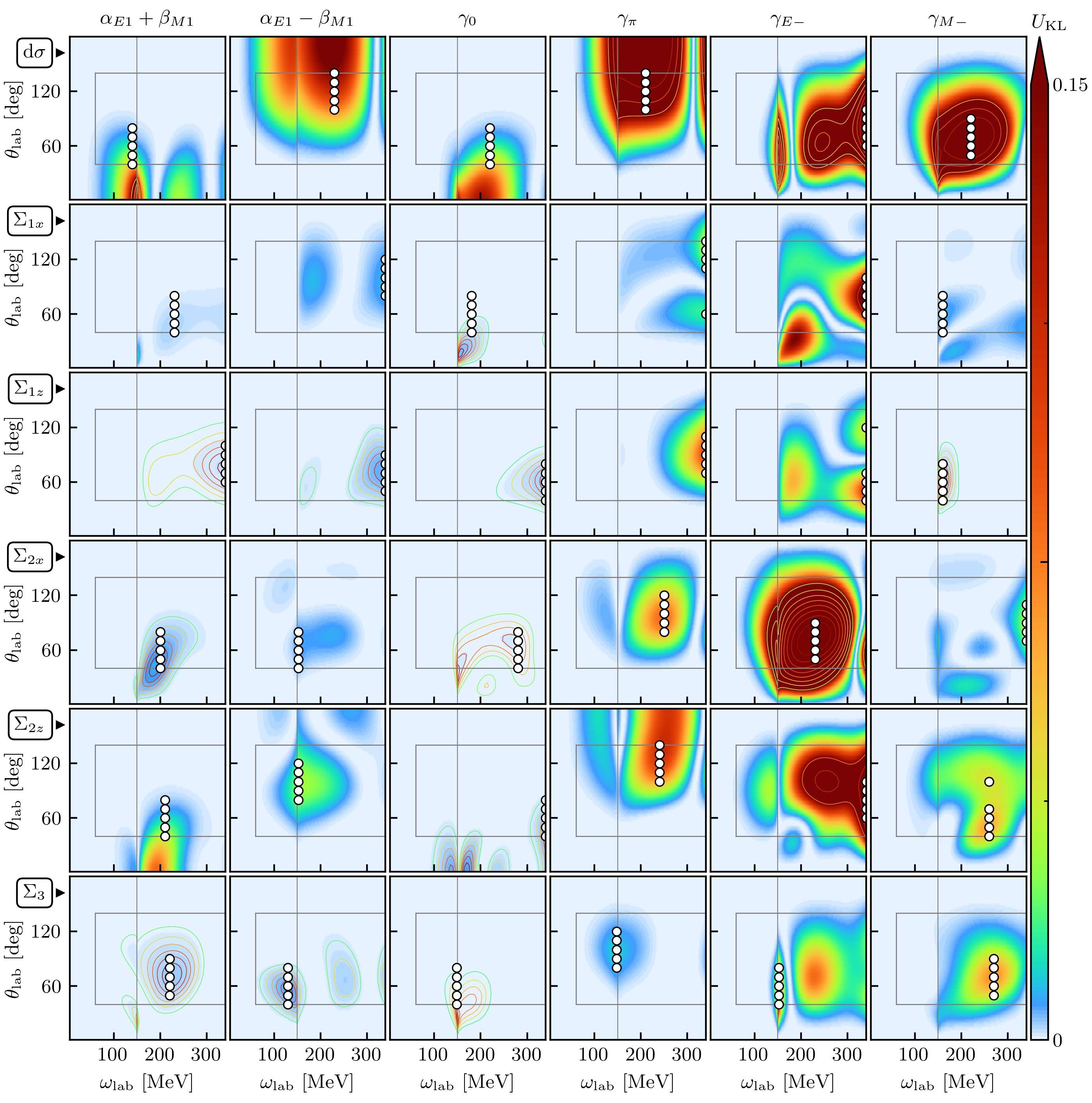}
    \caption{(Colour online) As in fig.~\ref{fig:utility_grid_subset_polarizabilities_observable_set_1}, with ``doable'' experimental precision, but without including an estimate of truncation error.
    }
    \label{fig:utility_grid_subset_polarizabilities_observable_set_1_doable_no-truncation}
\end{figure*}

\begin{figure*}[p]
    \centering
    \includegraphics[width=\textwidth]{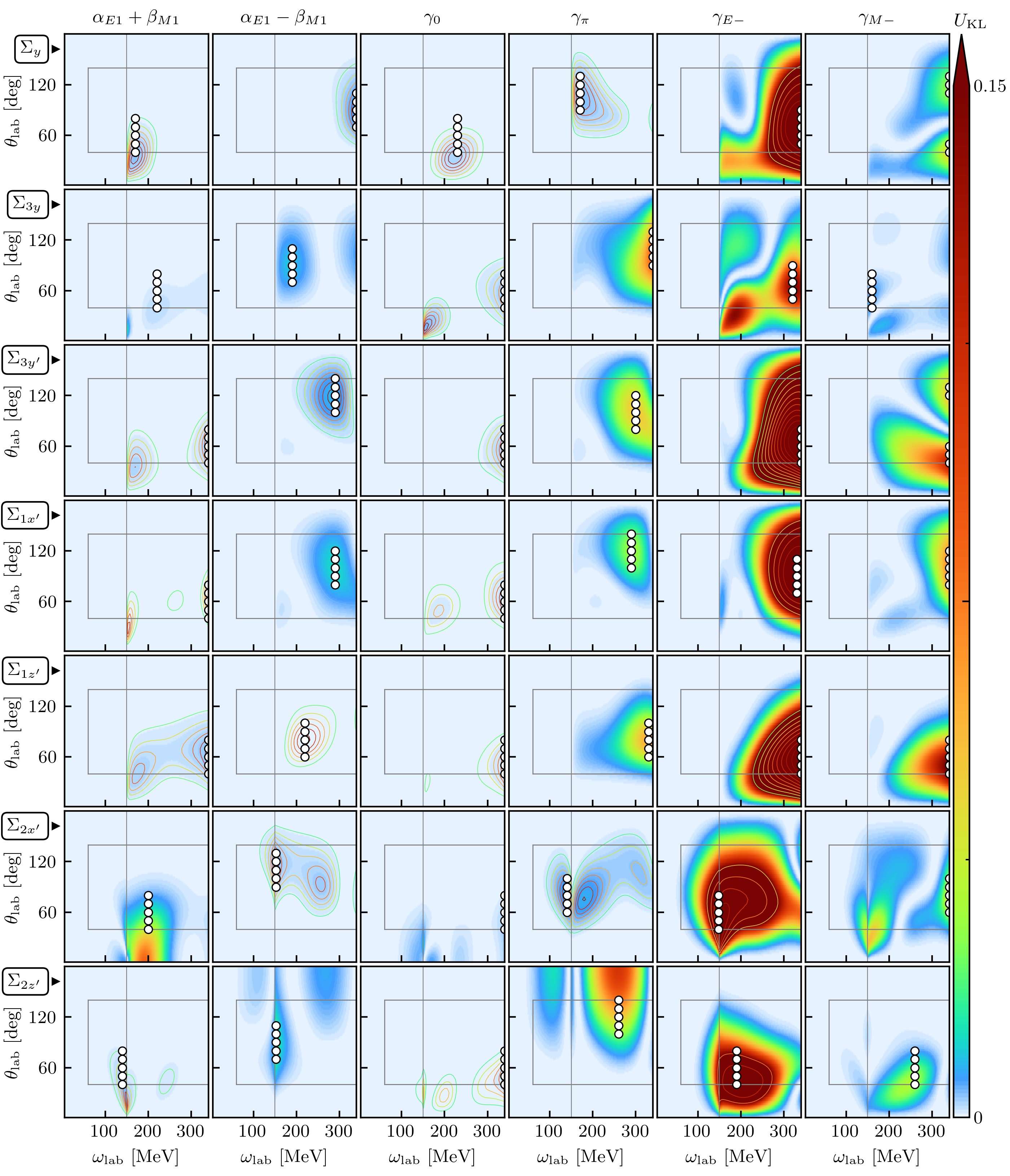}
    \caption{(Colour online) As in fig.~\ref{fig:utility_grid_subset_polarizabilities_observable_set_2}, with ``doable'' experimental precision, but without including an estimate of truncation error.
    }
    \label{fig:utility_grid_subset_polarizabilities_observable_set_2_doable_no-truncation}
\end{figure*}

\begin{figure*}[p]
    \centering
    \includegraphics[width=\textwidth]{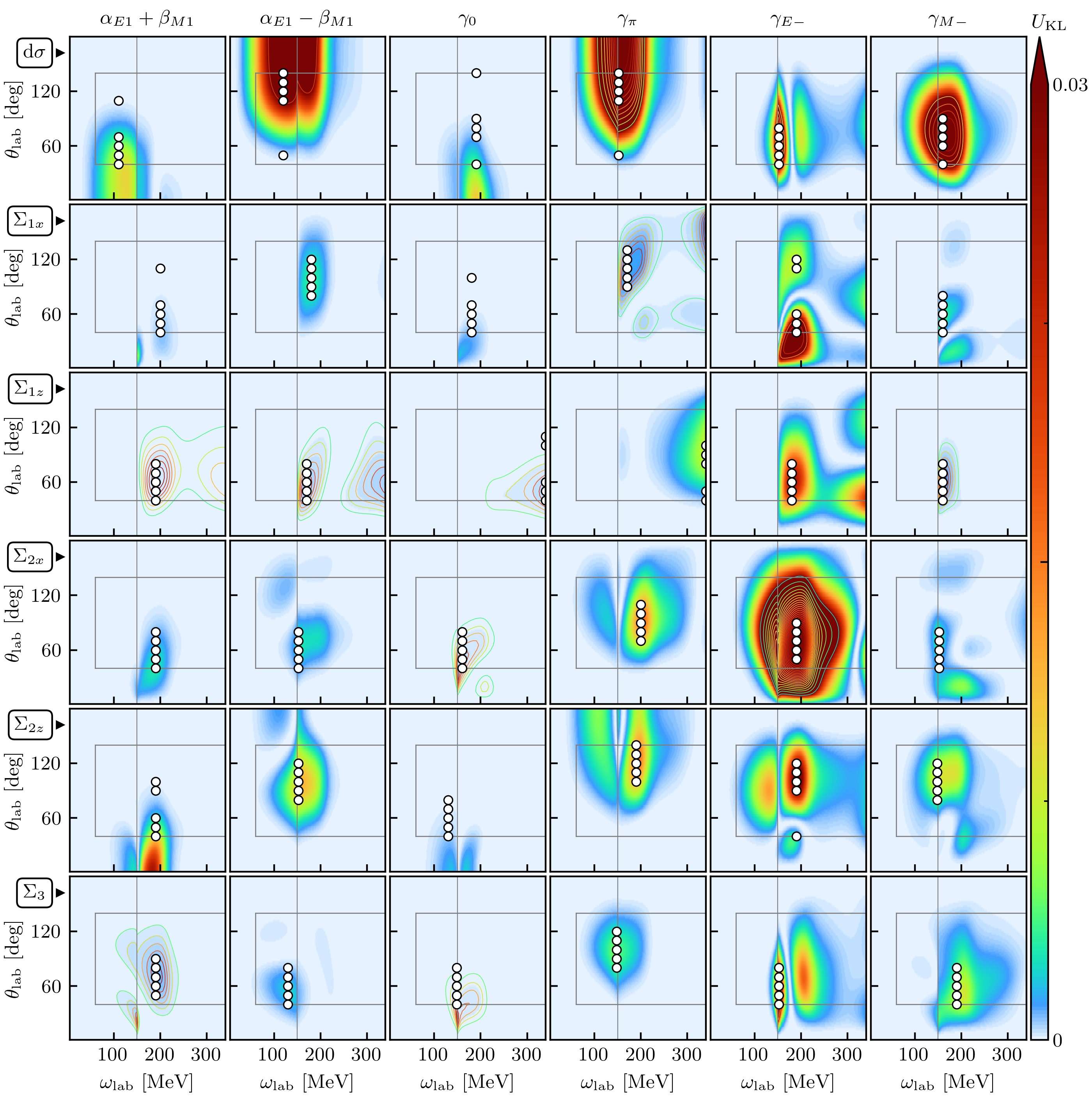}
    \caption{(Colour online) As in fig.~\ref{fig:utility_grid_subset_polarizabilities_observable_set_1}, but with the ``standard'' level of experimental precision; see table~\ref{tab:experimental_precision_levels}.
    }
    \label{fig:utility_grid_subset_polarizabilities_observable_set_1_standard}
\end{figure*}

\begin{figure*}[p]
    \centering
    \includegraphics[width=\textwidth]{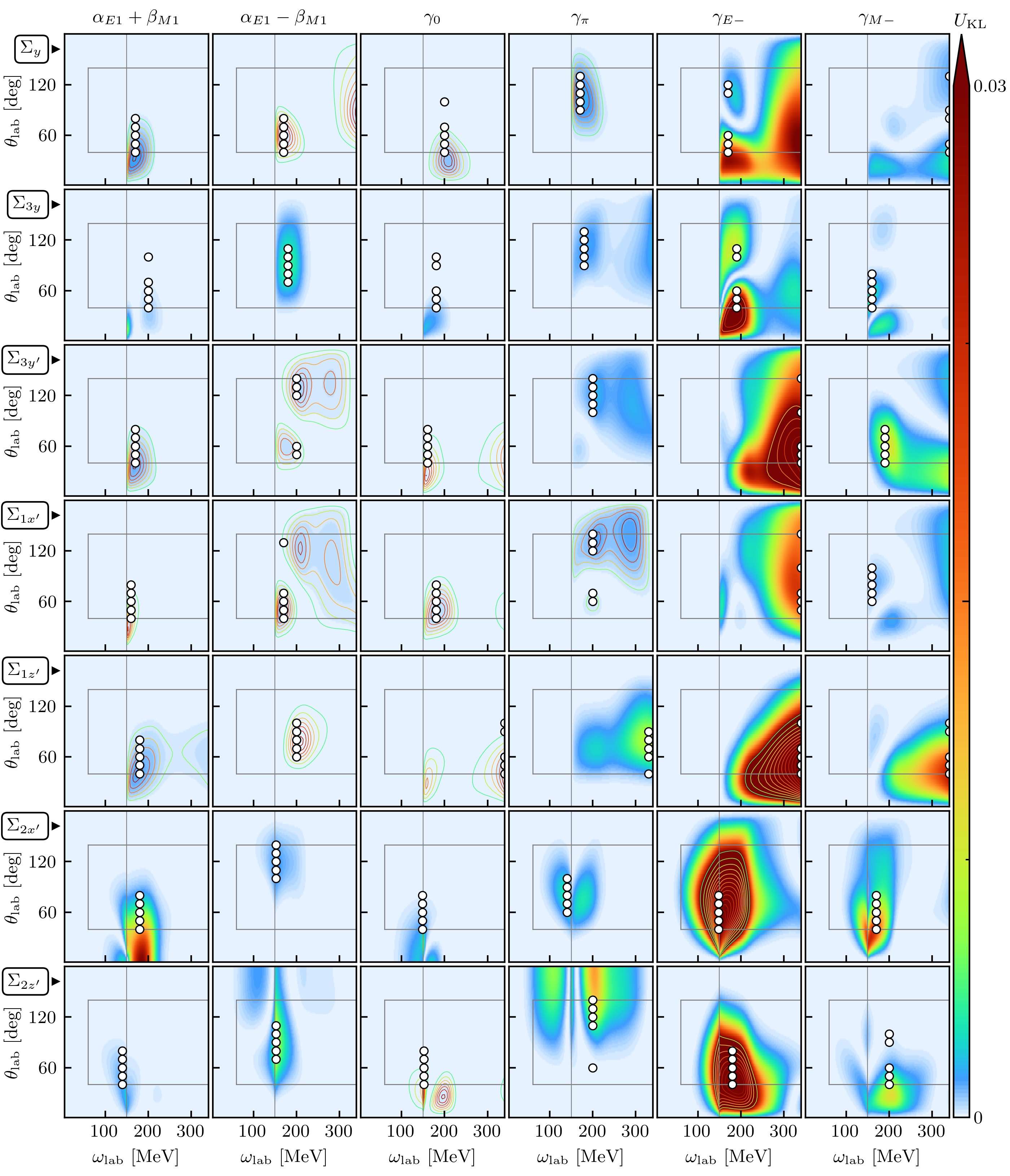}
    \caption{(Colour online) As in fig.~\ref{fig:utility_grid_subset_polarizabilities_observable_set_2}, but with the ``standard'' level of experimental precision; see table~\ref{tab:experimental_precision_levels}.
    }
    \label{fig:utility_grid_subset_polarizabilities_observable_set_2_standard}
\end{figure*}

\begin{figure*}[p]
    \centering
    \includegraphics[width=\textwidth]{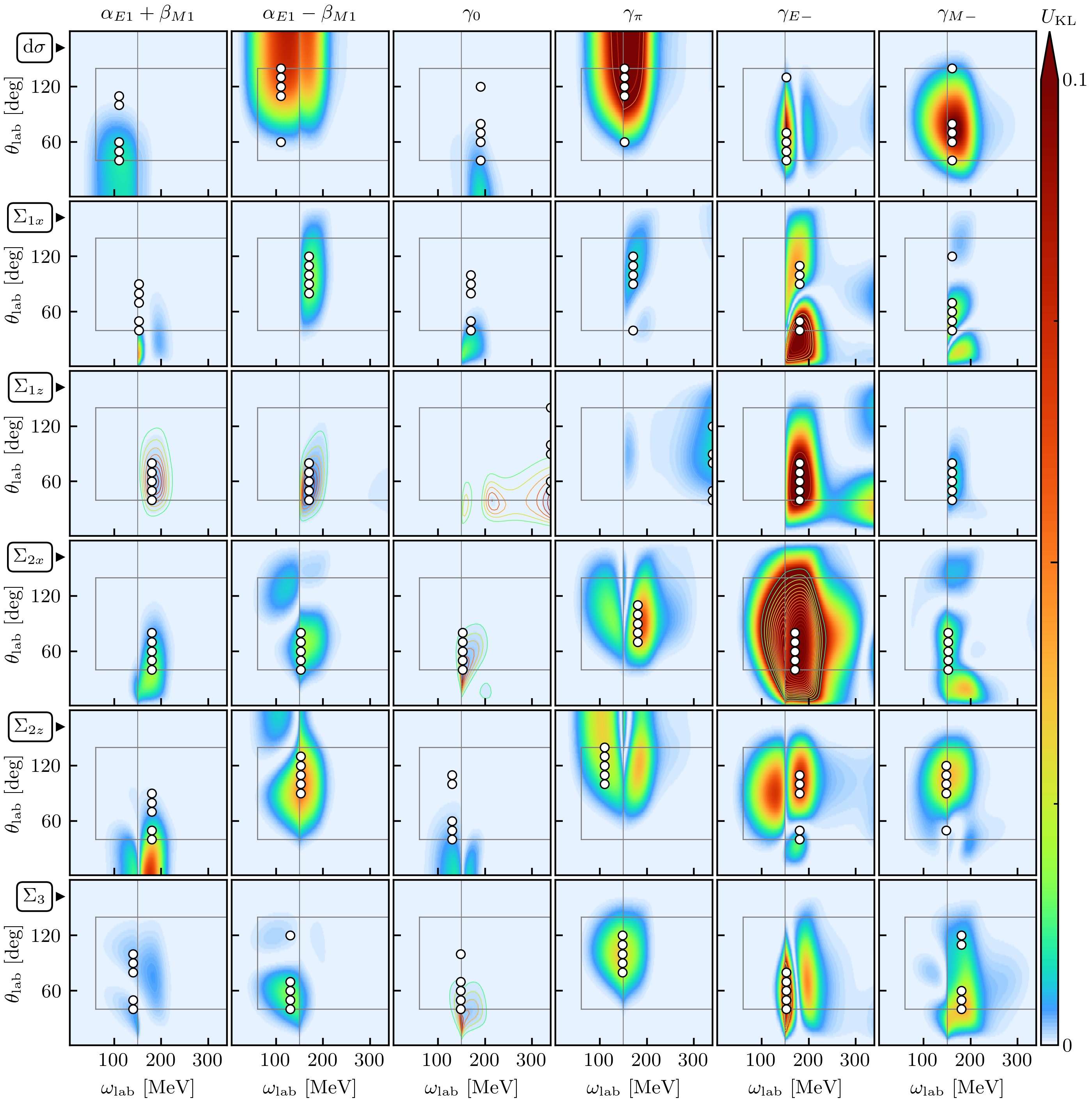}
    \caption{(Colour online) As in fig.~\ref{fig:utility_grid_subset_polarizabilities_observable_set_1}, but with the ``aspirational'' level of experimental precision; see table~\ref{tab:experimental_precision_levels}.
    }
    \label{fig:utility_grid_subset_polarizabilities_observable_set_1_aspirational}
\end{figure*}

\begin{figure*}[p]
    \centering
    \includegraphics[width=\textwidth]{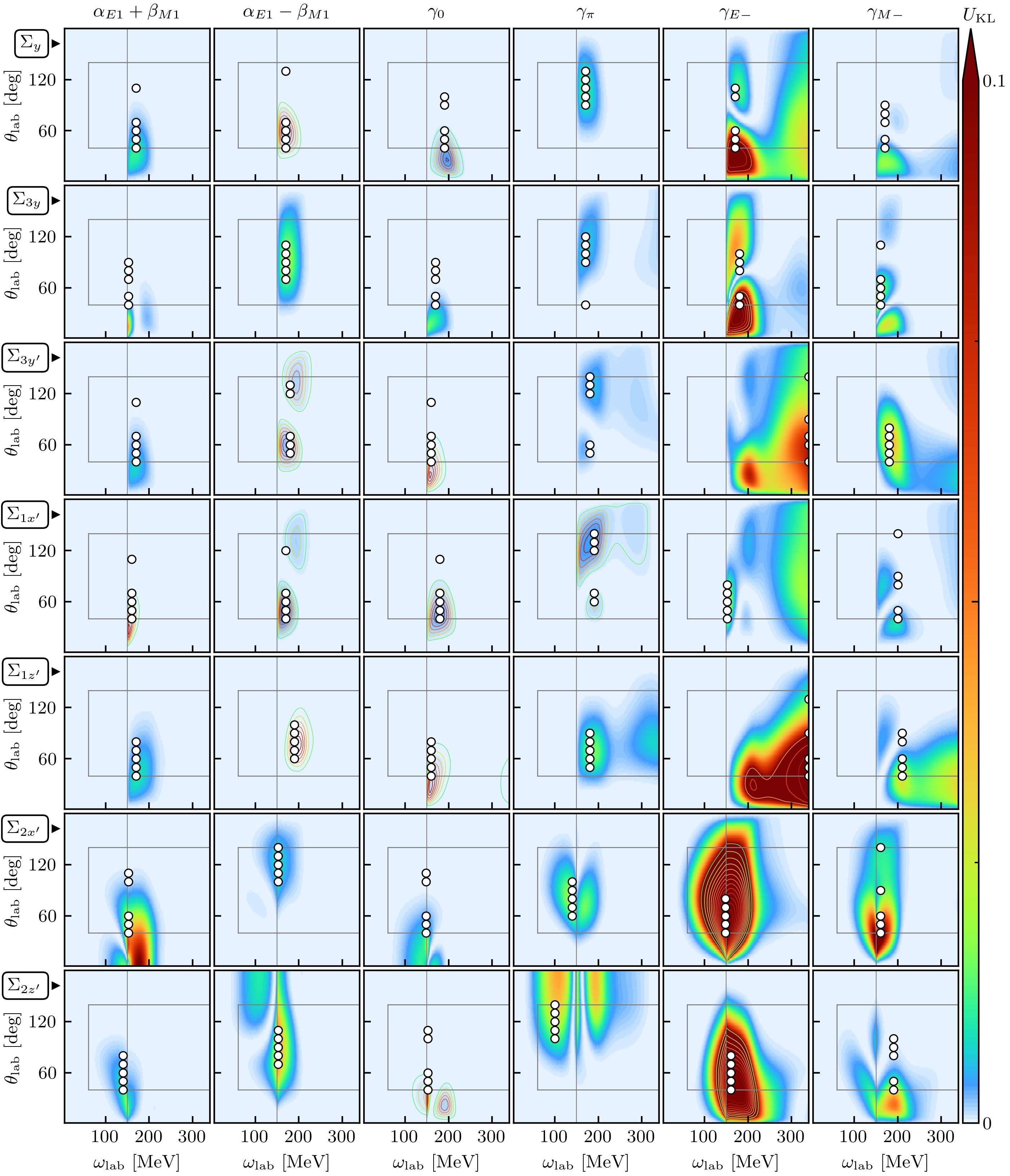}
    \caption{(Colour online) As in fig.~\ref{fig:utility_grid_subset_polarizabilities_observable_set_2}, but with the ``aspirational'' level of experimental precision; see table~\ref{tab:experimental_precision_levels}.
    }
    \label{fig:utility_grid_subset_polarizabilities_observable_set_2_aspirational}
\end{figure*}

\begin{figure*}[tbp]
    \centering
    \includegraphics[width=\textwidth]{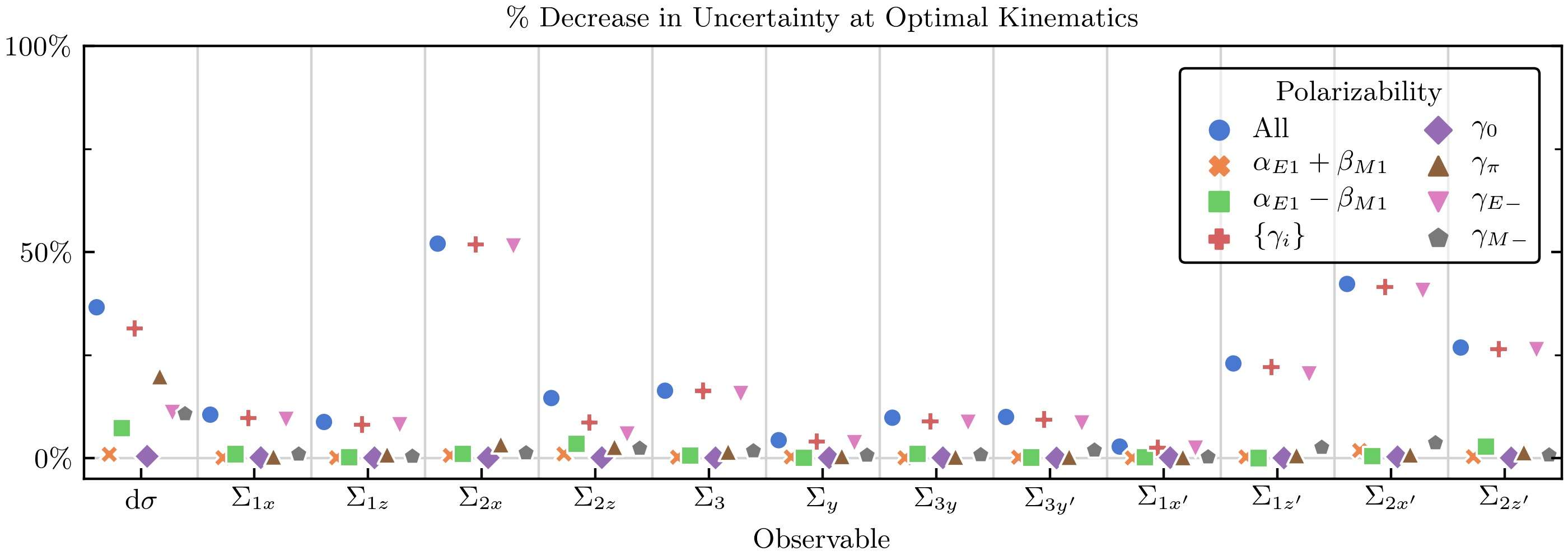}
    \caption{(Colour online) As in fig.~\ref{fig:shrinkage_per_subset}, but with the ``standard'' level of experimental precision; see table~\ref{tab:experimental_precision_levels}.
    }
    \label{fig:shrinkage_per_subset_standard}
\end{figure*}

\begin{figure*}[tbp]
    \centering
    \includegraphics[width=\textwidth]{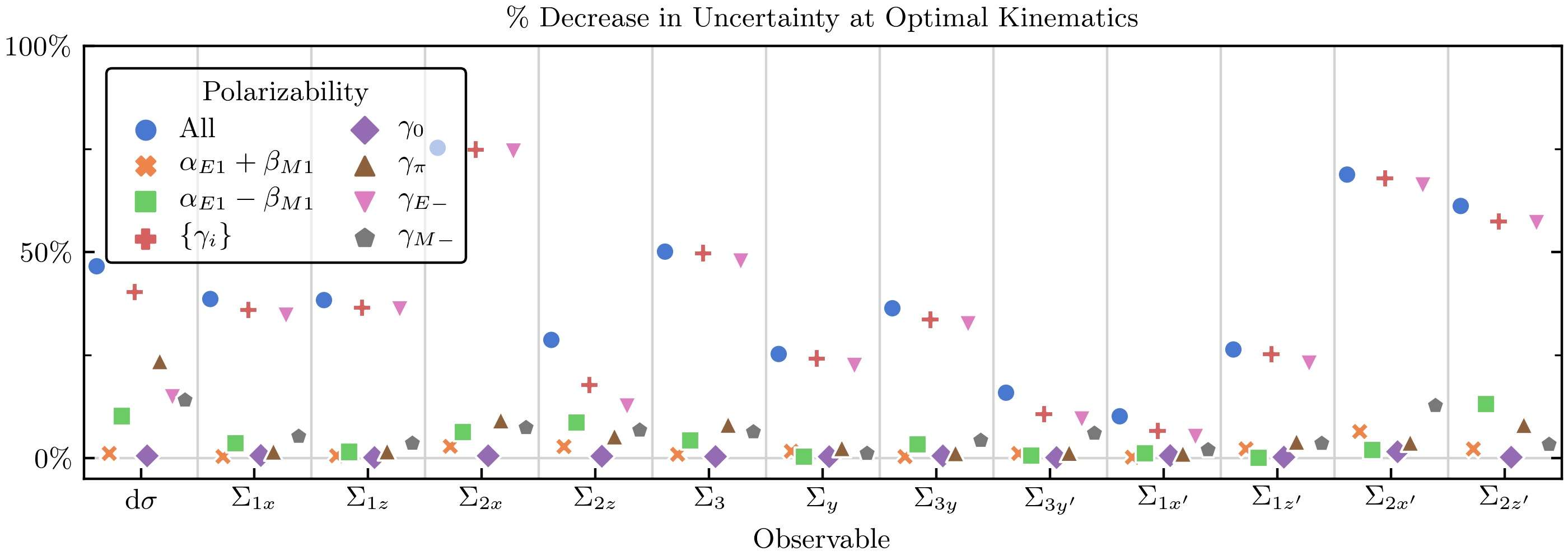}
    \caption{(Colour online) As in fig.~\ref{fig:shrinkage_per_subset}, but with the ``aspirational'' level of experimental precision; see table~\ref{tab:experimental_precision_levels}.
    }
    \label{fig:shrinkage_per_subset_aspirational}
\end{figure*}

Figures~\ref{fig:utility_grid_subset_polarizabilities_observable_set_1_doable_no-truncation}
and~\ref{fig:utility_grid_subset_polarizabilities_observable_set_2_doable_no-truncation}
are the analogs of
figs.~\ref{fig:utility_grid_subset_polarizabilities_observable_set_1}
and~\ref{fig:utility_grid_subset_polarizabilities_observable_set_2} in the
main text, but this time without any \chiEFT\ truncation error included.
Figures~\ref{fig:utility_grid_subset_polarizabilities_observable_set_1_standard},~\ref{fig:utility_grid_subset_polarizabilities_observable_set_2_standard},
and~\ref{fig:shrinkage_per_subset_standard} are the analogs of
figs.~\ref{fig:utility_grid_subset_polarizabilities_observable_set_1},
\ref{fig:utility_grid_subset_polarizabilities_observable_set_2}, and
\ref{fig:shrinkage_per_subset} but for the ``standard'' level of precision,
rather than the ``doable'' one employed for results in the main text.
Figures~\ref{fig:utility_grid_subset_polarizabilities_observable_set_1_aspirational},~\ref{fig:utility_grid_subset_polarizabilities_observable_set_2_aspirational},
and~\ref{fig:shrinkage_per_subset_aspirational} show the corresponding results
for the ``aspirational'' precision level.

\clearpage

\subsection{Neutron Observables}

Here we show the corresponding results for the neutron observables.  Because
such experiments are difficult, only the ``standard'' level of precision is
used (see table~\ref{tab:experimental_precision_levels}). Even this is likely
optimistic for such measurements, as discussed in the main text.

\begin{figure*}[!htb]
    \centering
    \includegraphics[width=\textwidth]{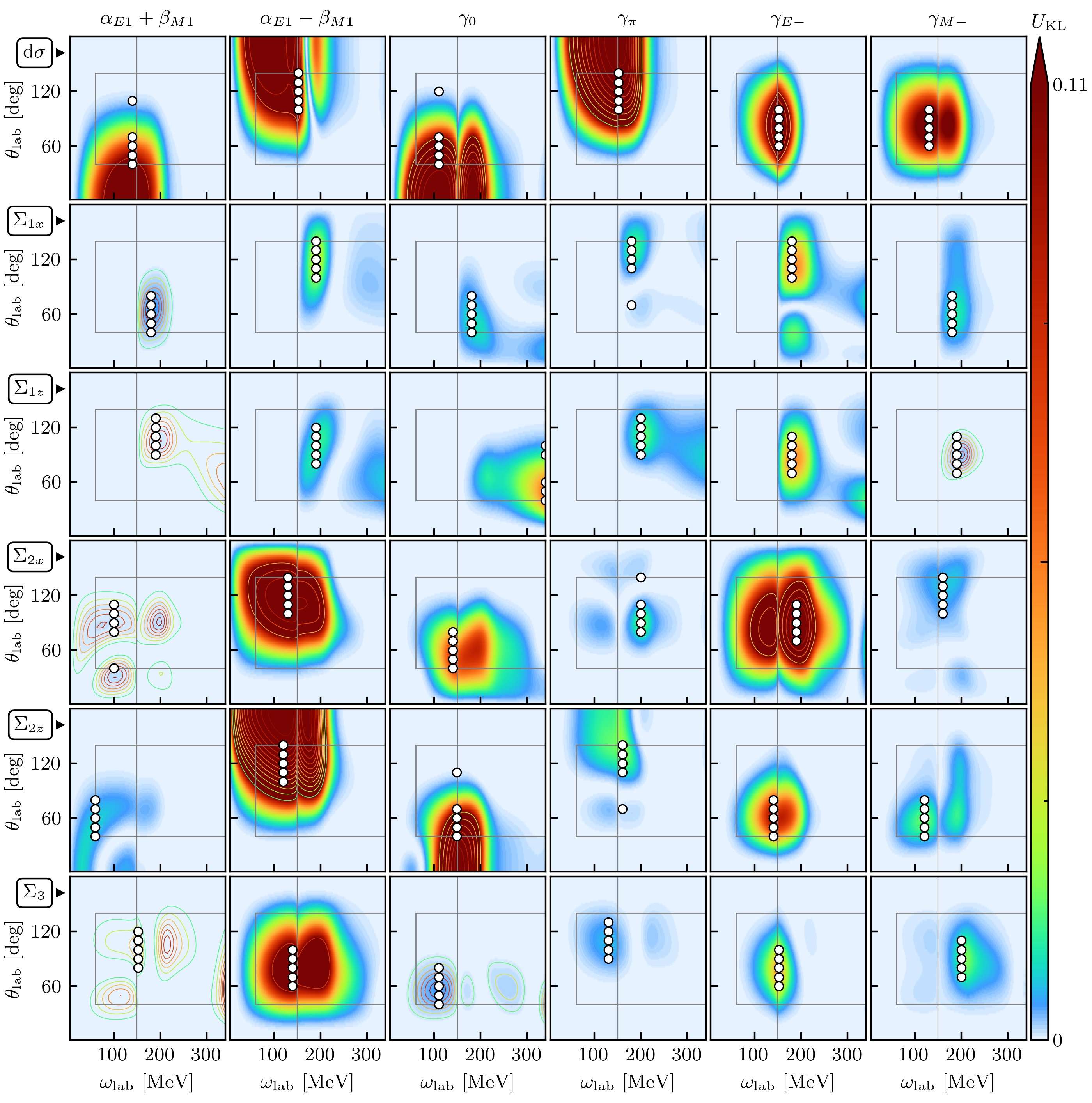}
    \caption{(Colour online) As in fig.~\ref{fig:utility_grid_subset_polarizabilities_observable_set_1}, but for neutron observables with the ``standard'' level of experimental precision; see table~\ref{tab:experimental_precision_levels}.
    }
    \label{fig:utility_grid_subset_polarizabilities_observable_set_1_standard_neutron}
\end{figure*}

\begin{figure*}[p]
    \centering
    \includegraphics[width=\textwidth]{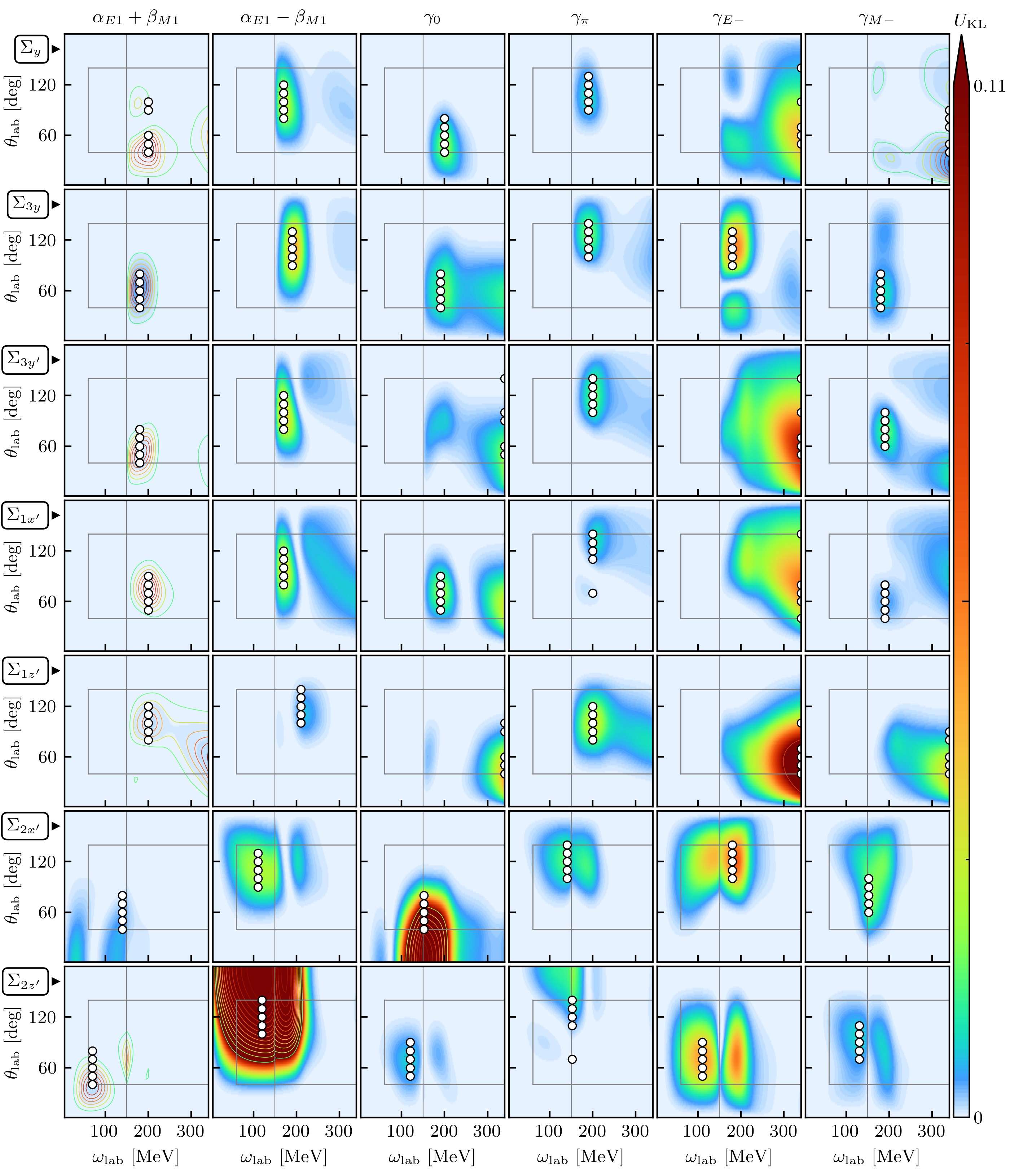}
    \caption{(Colour online) As in fig.~\ref{fig:utility_grid_subset_polarizabilities_observable_set_2}, but for neutron observables with the ``standard'' level of experimental precision; see table~\ref{tab:experimental_precision_levels}.
    }
    \label{fig:utility_grid_subset_polarizabilities_observable_set_2_standard_neutron}
\end{figure*}

\begin{figure*}[tbp]
    \centering
    \includegraphics[width=\textwidth]{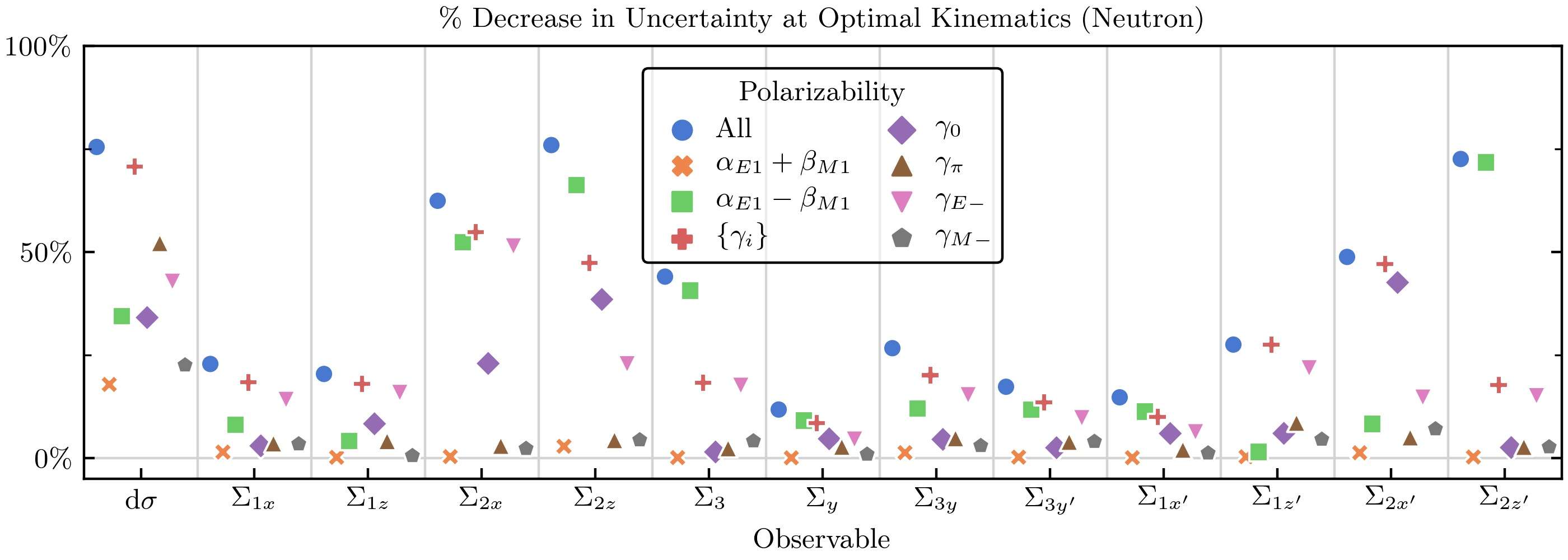}
    \caption{(Colour online) As in fig.~\ref{fig:shrinkage_per_subset}, but
      for neutron observables with the ``standard'' level of experimental
      precision; see table~\ref{tab:experimental_precision_levels}.  }
    \label{fig:shrinkage_per_subset_standard_neutron}
\end{figure*}

\clearpage
\bibliography{bayesian_refs,more-bayesian-refs}

\end{document}